\documentclass{aa} 

 \newcommand{\gc}{\bf{}}
 \renewcommand{\gc}{} 

\usepackage{graphicx}
\usepackage{amsmath}
\usepackage{txfonts}
\usepackage{epsfig}
\usepackage{ulem}
\usepackage[]{natbib}
\usepackage{tabularx}

\citeindextrue
\newcolumntype{R}{>{\raggedleft\arraybackslash}X}

\bibpunct{(}{)}{;}{a}{}{,} 
\usepackage{subcaption}
\begin{document}
\title{The structure of disks around Herbig Ae/Be stars as traced by CO ro-vibrational emission.\thanks{Based on observations collected at the European Southern Observatory, Paranal, Chile. (Program IDs 079.C-0349(A) and 081.C-0833(A))}}

\author{G. van der Plas\inst{1,2, 3}
  \and M. E. van den Ancker\inst{3}
  \and L.B.F.M. Waters\inst{2,4}
  \and C. Dominik\inst{2,5}
}

\offprints{G. van der Plas, \email{info@gerritvanderplas.com}}

\date{Received \texttt{26 09 2014} / Accepted \texttt{25 11 2014}}

\institute{ Departamento de Astronom\'{\i}a, Universidad de Chile, Casilla 36-D, Santiago, Chile, 
  \and Astronomical Institute Anton Pannekoek, University of Amsterdam, PO box 94249, 1090 GE, Amsterdam, The Netherlands
  \and European Southern Observatory, Karl-Schwarzschild-Str.2, D 85748 Garching bei M\"unchen, Germany
  \and SRON Netherlands Institute for Space Research, Sorbonnelaan 2, 3584 CA Utrecht, The Netherlands 
  \and Institute for Astrophysics, Radbout University, Heyendaalseweg 135, 6525 AJ Nijmegen, The Netherlands
}


  \abstract
   {}
  {We study the emission and absorption of CO ro-vibrational lines in the spectra of intermediate mass pre-main-sequence stars with the aim to determine both the spatial distribution of the CO gas and its physical properties. We also aim to correlate CO emission properties with disk geometry.}
   {Using high-resolution spectra containing fundamental and first overtone CO ro-vibrational emission, observed with CRIRES on the VLT, we probe the physical properties of the circumstellar gas by studying its kinematics and excitation conditions.}
   {We detect and spectrally resolve CO fundamental ro-vibrational emission in 12 of the 13 stars observed, and in two cases in absorption.} 
   {Keeping in mind that we studied a limited sample, we find that the physical properties and spatial distribution of the CO gas correlate with disk geometry. Flaring disks show highly excited CO fundamental emission up to v$_u$ = 5, while self-shadowed disks show CO emission that is not as highly excited. Rotational temperatures range between ~250-2000 K. The $^{13}$CO rotational temperatures are lower than those of $^{12}$CO.   The vibrational temperatures in self-shadowed disks are similar to or slightly below the rotational temperatures, suggesting that thermal excitation or IR pumping is important in these lines. In flaring disks the vibrational temperatures reach as high as 6000 K,  suggesting fluorescent pumping. Using a simple kinematic model we show that the CO inner radius of the emitting region is $\approx$10 au for flaring disks and $\leq$ 1 au for self-shadowed disks.
Comparison with hot dust and other gas tracers shows that CO emission from the disks around Herbig Ae/Be stars, in contrast to T Tauri stars, does not necessarily trace the circumstellar disk up to, or inside the dust sublimation radius, $R_{subl}$. Rather, the onset of the CO emission starts from $\approx$R$_{subl}$ for self-shadowed disks, to tens of R$_{subl}$ for flaring disks. It has recently been postulated that group I Herbig stars may be transitional disks and have gaps. Our CO observations are qualitatively in agreement with this picture. We identify the location of the CO emission in these group I disks with the inner rim of the outer disk after such a gap, and suggest that the presence of highly vibrationally excited CO emission and a mismatch between the rotational and vibrational temperature may be a proxy for the presence of moderately sized disk gaps in Herbig Ae/Be disks. }

   \titlerunning{CO rovibrational line survey of HAeBe disks}
   \maketitle

\section{Introduction}

Accretion disks around young stars are currently one of the most widely studied phenomena in astronomy. These gas- and dust-rich disks form as a natural consequence of the conservation of angular momentum in a collapsing cloud core, from which the star forms. As the disk itself evolves, planets are born in the dense dusty midplane, hence they are often called protoplanetary (PP) disks. The evolution, and ultimately the method of dispersion of these passive disks, holds the key to many open questions related to planet formation \citep[e.g.][and references therein]{alexander-disk-review}.

Protoplanetary disks have historically been studied predominantly using dust, which presents only 1$\%$ of the total circumstellar mass but carries most of the opacity. Whereas dust emission dominates the infrared (IR) spectra of PP disks, and plays an important role in the evolution of PP disks, gas and dust do not necessarily evolve co-spatially or on similar timescales. Hence, gas observations are both complementary to, and essential for, our understanding of PP disks.  With the advent of 10-meter class telescopes, increasing amounts of gas tracers have become available with which to study the PP disks. Each of these tracers is sensitive to a different set of physical parameters, and we can tie these parameters to a specific radial location in the disk by using the kinematic information in the line profiles: from inside the dust sublimation radius ($\approx$ 0.1 au) up to far in the outer disk (100s of au), and from the mid-plane to the disk atmosphere. For a review on the diagnostic power of gas tracers see e.g. \citet{2010EM&P..106...71C}.

{Carbon monoxide (CO) is the second most abundant molecule in the universe after H$_2$. In contrast to H$_2$, the ro-vibrational transitions of CO are much stronger, and hence CO is easier to detect. Carbon monoxide is commonly seen in emission from disks and is a versatile tracer because of its many rotational and ro-vibrational transitions that are sensitive to a wide range of physical environments. The rotational ($\Delta$v = 0, sub-mm CO lines \citep[e.g.][]{1989ApJ...344..915W, 1993ApJ...402..280B, 1997ApJ...490..792M, 2000ApJ...529..391M} trace the cold outer disk and disk midplane. Hot (T $>$ 2000 K) gas close to the star enables the ro-vibrational first overtone ($\Delta$v=2, 2.3 $\mu$m, \citep[e.g.][]{1983ApJ...275..201S, 1989ApJ...345..522C, 1993ApJ...411L..37C, 1996ApJ...462..919N}) transitions to be detected, and the warm (T $\approx$ 1000 K) gas in the inner disk and disk surface are excited via the ro-vibrational fundamental transitions \citep[$\Delta$v = 1, 4.6 $\mu$m, e.g.][]{1990ApJ...363..554M, 2003ApJ...589..931N, 2004ApJ...606L..73B,  2007ApJ...659..685B, 2009ApJ...699..330S, 2011A&A...527A.119B, 2013ApJ...770...94B}.

Many recent studies of PP disks have focused on the so-called Herbig Ae/Be (HAeBe) stars, because their disks are often more massive and more extended (and hence brighter) than the disks around their lower-mass counterparts, the T Tauri stars. In recent years, it has become clear that dust in passive disks around these HAeBe stars comes in two varieties: \textit{flaring} (group I) and \textit{self-shadowed} (group II) disks \citep{2001A&A...365..476M}. This classification is based on the topology of the dust in the PP disk, and the amount of small dust grains in direct view of the central star that can reprocess stellar light to IR photons, and so cause the infrared excess. Disks around HAeBe stars have a 'puffed up' inner rim that casts a shadow over the outer disk. In flaring disks, the dust in the outer disk eventually rises out of this shadow, while the dust in self-shadowed disks does not. \citet{2013A&A...555A..64M} found that the absence of silicate emission in the infrared spectrum of some group I sources can only be explained by the presence of large gaps in the disks of these sources, leading these authors to postulate that group I and group II sources do not form an evolutionary sequence, but may instead be separate evolutionary pathways of the PP disk. At present it is unclear whether the gas has a similar morphology as the dust for both Meeus et al. group I and group II sources, or whether there are qualitative differences in the behavior of the gas between group I's and group II's as well.

In this paper, we use the CO fundamental and first overtone ro-vibrational transitions as a tool to trace the inner disk (inner tens of au) regions of thirteen PP disks surrounding young intermediate-mass stars. We present the studied sample, the observations, and the data reduction method in Sections \ref{sec:sample}, \ref{sec:obs} and \ref{sec:method}, and show the results in Section \ref{sec:results}. In Sections \ref{sec:whereco?} and \ref{sec:discussion} we discuss the location of the emitting CO gas and the  found correlations between disk shape and CO excitation conditions as a probe  for the influence of disk structure on the emission properties of the molecular gas. We summarize our conclusions in Section \ref{sec:conclusion}.

\section{Sample description}
\label{sec:sample}

\begin{table*}
\small
\begin{minipage}[t]{\textwidth}
\caption{Astrophysical parameters of the programme stars. The references for the values for the effective temperature of the central star $\log T_{\rm eff}$, the observed bolometric luminosity log $L_{\rm bol}$, and the distance $d$ can be found in \citet{2005A&A...436..209A} and references therein, unless indicated otherwise. $^\star$ Inclination in degrees from face-on. $^\clubsuit$ Derived from the PAH 8.6\,$\mu$m image of \citet{2006Sci...314..621L}.  $^\dag$ We do not classify this source in group I/II (Section \ref{sec:sample}). $^\ddag$ Ambiguity about the distance exists \citep{2006A&A...456.1045B}} 
\label{table:stellar_parameters}
\centering
\renewcommand{\footnoterule}{}  
\noindent\begin{tabularx}{\columnwidth}{@{\extracolsep{\stretch{1}}}*{15}{l}@{}}

\hline \hline
ID & Name &  Sp. Type & Group & $\log T_{\rm eff}$ & log $L_{\rm bol}$ & M            & $d$ &  i$^\star$    &  PA       \\
     &   &          &       & log [K]          & log [$L_{\odot}$]  & [$M_{\odot}$]& [pc]& [$^{\circ}$]  & [$^{\circ}$] \\
\hline

1 &HD\,100546 &  B9Vne    & I   & 4.02 & 1.62          & 2.4$\pm 0.1^{~ \it o}$      & 103$^{+7, {~ \it y}}_{-6}$        & 42$\pm 5 ^{~{\it d}}$     & 145 $\pm 5 ^{{~ \it d}}$   \\
2 &HD\,97048  &  A0pshe   & I   & 4.00 & 1.42          & 2.5$\pm 0.2^{~ \it o}$      & 180$^{+30, {~ \it y}}_{-20}$        & 42.8$^{+0.8}_{-2.5}$ \footnote{\citet{2006Sci...314..621L}, $^b$\citet{1981AJ.....86...62M}, $^c$\citet{2006A&A...449..267A} $^d$\citet{2007ApJ...665..512A}, $^e$ \citet{2008ApJ...684.1323P}, $^f$ \citet{1999AJ....117..354D}, $^g$ \citet{2004A&A...426..151A}, $^h$ \citet{1998A&A...329..131M}, $^i$ \citet{2008A&A...485..487V}, $^k$ \citet{2004ApJ...608..809G}, $^l$ \citet{2007A&A...462..293C}, $^m$ \citet{2004ApJ...613.1049E}, $^n$ \citet{2010A&A...516A..48V},  $^o$ \citet{2005A&A...437..189V}, $^{\it q}$ \citet{2009A&A...508..787K}, $^{\it r}$ \citet{2006A&A...446..155F}, $^{\it s}$ \citet{2004A&A...419..301M}, $^{\it t}$ \citet{1999ApJ...525L..53W}, $^{\it u}$ \citet{2008A&A...491..809F}, $^{\it v}$ \citet{2003ApJ...590L..49F}, $^{\it w}$ \citet{2008A&A...489.1157K}, $^{\it x}$ \citet{2006A&A...456.1045B}, $^y$\citet{1998A&A...330..145V} } & 175$\pm 1^{\clubsuit}$ \\
3 &HD\,179218 &  B9e      & I   & 4.02 & 1.88          & 2.7$\pm 0.2 ^{~ \it i}$ & 240$^{+70, {~ \it y}}_{-40}$        & 57$\pm 2^{ ~ \it u}$       & 23$\pm 3^{ {~ \it u}}$    \\
4 &HD\,101412 &  B9.5V    & II  & 4.02 & 1.40          & 2.3$\pm 0.2^{~ \it i}$ & 160$^{~ \it g}$        & 80$\pm 7^{ ~ \it u}$       & 38$\pm 5^{ {~ \it u}}$   \\
5 &HD\,141569 &  B9.5e    &     & 3.98 & 1.10          & 2.00$^{~ \it s}$       & 99$^{+9, {~ \it y}}_{-8}$         &  51$\pm 3^{ ~ \it t}$    & 356$\pm 5^{ {~ \it t}}$   \\
6 &HD\,190073 &  A2IVpe   & II  & 3.95 & 1.92          & 2.85$^{~ \it l}$       & $\geq$ 290$^{~ \it y}$ & 23$^{+15, {~ \it m}}_{-23}$ & 10$^{+170, {~ \it m}}_{-10}$  \\
7 &HD\,98922  &  B9Ve     & II  & 4.02 & 2.95$^\ddag$   & 2.2$\pm 0.8^{ ~ \it x}$ & $\geq$ 540 $^\ddag$ $^{~ \it y}$& 45$^{~ \it x}$    & -                      \\
8 &HD\,95881  &  A2III/IV & II  & 3.95 & 1.34$^{~ \it n}$ &2.0$\pm 0.3^{ ~ \it n}$& 170$\pm 30^{ {~ \it n}}$ & 55$^{~ \it n}$  & 103$^{~ \it n}$         \\
9 &R CrA     &  A5IIev   & II  & 3.86 &  -            & 3.5$^{~ \it r}$        & 130$^{~ \it b}$        & 65$^{~ \it q}$            & 180$^{~ \it q}$          \\
10&HD\,135344B&  F8V      &     & 3.82 & 1.01          & 1.7$\pm 0.2^{ ~ \it i}$& 140$\pm2^{~ \it f}$        & 14$\pm 3^{ {~ \it e}}$     & 56$\pm 2^{ { ~ \it e}}$    \\
11&HD\,150193 &  A1Ve     & II  & 3.95 & 1.19          & 2.3$\pm 0.2^{~ \it o}$        & 150$^{+50, {~ \it y}}_{-30}$        & 38$\pm 9^{ {~ \it v}}$     & 358$\pm 6^{ {~ \it v}}$    \\
12&HD\,104237 & A4Ve+sh$^{~ \it o}$&II&3.92& 1.53         & 1.96$^{~ \it k}$       & 116$^{+7, {~ \it y}}_{-6}$        & 18$^{+14, {~ \it k}}_{-11}$ & -                   \\
13&HD\,142666 &  A8Ve     & II  & 3.88 & 1.03          & 1.8$\pm 0.3^{~ \it o}$        & 145$\pm2 ^{~ \it f}$        & almost edge-on$^{~ \it h}$       & -                  \\
\hline
\end{tabularx}
\end{minipage}
\end{table*}

The observed sample consists of 12 HAeBe stars and one F8 star (HD\,135344B), selected from \citet{1994A&AS..104..315T}. Our targets have been selected to be homogeneous in mass of the central star and disk accretion activity \citep[all have passive disks; {\it \.M} $\la 2 x 10^{-7}$ M$_\odot$~yr$^{-1}$ as measured from their Br $\gamma$ luminosity listed in][]{2005hris.conf..309V}, but to be heterogeneous in disk structure (as determined by the spatial distribution of the dust in the disk: four flaring disks and  nine self-shadowed disks). The stellar parameters of the program stars are listed in Table \ref{table:stellar_parameters}. 

Because the group I/II classification of the disk structure is based on the shape of the mid-IR emission, it is sensitive to the dust-emission from the circumstellar disks. Hence the presence of a disk hole or, to a lesser extent, a disk gap,  may influence this classification. HD\,141569 has a large inner dust hole extending up to 30 au  \citep[e.g.][]{2002ApJ...573..425M}, and therefore we do not classify it as a group I/II source. HD\,97048, HD\,100546 and HD\,135344B are known to have disk gaps. The gap in the disk around HD\,97048 extends out to $\approx$ 34 au \citep{2013A&A...555A..64M} and in HD\,100546 extends between a few  au and  $\approx$ 13 au \citep{2005ApJ...620..470G, 2010A&A...511A..75B}. {\gc B}ecause both disks show ample signs of a flaring disk structure, we maintain the group I classification. Based on its SED, the disk around HD\,135344B is also flaring. However, because of the large \citep[$\approx$ 45 au;][]{2007ApJ...664L.107B} gap in the dust disk, and because of its late spectral type (F8), we do not classify it as a group I/II source.

\section{Observations and data reduction}
\label{sec:obs}

\begin{figure*}
   \centering
   \includegraphics{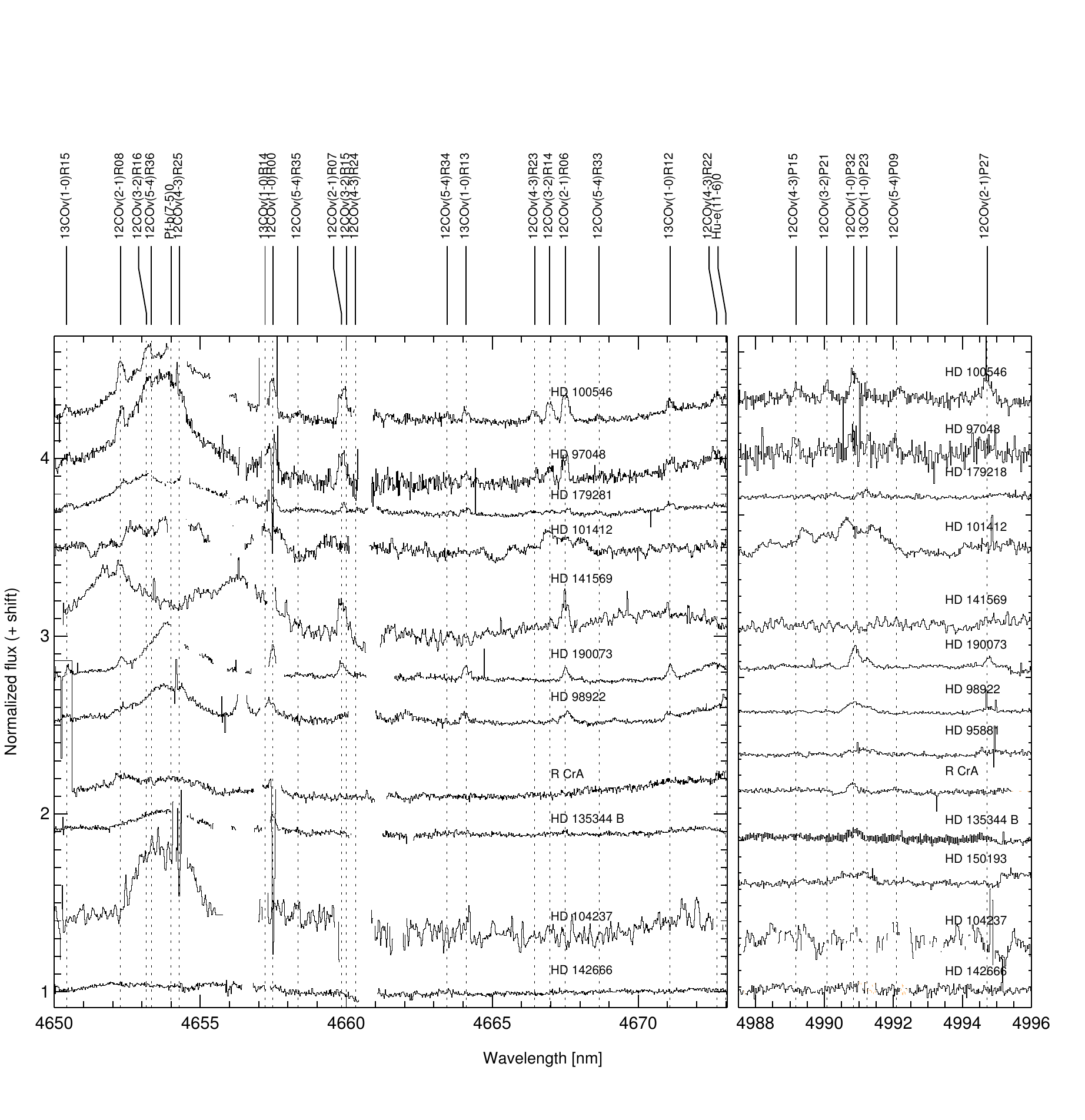}
   \caption{Example of CRIRES spectra of our targets in the between 4650 - 4673 nm and  4988 - 4996 nm. The continuum in all spectra is normalized to 1 and subsequently vertically shifted for clarity. Regions with low transmittance have been omitted from the spectra, and the spectra of HD\,95881 and HD\,150193  have been omitted completely from the left panel because of problems with the telluric correction in this wavelength range.}
   \label{fig:4662_5034_all}
\end{figure*}
   
\begin{figure*}
   \centering
   \includegraphics{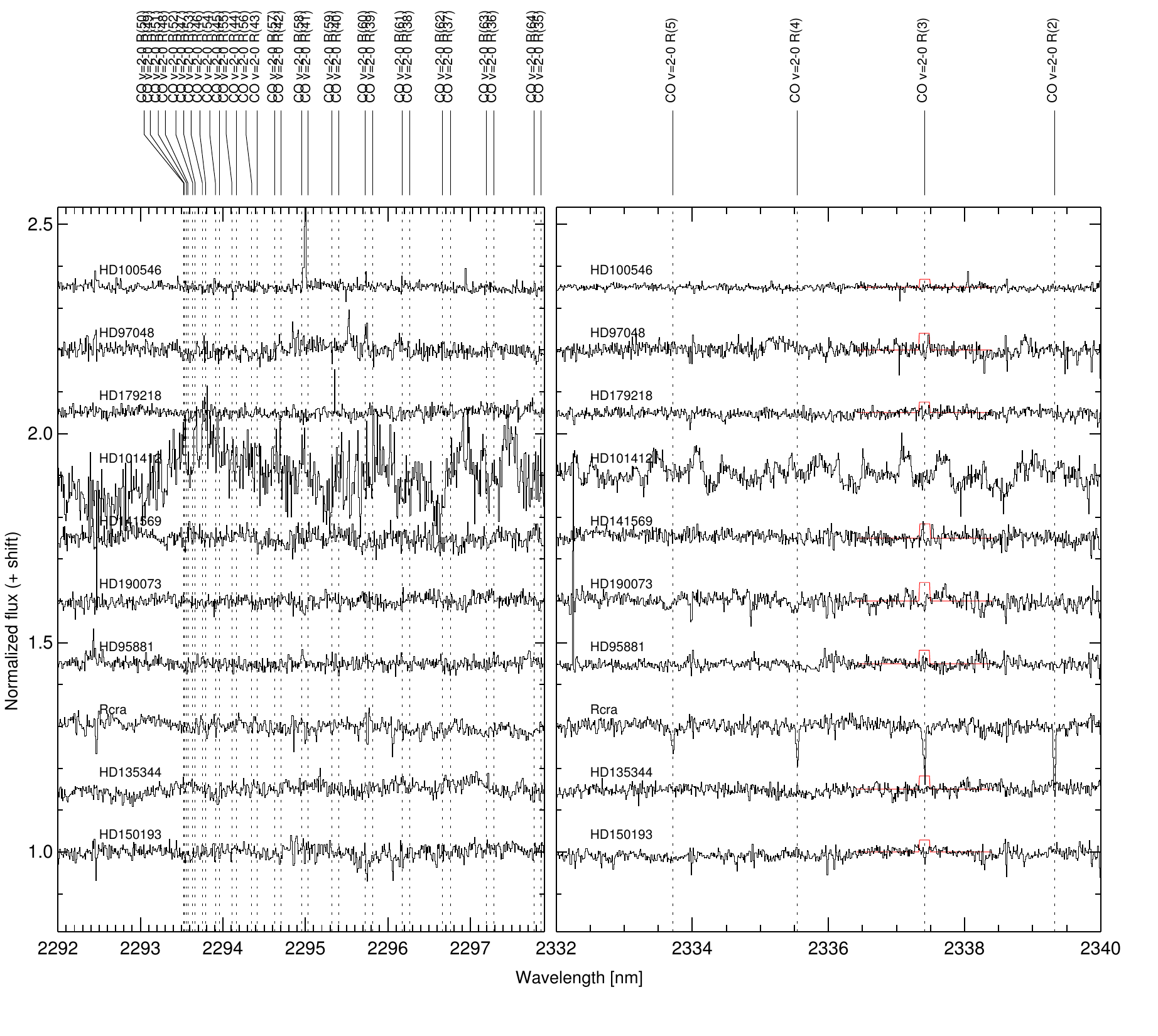}
   \caption{Example of CRIRES spectra of our targets in the between 2292 and 2298 nm (the v = 2-0 band head, left panel) and between 2332 and 2340 nm (the low rotational lines, right panel). The continuum in all spectra is normalized to 1 and subsequently vertically shifted for clarity. The red line in the right panel denotes the upper limit for the line fluxes noted in Section \ref{sec:overtone} and Table \ref{table:2mic}}
         \label{fig:2mic}
   \end{figure*}

\textbf{Observations: }High-spectral-resolution (R $\approx$ 94000, determined from telluric lines) spectra were taken with the VLT Cryogenic high-Resolution InfraRed Echelle Spectrograph (CRIRES\footnote{http://www.eso.org/sci/facilities/paranal/instruments/crires/}, \citet{2004SPIE.5492.1218K}) between June 14$^{th}$ and June 16$^{th}$ 2007 and July 2$^{nd}$ and July 5$^{th}$ 2008. 
Adaptive Optics were used to optimize the signal-to-noise ratio and spatial resolution of the observations. Our observations contain the wavelength ranges between 2.011 - 2.138 and 2.199 - 2.450  $\mu$m (covering the CO first overtone ro-vibrational emission), and between 4.588 - 4.809 and 4.905 - 5.094 $\mu$m (covering many fundamental ro-vibrational transitions of the CO molecule). The observations were made with a slit width of 0.2", and during the observations the slit was allowed to rotate to keep alignment with the parallactic angle.
\\

\textbf{Source acquisition and confidence of centering: }
During our observations the slit was allowed to rotate to keep aligned with the parallactic angle, which can smear out the signal of those sources with spatially resolved continuum and/or CO emission. The time on target varies between targets, but is on average $\pm$1 hour per target, and $\pm$10 minutes per spectral window. This translates into an average rotation of the slit parallactic angle of 15$^\circ$ (with a maximum of 29.5$^\circ$ for HD\,135344B) per target, or a couple of degrees per wavelength setting. Compared to the PSF (0.193$\arcsec$) this is negligible within each spectral window.\\
Centering of the slit on our objects was performed on the photo-center of the K band flux of our objects. Some of our targets have disks with a considerable contribution to the K band flux, but that excess emission comes from distances very close (within $\approx$0.5 au) to the star and has little influence on the photo center of the K band emission. For example, $\geq$ 95\% of the K band excess of HD\,100546 (total excess: 1.24 magnitudes) comes from the inner disk rim at 0.25 au of the central star, while just a small percent comes from photons scattered off the inner wall of the outer disk at $\approx$13 au (cf. Fig 8 of \citet{2013A&A...549A.112M} or Fig. 4 of \citet{2011A&A...531A...1T}). An offset between the stellar position of HD\,100546 and the center of the cavity of $\approx$ 5 au (1/4 slit width) to the NW of the center of the cavity has been reported by \citet{2005ApJ...620..470G}. Because the CO emission extends up to at least 35 au (i.e. beyond the slit, cf. paragraph \ref{sec:spatial}), this introduces asymmetries in both the line spatial and spectral profile. For a complete discussion of the influence of a slit offset on spatially resolved line profiles we refer to Hein Bertelsen et al. (2014).\\

\textbf{Data reduction: }The data were reduced using the CRIRES pipeline V1.7.0\footnote{www.eso.org/sci/data-processing/software/pipelines/index.html}, which performs wavelength calibration, background subtraction and flatfield correction. For the targets without spatially resolved CO emission, we use the "optimal" extraction method, which uses a weighting function derived from a fit to the spatial profile.  Because this method does not correctly account for extended emission, we use the "rectangular" extraction method for the spectra of the spatially resolved disks around HD\,97048, HD\,100546, and HD\,141569.  

To correct for telluric absorption we observed a telluric standard star directly after each science observation. Each standard was chosen to be as close as possible on the sky so that their spectra are affected by similar atmospheric conditions, and compared to an appropriate \citet{1991ppag.proc...27K} stellar atmosphere model to determine the instrumental response. The optical depth of the telluric lines of the standard was scaled to that of the science target, and the two were ratioed using a manually determined sub-pixel wavelength shift to reduce telluric residuals (i.e. spikes due to minor wavelength mismatches). Some telluric absorption lines are fully saturated, causing problems with the division of the spectra. We inspect all resulting spectra visually for residuals of an (imperfect) telluric correction, and make a conservative estimate for the cut-off value of areas of low transmittance on a per-spectrum basis. In practice this means that we ignore all areas below a transmittance of $\approx$ 15\%  in our analysis. In the following data reduction, we will disregard these regions. 

CO lines in our spectra were identified using the line list of \citet{1996A&AS..117..557C}. We show typical spectra for all sources with the CO centered on its rest wavelength in Figs. \ref{fig:4662_5034_all} and \ref{fig:2mic}. The CO overtone spectra are flux calibrated using the continuum fluxes for our sources as published in the 2MASS catalog \citep{2006AJ....131.1163S}. The CO fundamental spectra are flux calibrated with the 4.77 $\mu$m flux, as determined with a spline interpolation to the M band data points, from e.g. \citet[][]{1985SAAOC...9...55K, 1992ApJ...397..613H, 2001A&A...380..609D} (see Table \ref{tab:contflux}). We fit a 2nd degree polynomial to either side of each unblended CO line to determine the continuum baseline. We define the FWHM as the width of the line half way between the continuum and the line maximum, and obtain the line flux by integrating the line where it is more than 1$\sigma$ above the continuum. We note that the 0.2'' slit of CRIRES may obscure parts of the emitting region of the spatially resolved disks. CO emission, extended out to many tens of au has been detected around HD\,97048 \citep{2009A&A...500.1137V}, HD\,100546 \citep{2009A&A...500.1137V,2009ApJ...702...85B,2012A&A...539A..81G,2014A&A...561A.102H}, and HD\,141569 \citep{2006ApJ...652..758G}.

\section{Method}
\label{sec:method}

When in Local Thermodynamic Equilibrium (LTE), the strength and ratio of all rotational transitions within and between vibrational bands dictates an unique temperature of the gas via the Boltzmann equation. We can estimate the total amount of emitting CO gas (N$_\mathrm{tot}$) and its rotational temperature ($T_\mathrm{rot}$) using this Boltzmann equation and the measured line fluxes 
\begin{equation}
  \frac{F_{ij}}{g_i \nu_{ij} A_{ij} } = \frac{1}{4 \ \pi \ d^2 } \ \frac{h \ B \ N_\mathrm{tot}}{ T_\mathrm{rot} } \  e^{E_{i} / kT_\mathrm{rot}}
\end{equation}
where $F_{ij}$ is the line flux, $\nu_{ij}$ the frequency of the transition, $A_{ij}$ the Einstein A-coefficient, $g_i$ the degeneracy of the upper level, $d$ the distance to the source, $B$ the rotational constant, and $E_{i}$ the energy of the transition.

We present our data in a so-called ``Boltzmann plot'' in Figure \ref{fig:rovib_all}. Inspection of  Figure \ref{fig:rovib_all} shows that some rotational transitions, most notably those of the  $^{12}$CO v$_u$ = 1 vibrational bands, deviate from linearity and curve upwards for lower energies. This behavior is observed frequently in CO emission coming from disks \citep[e.g.][]{2003ApJ...589..931N, 2004ApJ...606L..73B, 2007ApJ...659..685B, 2009ApJ...699..330S, 2014A&A...561A.102H}, and is commonly explained by either a radial gradient in the CO temperature or by optically thick gas. Fitting our data with a two-temperature CO gas model yields acceptable fits, but requires for the cold component both unrealistically low temperatures ($\approx$ 65 K), and a CO column that is over 17 orders of magnitude larger that the amount of CO needed to explain the warmer CO. This is incompatible with the (much broader) observed line widths in disks, and we thus interpret the curvature seen in the Boltzmann plots as originating from optical depth effects. See \citet[][]{2009ApJ...699..330S} for a similar argument regarding T Tauri stars. 

\subsection{Rotational and vibrational temperatures}\label{sec:rot-vib-temp}

The temperature of a gas in LTE dictates the distribution of its emission over the various rotational transitions and vibrational bands. The temperatures derived via the Boltzmann equation from the fractional distribution between the different rotational transitions within each vibrational band ($T_\mathrm{rot}$), and the temperature derived from the ratio of the total population of each vibrational band ($T_\mathrm{vib}$), are thus the same for gas in LTE. Deviations between these two temperatures point to alternative, non-LTE (de)excitation mechanisms. For example, sub-thermal vibrational level populations are expected in low-density environments where the higher J rotational transitions are depopulated relative to LTE, while super-thermal level populations can be caused by e.g. IR pumping \citep{1980ApJ...240..929S} or UV fluorescence \citep{1980ApJ...240..940K} .

To determine the vibrational temperature we use the cumulative best fit CO rotational temperature and column per source to calculate the total amount of CO molecules needed to explain the observed line fluxes per vibrational band. We then calculate the vibrational temperature according to the relative populations of the vibrational bands following \citet{2007ApJ...659..685B}. One $\sigma$ errors on the vibrational temperature are determined by adopting a relative error in $N$(CO) of 25$\%$ between the vibrational bands for each individual source. 

\subsection{Isothermal slab model}

We interpret the curvature in the Boltzmann plots in the subsequent analysis to be due to opacity effects. We model the emitting CO gas as to be originating from an isothermal slab of CO gas that is in LTE. Free parameters in this slab are the excitation temperature of the CO gas, $T_{ex}$, the CO column, $N$(CO), and the slab surface. A similar approach has been taken before by e.g. \citet[][]{1999ApJ...517..209G, 2009ApJ...699..330S, 2012A&A...539A..81G, 2013A&A...551A..49T, 2013ApJ...770...94B}.

 We correct for optical thickness effects by defining an optical depth correction factor for each upper level column density: C$_{\tau}$ (See \citet{1999ApJ...517..209G} equations 6, 16 and 17). We set the line width $\Delta$v as the FWHM value of the combined thermal and turbulent velocities. The turbulent velocity is fixed at one tenth of the sound speed \citep[see e.g.][]{2011ApJ...727...85H, 2013ApJ...774...16R}. Within our parameter space $\Delta$v varies between $\approx$ 0.5 and 3 km s$^{-1}$, and the contribution to line broadening of turbulent motions is minor. For each pair of $T_{ex}$ and $N$(CO) we find a best fit surface assuming the distance to the source listed in Table \ref{table:stellar_parameters}, and compare our model and observations using this best fit surface. We also use this surface calculate the vibrational temperature through the total number of emitting CO molecules (Figure \ref{fig:vib_12co}).

To find the best fit CO column and rotational temperature we create a grid of models and explore the parameter space between 100 K $\leq$ $T_{ex}$ $\leq$ 2500 K and 10$^{16}$ cm$^{-2}$ $\leq$ $N$(CO) $\leq$ 10$^{20}$ cm$^{-2}$ for $^{12}$CO emission, and 100 K $\leq$ $T_{ex}$ $\leq$ 1200 K and 10$^{17}$ cm$^{-2}$ $\leq$ $N$(CO) $\leq$ 10$^{21}$ cm$^{-2}$ for $^{13}$CO emission, in steps of $\Delta$$T_{ex}$ = 25 K and $\Delta$$^{10}$log($N$(CO)) = 0.1.  We define our best fit and 1 $\sigma$ error following the reduced $\chi$$^2$ formalism. We remind the reader that the number of degrees of freedom can only be estimated for linear models \citep[see e.g.][]{2010arXiv1012.3754A}, and thus that our reported 1 $\sigma$ errors likely do not represent 68\% confidence intervals. We use the $\chi _r$$^2$ value in this paper to facilitate comparison with other, similar works.

Using an isothermal slab of CO gas in LTE to model the emission coming from a disk is a simplified approach.  There exists a degeneracy between the line width $\Delta$v and $N$(CO) ( $\tau$ $\propto$ $\frac{1}{\Delta v}$) which becomes significant when the line emission is optically thick. It also ignores any vertical or radial temperature gradient and non LTE excitation mechanisms such as UV fluorescence and non resonant scattering of IR photons which we will discuss in section \ref{sec:co-excitation}. The presence of such non-thermal excitation mechanisms can be recognized by a mismatch between the rotational and vibrational temperatures derived from the Boltzmann diagrams.

We present the results of our modeling of the rotational temperatures and CO column in Table \ref{table:ex_properties}, and the best fit contours in Figures \ref{fig:cont} and \ref{fig:cont13}. The modeling results for the vibrational temperatures are listed in Table \ref{table:vibtemp} and shown in Figure \ref{fig:vib_12co}.

\section{Results}
\label{sec:results}

\subsection{General description of the spectra}
\label{sec:general_description}

Twelve out of thirteen surveyed HAeBe stars show spectrally resolved fundamental $^{12}$CO emission, eight show fundamental $^{13}$CO emission, and one shows first overtone $^{12}$CO emission (Figures \ref{fig:4662_5034_all} and \ref{fig:2mic}). For the source in which we did not detect $^{12}$CO emission, HD\,142666, we calculate a 3 sigma upper limit on the line flux of 1.2 $\times$ 10$^{-15}$ erg cm$^{-2}$ for the CO lines at 4.667 $\mu$m assuming a line width of 20 km s$^{-1}$ and the continuum flux reported in \citet{1996MNRAS.279..915S}. There is a large spread in the line widths and distribution of CO molecules over vibrational transitions. The v$_{\rm u}$ = 1 bands in all stars but HD\,141569 are rotationally excited up to high (J$_\mathrm{up} > $ 30) transitions (Table \ref{table:CO_properties}).  The $\frac{^{12}CO ~P12}{^{13}CO ~R12}$ ratio for HD\,179218, and the $\frac{^{12}CO ~R04}{^{13}CO ~R04}$ ratio for the other targets, varies between 3 and 7.

We have been unable to characterize the spectra of two stars for which we have detected CO emission.  The combination of broad lines (FWHM $\approx$ 80 km~s$^{-1}$) and multiple (at least five: $^{12}$CO v$_\mathrm{u}$ = 1, 2, 3 and 4, and $^{13}$CO v$_\mathrm{u}$ = 1) vibrational transitions in the spectrum of HD\,101412 caused the emission lines to be highly blended. A combination of poor SN and telluric absorption partially overlapping with the CO emission made it possible to identify CO in the spectrum of HD\,104237, but we have been unable to reliably measure line fluxes.

The CO emission in HD\,97048, HD\,100546, HD\,141569 and R CrA is spatially resolved, and the 4.6 $\mu$m continuum emission is resolved in  HD\,97048, HD\,100546, and HD\,179218. We discuss detected CO absorption towards HD\,97048 and R CrA in Section \ref{sec:absorption}, and the detected emission lines other than CO in Section \ref{sec:others}.

\begin{table*}
\small
\begin{minipage}[t]{\textwidth}
\caption{Detected CO lines, their radial velocities with a typical error of 2 km s$^{-1}$ as determined in Section \ref{sec:kinematics}, and the stellar radial velocities derived from photospheric absorption lines in columns 1--3. References for the stellar radial velocity are: $^a$ \citet{2005A&A...436..209A}, $^b$ \citet{2008A&A...485..487V}, $^c$\citet{2013MNRAS.429.1001A} and $^d$ \citet{1996A&AS..120..229R}. Emission lines are listed in the upper panel, and absorption lines in the lower. $\ddag$Even though we do not detect these CO transitions in our spectra, they have been detected by \citet{2007ApJ...659..685B} in this source.  $\dagger$ Not all spectra have s/n sufficient to identify the ro-vibrational transitions, accordingly this may be a subset of all transitions. {\gc $^{\Diamond}$The CO emission lines from R CrA most likely originate from an outflow as discussed in section \ref{sec:spatial}. $^\blacksquare$Radial velocity determined based on CaII lines.} $^\star$No observations available}
\label{table:CO_properties}
\centering
\renewcommand{\footnoterule}{}  
\noindent\begin{tabularx}{\columnwidth}{@{\extracolsep{\stretch{1}}}*{10}{l}@{}}

\hline 

Name           &  v$_\mathrm{CO}$  & v$_\mathrm{\star}$ & $^{12}CO$ & $^{12}CO$ & $^{12}CO$ & $^{12}CO$ & $^{12}CO$ & $^{13}CO$ & $^{12}CO$ \\   
       & km s$^{-1}$    & km s$^{-1}$      & v$_u$ = 1    & v$_u$ = 2   & v$_u$ = 3  & v$_u$ = 4   & v$_u$ = 5  & v$_u$ = 1   & v = (2-0)  \\
  
\hline
Emission & & & & & & & & & \\
\hline
HD\,100546 & 14 &  18$^a$    & R08 - P37 & R17 - P28 & R25 - P21 & R32 - P17 & R39 - P09 & R18- P23 & no  \\
HD\,97048  & 16 &  21$^a$    & R08 - P36 & R17 - P07 & R26 - R05 & R27 - P17 & R33 - P09 & R15 - R04 & no  \\
HD\,179218 & 17 &  15.4$\pm 2.3^b$  & R06 - P36 & R14 - P28 & R23 - P21 & R27 - P20 &     -     & R21 - P21 & no  \\
HD\,101412 & 16 &  16.9$\pm 0.2^b$  & R09 - P37 & R17 - P32 & R26 - P26 & R35 - P20 &     -     & R23 - P29 & yes\\
HD\,141569 & -11 &  $-$12$\pm 7^c$   & R09 - P13$^\dagger$& R14 - P01$^\dagger$& $^\ddag$& $^\ddag$   & $^\ddag$  & R09 - R00$^\dagger$ & no  \\
HD\,190073 & -1 &  0.21$\pm 1.0^3$     & R08 - P38 & R13 - P32 &         - &         - &         - & R23 - P23 & no  \\
HD\,98922  & 3  &  0.2$\pm 2.2^c$          & R07 - P38 & R13 - P28 & -         &         - &         - & R16 - P27 & $^\star$  \\
HD\,95881  & 18 &   36$^{a, \blacksquare}$         & R06 - P37 & - 	& -         &         - &         - & R16 - P25 & no  \\
R CrA      & -13$^{\Diamond}$&  0$^d$     & P04 - P37 & - 	& -         &         - &         - &         - & no  \\
HD\,135344B& 2  &  1.6$\pm 1.3^b$   & R08 - P38 & - 	& -         &         - &         - &         - & no \\
HD\,150193 & -6 &  $-$4.9$\pm 3.9^c$    & R06 - P36 & - 	& -         &         - &         - &         - & no  \\
HD\,104237 & 14 &  13$^a$    & R08 - P10$^\dagger$ &- & -         &         - &         - &         - & $^\star$  \\
HD\,142666 & -  &      $-$7.0$\pm 2.7^c$      & -         & - 	& -         &         - &         - &         - & $^\star$  \\
\hline              
Absorption&    &             &           &           &           &           &           &           &    \\
\hline              
HD\,97048  & 16 &  21$^a$    & R04 - P04 &         - &         - &         - &         - & R02 - P01 & no  \\
R CrA      & -3 &  0$^d$     & R08 - P15 &         - &         - &         - &         - & R08 - P04 & R08 - P08\\

\hline
\end{tabularx}
\end{minipage}
\end{table*}

\subsection{CO rotational temperatures}
\label{sec:excitation_diagram}

We show the best fit rotational temperatures in Table \ref{table:ex_properties} and show the corresponding Boltzmann plots in Figure \ref{fig:rovib_all}. The temperature errors reported in Table \ref{table:ex_properties} represent a 1 $\sigma$ deviation of the best fit value. The complete best fit contours are shown in Figure \ref{fig:cont}. 

For HD\,101412, we are unable to determine $T_\mathrm{rot}$ as individual lines are heavily blended, but since multiple vibrational bands of both CO fundamental and 1$^{st}$ overtone emission are detected we estimate a lower limit to $T_\mathrm{rot}$ of 2000 K. This value is in agreement with the temperature of 2500~K found for the CO emission in HD\,101412 by \citet{2012A&A...537L...6C}.

Some trends can be seen. \textit{First}, the derived temperatures for the CO temperatures in all disks but HD\,141569 is above 500 K, and on average around 1000 K. \textit{Second}, the $^{12}$CO temperatures are higher than the $^{13}$CO temperatures, except for HD 190073, where they are similar. %

   \begin{figure*}
   \centering
   \includegraphics[]{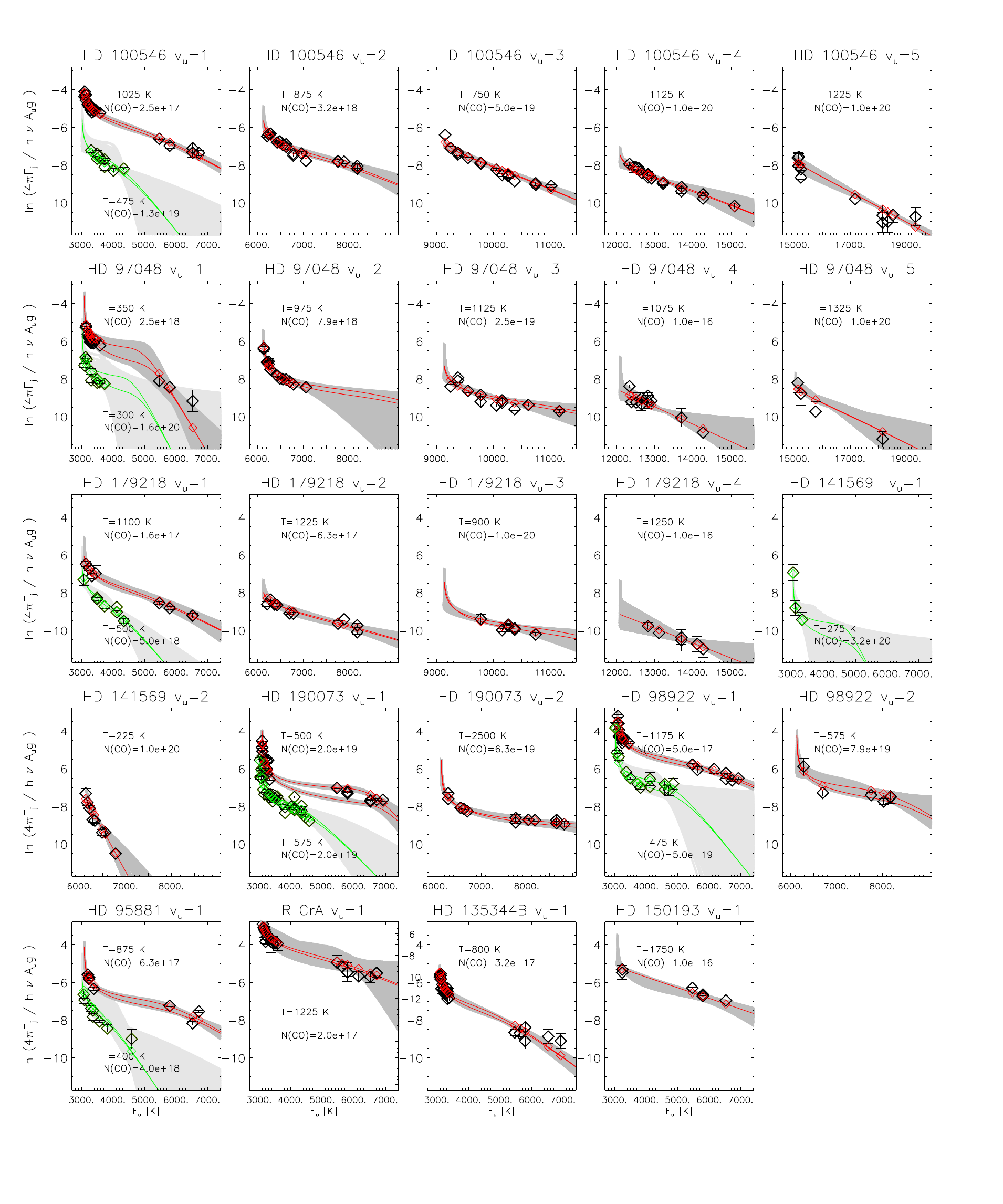}
   \caption{Boltzmann plots for detected $^{12}$CO and, if detected, $^{13}$CO emission (black plot symbols). The best fit for the $^{12}$CO rotational temperature $T_\mathrm{rot}$ is shown by the red lines (upper line: P branch, lower: R branch), whereas the best fit $T_\mathrm{rot}$ for $^{13}$CO is shown by the green lines. Model predictions for the best fit ($T$, $N$) are shown with the smaller open red/green diamonds. The gray area around the best fit lines show the model predictions for parameters within the 1 $\sigma$ areas shown in Figs. A.1 and A.2. The Boltzmann plots for HD\,101412 and HD\,104237 are absent for reasons described in section \ref{sec:general_description}}
          \label{fig:rovib_all}
   \end{figure*}

\bgroup
\def\arraystretch{1.2}
\tabcolsep=0.10cm
\begin{table*}
\small
\begin{minipage}[t]{\linewidth}
\centering
\caption{Derived physical properties (N$_\mathrm{CO}$, $T_\mathrm{rot}$, emitting surface, best fit reduced $\chi^2$ value and FWHM) for the CO emission in our sample.  Errors reported are within the 1$\sigma$ confidence intervals. When the reported error exceeds one border of the explored parameter space, we mark this with  '$\infty$'. Unconstrained quantities are marked with '-' (cf. Figures \ref{fig:cont} and \ref{fig:cont13}). In the case of an unconstrained upper (lower) temperature limit, we also omit the upper (lower) limit on the emitting surface. The FWHM value of transitions marked with a star $(^*)$ is affected by imperfect telluric correction and/or absorption.}  
\noindent\begin{tabularx}{\columnwidth}{@{\extracolsep{\stretch{1}}}*{10}{l}@{}}
Target & v$_u$ & $^{10}log$ $N\mathrm{(CO)}$  & $T_\mathrm{rot}$  & Surface & Surface & $\chi_r^2$ & FWHM & FWHM   & FWHM   \\
   &  &  &  & min & max & min & (J$<$10) & (10$\leq$J$<$20) & (J$\geq$20) \\
  &   & [cm$^{-2}$] &  [K] & [au$^2$] & [au$^2$] & & [km s$^{-1}$] & [km s$^{-1}$] & [km s$^{-1}$] \\
\hline
HD 100546 & 1 & 17.40$^{+0.1}_{-0.1}$ & 1025$^{+100}_{-100}$ &    4.01e-01 &   9.10e-01 &6.09 & 16.1(1.3) & 16.2(1.2) & 16.0(1.9)\\
$^{13}$CO & 1 & - & 475$^{+\infty}_{-325}$ & - &   7.17e+09 &2.97 &  -  & 15.6(1.3) &  - \\
HD 100546 & 2 & 18.50$^{+0.9}_{-0.4}$ & 875$^{+275}_{-275}$ &    1.41e-01 &   2.29e+00 &10.24 & 18.1(0.9) & 16.5(1.3) & 17.2(3.0)\\
HD 100546 & 3 & - & 750$^{+175}_{-75}$ &    5.50e-01 &   5.13e+03 &4.41 & 16.5(1.3) & 16.4(2.6) & 15.7(2.4)\\
HD 100546 & 4 & - & 1125$^{+575}_{-200}$ &    3.15e-02 &   4.02e+03 &0.84 & 13.8(3.0) & 17.0(2.0) & 17.5(3.0)\\
HD 100546 & 5 & - & 1225$^{+225}_{-125}$ &    8.03e-02 &   1.36e+04 &4.17 & 19.5(3.7) &  -  & 19.3(4.2)\\
HD 97048 & 1 & 18.40$^{+\infty}_{-1.4}$ & 350$^{+350}_{-125}$ &    3.69e+00 &   7.88e+04 &1.38 & 18.2(2.4) & 16.0(1.4) & 16.1(1.8)\\
$^{13}$CO & 1 & - & - & - &- &6.82 & 15.5(0.9) & 19.2(2.4) &  - \\
HD 97048 & 2 & 18.90$^{+\infty}_{-0.4}$ & 975$^{+\infty}_{-550}$ & - &   2.94e+01 &2.06 & 18.1(0.9) & 17.3(1.8) &  - \\
HD 97048 & 3 & 19.40$^{+\infty}_{-1.2}$ & 1125$^{+\infty}_{-350}$ & - &   1.36e+00 &8.01 & 19.7(0.5) & 19.9(2.9) & 17.7(0.3)\\
HD 97048 & 4 & - & 1075$^{+\infty}_{-325}$ & - &   2.22e+05 &4.67 & 23.4(2.9) & 19.9(4.5) & 18.6(3.6)\\
HD 97048 & 5 & - & 1325$^{+600}_{-475}$ &    3.11e-02 &   1.38e+06 &1.39 & 16.4(1.7) & 20.5(4.3) & 26.1(6.9)\\
HD 179218 & 1 & 17.20$^{+0.4}_{-\infty}$ & 1100$^{+325}_{-375}$ &    3.27e-01 &   4.28e+00 &0.67 & 16.0(5.1)$^*$ & 13.7(10.3)$^*$ & 17.7(5.4)\\
$^{13}$CO & 1 & 18.70$^{+0.6}_{-\infty}$ & 500$^{+350}_{-250}$ &    1.18e+00 &   3.35e+05 &1.29 & 15.0(7.9)$^*$ & 19.5(1.7)$^*$ & 19.0(0.9)\\
HD 179218 & 2 & 17.80$^{+0.7}_{-\infty}$ & 1225$^{+500}_{-375}$ &    5.53e-02 &   7.06e+00 &2.71 & 19.3(0.7) & 21.7(2.1) & 21.0(4.8)\\
HD 179218 & 3 & 20.00$^{+\infty}_{-1.8}$ & 900$^{+\infty}_{-150}$ & - &   1.07e+00 &1.09 &  -  & 20.6(4.6) & 22.1(1.3)\\
HD 179218 & 4 & - & 1250$^{+\infty}_{-400}$ & - &   5.39e+04 &0.21 &  -  & 16.6(1.4) & 23.2(4.2)\\
HD 141569 & 2 & - & 225$^{+125}_{-50}$ &    2.11e+01 &   6.71e+12 &1.73 & 16.1(2.4) & 16.3(1.6) &  - \\
$^{13}$CO & 1 & 20.50$^{+\infty}_{-2.2}$ & - & - &- &0.14 & 15.1(2.1) & 13.1(1.4) &  - \\
HD 190073 & 1 & 19.30$^{+0.6}_{-0.7}$ & 500$^{+225}_{-100}$ &    4.40e+00 &   1.91e+02 &13.61 & 15.6(2.6) &  -  & 16.4(2.4)\\
$^{13}$CO & 1 & 19.30$^{+0.3}_{-0.2}$ & 575$^{+375}_{-175}$ &    1.36e+00 &   6.45e+02 &11.12 & 12.7(0.5) & 12.3(2.1) & 11.5(2.4)\\
HD 190073 & 2 & 19.80$^{+\infty}_{-1.3}$ & 2500$^{+\infty}_{-1600}$ & - &   7.00e-01 &5.98 & 15.5(3.1) & 15.8(1.0) & 17.3(3.7)\\
HD 98922 & 1 & 17.70$^{+0.1}_{-0.1}$ & 1175$^{+150}_{-125}$ &    1.06e+01 &   2.05e+01 &14.18 & 18.1(0.9) & 18.6(8.8) & 26.1(4.1)\\
$^{13}$CO & 1 & 19.70$^{+\infty}_{-0.3}$ & 475$^{+\infty}_{-275}$ & - &   5.40e+08 &2.16 & 16.5(0.2) & 17.8(1.8) & 21.3(1.7)\\
HD 98922 & 2 & - & 575$^{+\infty}_{-75}$ & - &   3.86e+02 &0.87 & 26.7(1.1) & 22.5(5.1) & 29.6(5.2)\\
HD 95881 & 1 & 17.80$^{+0.3}_{-0.3}$ & 875$^{+400}_{-200}$ &    3.40e-01 &   4.09e+00 &16.42 & 34.4(4.8) & 29.2(6.9) & 47.2(4.4)\\
$^{13}$CO & 1 & - & 400$^{+625}_{-250}$ &    5.64e-01 &   1.84e+11 &1.04 & 22.9(3.9) & 24.5(4.0) & 17.5(12.4)\\
R CrA  & 1 & - & 1225$^{+\infty}_{-850}$ & - &   2.33e+03 &1.36 & 10.3(0.5)$^*$ & 11.8(0.6)$^*$ & 18.8(3.9)\\
HD 135344B & 1 & 17.50$^{+0.1}_{-0.2}$ & 800$^{+100}_{-75}$ &    2.78e-01 &   8.84e-01 &5.26 & 16.6(2.6) & 12.1(1.5) & 27.7(3.4)\\
HD 150193 & 1 & 16.00$^{+1.9}_{-\infty}$ & 1750$^{+250}_{-925}$ &    1.75e-01 &   5.84e+00 &5.77 & 54.1(1.8) &  -  & 60.0(2.2)\\
 \hline
\end{tabularx}
\label{table:ex_properties}
  \end{minipage}
\end{table*}
\egroup

\subsubsection{CO vibrational temperatures}

We show the vibrational diagrams in Figure \ref{fig:vib_12co} and summarize our results in Table \ref{table:vibtemp}. To investigate the dominant excitation mechanism of the CO gas we compare the vibrational and rotational temperatures of all sources with detections of at least the v$_u$ = 2 emission. 
We find that the distribution of CO molecules in the group II (self-shadowed) disks over the vibrational bands is consistent with the temperature derived from the rotational transitions, i.e. in agreement with collisions being the dominant excitation mechanism. The higher vibrational bands in the group I disks, however, show an overpopulation compared to the numbers expected for gas in LTE. 

Summarizing, we find (1) that $T_\mathrm{vib}$ $>$ $T_\mathrm{rot}$ for the flaring disks and (2) that $T_\mathrm{vib}$ $\approx$ $T_\mathrm{rot}$ for the self-shadowed disks.
\begin{table}
\centering
\caption{Vibrational temperatures and the averaged rotational temperatures of combined vibrational bands for sources with at least v$_\mathrm{up}$ = 1,2 detected. See also Figure \ref{fig:vib_12co}. Errors reported marked with $\infty$ have 1$\sigma$ best fit solutions that exceed the parameter space of the grid of models} 
\label{table:vibtemp}
\noindent\begin{tabularx}{\columnwidth}{@{\extracolsep{\stretch{1}}}*{6}{l}@{}}
 Target & n$_{vib}$ & $T_\mathrm{rot}$ & error & $T_{vib}$ & error\\
 \hline
 & & [K] & [K] & [K] & [K]\\
\hline
HD\,100546 & 5 & 1080 &  +180, -130 & 6400 & 2100 \\
HD\,97048 & 5 & 1330 & +$\infty$, -580 & 2100 & 330  \\
HD\,179218 & 5 & 1250 & +400, -300 & 6500 & 2700 \\
HD\,190073 & 2 & 1130 & +$\infty$, -450 & 900 & 90  \\
HD\,98922 & 2 & 1230 & +230, -180 & 1300 & 180 \\
\hline        
\end{tabularx}
\end{table}

\begin{figure}
  \includegraphics[width=\linewidth]{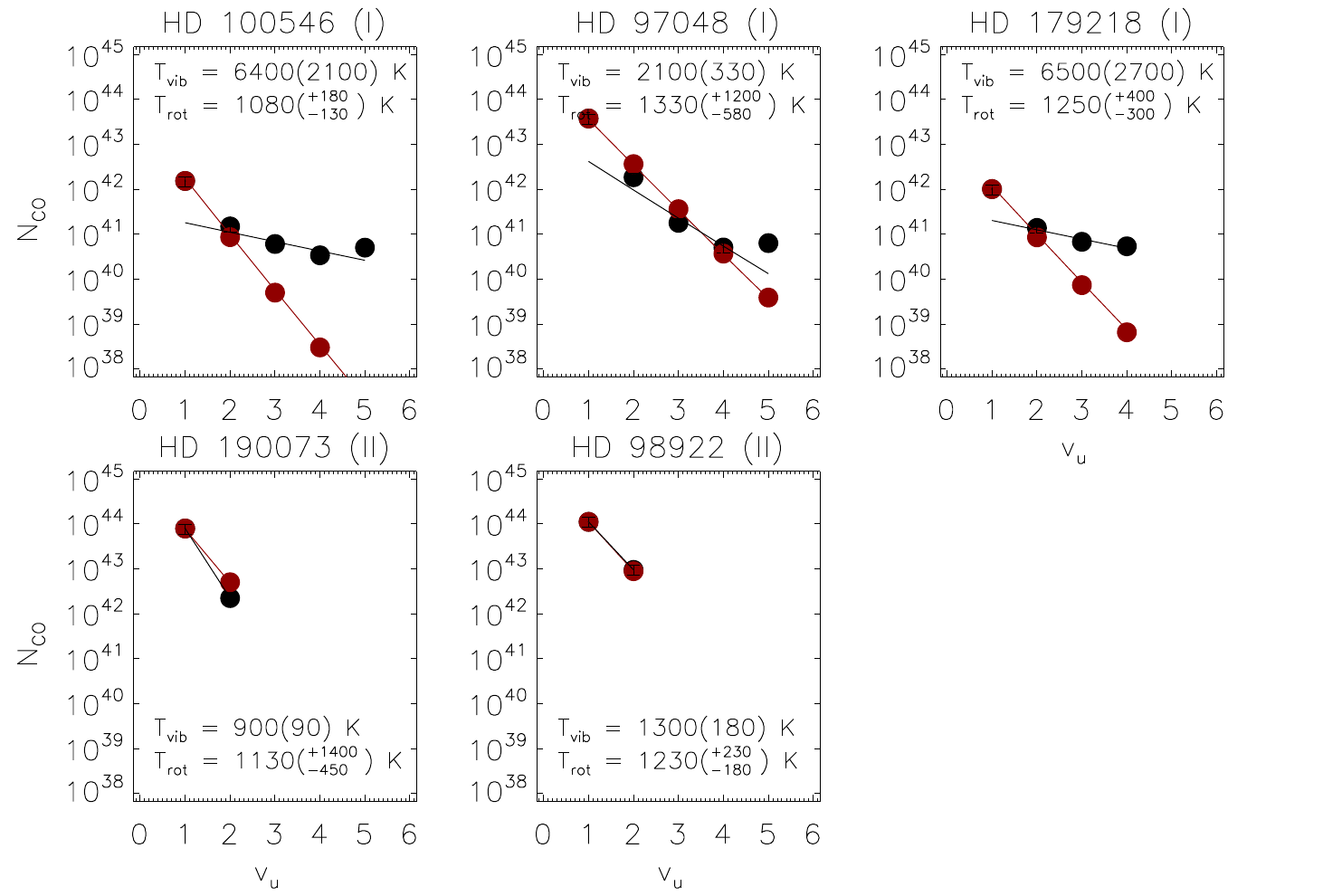}
  \caption{Best fit vibrational (black lines) and rotational (red lines) temperatures for sources with at least v$_\mathrm{up}$ = 1,2 detected. The target name and disk classification (group I/II) are given in each plot title. Black points show the total amount of CO molecules per vibrational band needed to explain the detected flux assuming the distance listed in Table \ref{table:stellar_parameters} and an averaged CO column and $T_\mathrm{rot}$. Red points show an LTE distribution calculated for the averaged $T_\mathrm{rot}$. Both sets are normalized on v$_u$ = 1. The errors for the rotational temperatures are 1$\sigma$ best fit contours of the summed best fit contour for all vibrational bands per source. Calculated vibrational temperature errors are 1$\sigma$ values adopting a relative error of 25\% in $N$(CO).}
  \label{fig:vib_12co}
\end{figure}

\subsection{Kinematics}
\label{sec:kinematics}

\textbf{Line width:} The shape of an emission line originating from a circumstellar disk is determined by the mass of the central star, the disk inclination, and the radial location of the emitting gas. Thus, a spectrally resolved emission line is an indicator for the distribution of the emitting gas throughout the disk. Comparing the line shape and width both for the rotational transitions within a vibrational band ($\Delta$E $\approx$ 10--100 K), and between the vibrational bands  ($\Delta$E $\approx$ 1000 K) is a useful exercise to compare the location of the emitting gas over a large range of excitation conditions.

As a proxy for line shape, we have measured the FWHM of each ro-vibrational line as function of  J$_\mathrm{up}$ (and thus excitation energy). We show the average FWHM values for low (J$_\mathrm{up}$$\leq$10), medium (10$<$J$_\mathrm{up}$$\leq$20) and high (J$_\mathrm{up}$$>$20) in Table \ref{table:ex_properties}. 
The FWHM of the low (J$_\mathrm{up} \leq 15$) rotational transitions is constant within our errors, and we construct composite line profiles for each detected vibrational band of these lines with sufficient S/N, to compare the line shapes. To construct these composite profiles we center each emission line at its line center, re-grid it on a velocity grid, subtract the continuum, and scale the line maximum to 1. We show the composite profiles in Figure \ref{fig:co_all}.  The line profiles of HD\,100546 and  HD\,97048 v = 3-2, and HD\,150193, HD\,179218 and R CrA v = 1-0, have been  made using the high  (J$_\mathrm{up} > 20$)  ro-vibrational transitions, because no suitable low J lines were available.

One would expect that line width correlates with the energy needed to excite the line - the higher excitation lines originating from closer to the star and so showing broader lines. Surprisingly, only two stars in our sample, HD\,98922 and HD\,135344B, show this trend (Table \ref{table:ex_properties}). In HD\,100546, there is no such change in line width between the different vibrational bands, but the line \textit{shape} of the $^{12}$CO transitions become more flat-topped with increasing excitation levels. In the remaining stars of our sample we find no correlation between line excitation and line shape, nor between self-shadowed and flaring disks. For all stars with self-shadowed disks the $^{13}$CO line width (FWHM) is smaller than that of the corresponding $^{12}$CO lines. This is not seen in the flaring disks. We will return to these trends in Section \ref{sec:1213}.

\textbf{Radial velocity}: To determine the radial velocity of the CO emission we fit a two-dimensional polynomial to the offset between the telluric absorption lines in the spectra and a HITRAN synthetic atmosphere model \citep{1992JQSRT..48..469R}, correct for the barycentric velocity, and measure the centroid of the emission line. We list these values in Table \ref{table:CO_properties}. Although missing error estimates for some stellar radial velocities from the literature make a quantitative interpretation for individual targets difficult, on average the observed CO emission line velocities appear to be similar to or slightly blue shifted from the stellar radial velocities from the literature. A slight blue-shift of the observed CO emission would be in line with a contribution to the CO line emission from a slowly rotating disk wind \citep[c.f. Section \ref{sec:whereco?};][]{2011A&A...527A.119B, 2011ApJ...733...84P}.

\begin{figure*}
   \centering
   \includegraphics{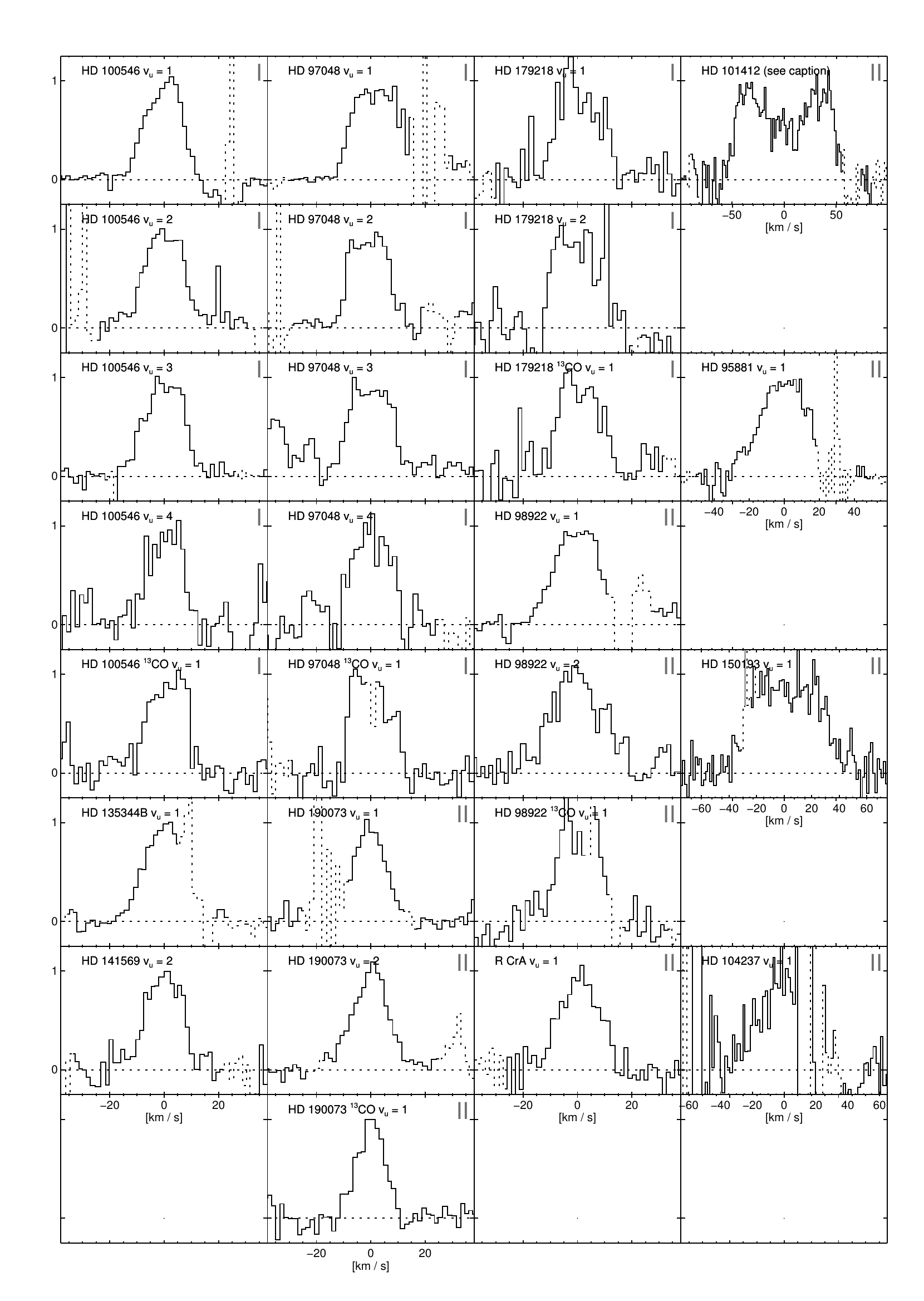}
   \caption{Averaged, normalized composite line profiles for all detected vibrational bands that have sufficient signal as described is Section \ref{sec:kinematics}. The composite profiles of different isotopologues and vibrational bands of the same source are presented below each other, the disk classification (group I/II) is noted on the top right. The line profile for HD\,101412 is a mix of the $^{13}$CO v = 1-0 R23 and $^{12}$CO v = 4-3 R35 lines. }
            \label{fig:co_all}
   \end{figure*}

\subsection{CO first overtone emission}
\label{sec:overtone}

We searched for first overtone CO emission in all our targets except HD\,98922, HD\,104237 and HD\,142666, for which no spectra covering the 2.2--2.3~$\mu$m range are available. We detect the v = 3-1 band head and confirm the presence of CO v = 2-0 overtone emission in the disk around HD\,101412 reported by \citet{2012A&A...537L...6C} and \citet{2014arXiv1409.4897I}. We detect no CO overtone emission in any of the other targets which exhibit CO fundamental emission{\gc , which is in agreement with the 7\% detection rate of CO overtone emission from disks around HAeBe stars by \citet{2014arXiv1409.4897I}}.  For these disks we define 3 $\sigma$ upper limits using a line width of 20 km s$^{-1}$ (a width typical for the fundamental CO emission) and continuum fluxes as published in the 2MASS catalog \citep{2006AJ....131.1163S}. We do detect first overtone CO foreground absorption towards R CrA (c.f. Section \ref{sec:absorption}). See Table \ref{table:2mic} and Figure \ref{fig:2mic} for the spectra, line fluxes and upper limits.

\subsection{Spatially resolved CO and continuum emission}
\label{sec:spatial}

The 4.6 $\mu$m continuum emission is spatially resolved in HD\,97048 and HD\,100546 (see \citet{2009A&A...500.1137V} for details) and, marginally, in  HD\,179218. The continuum emission of HD\,179218 has a FWHM of 0.209$\arcsec \pm$ 0.005$\arcsec$, or a deconvolved, deprojected size of 0.078$\arcsec \pm$ 0.026$\arcsec$ compared to the FWHM of the unresolved PSF of the telluric standards observed directly before (0.194$\arcsec$ $\pm$ 0.005$\arcsec$) and after (0.192$\arcsec$ $\pm$ 0.006$\arcsec$) the observations. 

The CO emission in HD\,97048, HD\,100546 and HD\,141569 is spatially resolved and shows equally resolved red- and blue- shifted components on spatially opposite sides of the star, compatible with a disk in Keplerian rotation. To calculate the spatial extent of the CO emission we sample and subtract the continuum on the red side of the line after scaling it to the intensity of the telluric spectrum at each corresponding spectral bin. We define the outer radius of the CO emission as the maximum extent of the 2 $\sigma$ contour of the continuum subtracted CO v$_u$ = 2 R05 emission line on the position-velocity diagram and the error as 1 pixel (0.086$\arcsec$). Using this procedure, in our spectra the CO emission is spatially resolved up to radii of 0.40$\arcsec$ $\pm$ 0.09$\arcsec$, 0.34$\arcsec$ $\pm$ 0.09$\arcsec$ and 0.30$\arcsec$ $\pm$ 0.09$\arcsec$ for HD\,97048, HD\,100546 and HD\,141569, respectively. For HD\,97048 this is the first published spatial extent of the CO emission. The values for HD\,100546 and HD\,141569 are smaller but consistent with the spatial CO extent published previously, which was based on similar or higher SN spectra \citep[cf.][]{2006ApJ...652..758G, 2012A&A...539A..81G, 2009A&A...500.1137V, 2014A&A...561A.102H}.

The CO emission in R CrA is only resolved in one spatial direction, and blueshifted 10 km s$^{-1}$ with respect to the surrounding Corona Australis molecular cloud (cf. Section \ref{sec:absorption}). This suggests the CO emission in the R CrA spectrum is likely to originate from an outflow. We note that the presence of an outflow is also suggested from the bow-shocks seen in shocked H$_2$ emission in the immediate vicinity of R CrA \citep{2011A&A...533A.137K}. The detected CO emission is spatially unresolved in the data for all other targets.

\subsection{CO absorption}
\label{sec:absorption}

We detect CO absorption in the spectra of R CrA and HD\,97048. In R CrA, we detect the $^{12}$CO v = 0-1, v = 0-2, and  $^{13}$CO v = 0-1 bands in the spectrum. We show two absorption lines, together with the line widths and a Boltzmann plot in Figure \ref{fig:rovib_overtone}.

The absorption lines in R CrA are spectrally resolved, and the radial velocity of the $^{12}$CO v = 1-0 emission is blue shifted 10 km~s$^{-1}$ with respect to the $^{12}$CO v = 0-1, v = 0-2, and  $^{13}$CO v = 0-1 absorption. We calculate a CO isotopologue ratio of $^{12}$CO v = 0-2 / $^{13}$CO v = 0-1 towards R CrA of 142, and a temperature ratio of 57 K / 44 K = 1.3. These values are comparable to values found toward local molecular clouds \citep{2003ApJ...598.1038G}, and are consistent with a scenario in which the absorption lines are formed in the surrounding Corona Australis molecular cloud. 

We detect $^{12}$CO and $^{13}$CO v = 0-1 absorption lines in the spectrum of HD\,97048. The absorption lines in HD\,97048 are unresolved and blended with emission lines, making it difficult to reliably determine the line strengths. The radial velocity of the absorption and emission lines is similar.

\begin{figure}
  \includegraphics[width=\linewidth]{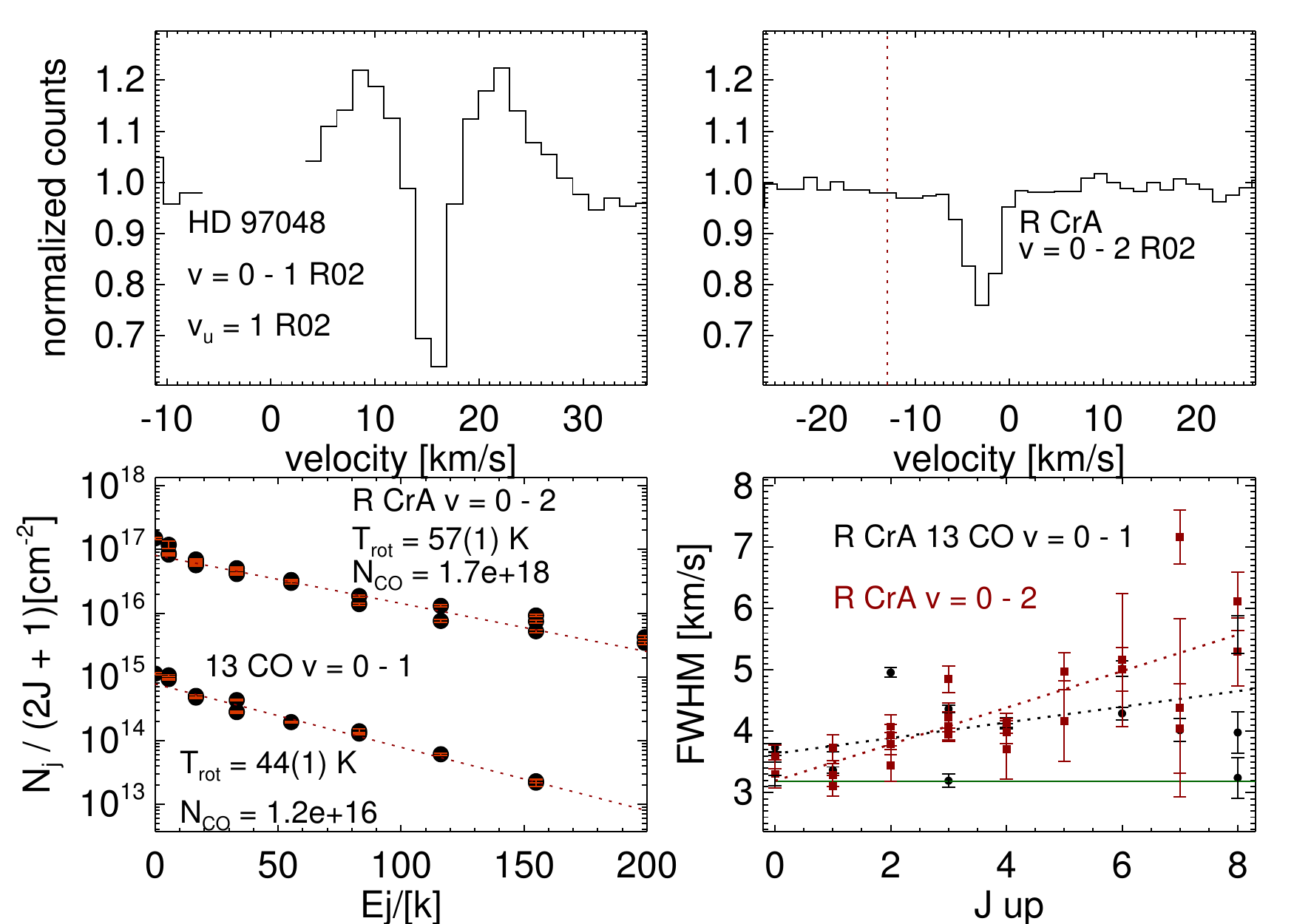}
  \caption{$^{12}$CO v = 1-0 R02 emission and v = 0-1 absorption in HD\,97048 (upper left panel). $^{12}$CO v = 0-2  R02 absorption in R CrA (upper right panel), the line center velocity of the $^{12}$CO v = 1-0 emission lines is shown by the red dashed line. The Boltzmann plot (lower left panel) and line widths (lower right panel) for the  $^{12}$CO v = 0-2 and $^{13}$CO v 0-1 ro-vibrational absorption lines in the spectrum of R CrA (notation similar to Figure \ref{fig:rovib_all}). The vertical green line in the bottom right panel at 3.18 km s$^{-1}$ shows the spectral resolution of our observations.}
  \label{fig:rovib_overtone}
\end{figure}

\subsection{Other emission lines}
\label{sec:others}

Except for HD\,101412, all targets with detected CO emission also show the H\,{ \sc i} recombination lines of Pf\,$\beta$ (4.654 $\mu$m) and Hu\,$\epsilon$ (4.672 $\mu$m) in emission. The line widths of the H\,{\sc i} recombination lines are much wider than those of the CO lines (Figure \ref{fig:4662_5034_all}){\gc , and has been used by \citet{2013ApJ...769...21S} as an accretion tracer}. The origin of these infrared H\,{\sc i} emission lines is likely identical to the optical H\,{\sc i} emission, which is also believed to be linked closely to accretion \citep[e.g.][]{2004AJ....128.1294C, 2006A&A...459..837G, 2011A&A...535A..99M}. The H\,{\sc i} lines we detect here are consistent with those presented in the literature so far. We list the line flux and FWHM values in Table \ref{tab:pfb} and refrain from discussing these in more detail in the current paper.

We detect the neutral Na\,{\sc i}\,D doublet (2.206 and 2.209 $\mu$m, 4p$^2$P$^0$ - 4s$^2$S) in emission in the spectrum of HD\,190073, with a FWHM of  34${\pm3}$ km s$^{-1}$, and a 2.206 $\mu$m / 2.209 $\mu$m line ratio of $\approx$ 2. The Na\,{\sc i} radial velocity in HD\,190073 is -3${\pm3 }$ km s$^{-1}$, and is in agreement with the radial velocity of the central star and the CO emission. The low first ionization potential of neutral sodium (5.1 eV, 243 nm), makes a non-thermal excitation mechanism such as fluorescent pumping by 330.3 nm photons \citep{1977ApJ...216L..75T}, or ionization and subsequent recombination likely. This doublet is more often detected in early-type high-luminosity stars \citep{1988ApJ...324.1071M}, as well as in EXor stars in concert with CO first overtone v = 2 - 0 emission, and in those cases is explained by disk emission \citep{2009ApJ...693.1056L}. We note that CO first overtone emission is not detected in HD\,190073.

HD\,98922 has an un-identified emission line at 4.7485 $\mu$m, similar in line shape and line flux (5.5 $\times$ 10$^{-14}$ erg cm$^{-2}$ s$^{-1}$), but with a FWHM of 14 km~s$^{-1}$ slightly narrower than the CO emission lines.

\subsection{Summary of observed trends in the CO emission}

We detect ro-vibrational fundamental $^{12}$CO emission in 12 out of 13 surveyed HAeBe stars. The temperature of the CO gas in these objects varies, with a few exceptions, between 500 and 2000 K. We also detect  $^{13}$CO emission in 8 of these objects, with temperatures between 200 and 1050 K, all lower than or similar to the  $^{12}$CO rotational temperature in the same disks. There is a correlation between line shape and excitation temperature in at least two disks. However, this correlation between line shape and excitation temperature is absent in the higher vibrational bands in the three flaring disks. In these disks, the CO fundamental emission is excited high up the vibrational ladder (at least up to v = 5-4), but no CO overtone emission is detected. We have spatially resolved the 4.6 $\mu$m continuum emission in all flaring disks and the CO emission in  the two out of three flaring disks, with outer radii for the CO emission of $\approx$ 70 au for HD\,97048 and 35 au for HD\,100546. We detect highly excited CO fundamental and CO overtone emission in the disk around HD\,101412, and CO absorption toward R CrA and HD\,97048.

\section{Location of the CO emitting gas}
\label{sec:whereco?}

Carbon monoxide fundamental emission appears not to be a direct proxy for gas in the disks around HAeBe stars. The CO emission around group I sources originates from larger radii than around group II sources, but in both types of disks there is circumstellar gas present closer to the central star. There is no correlation between the 4.77 $\mu$m continuum flux and radial CO location, ruling out contrast effects. 

 We wish to trace the CO as close as possible to its inner radius. Because of the quality of our data, it is difficult to assess where the line ends and the continuum starts. Therefore, we use the Half Width at 10$\%$ of the line Maximum (HW10M). We define the \textit{onset of the  CO emission} (R$_\mathrm{CO, 10\%}$) using this HW10M, the assumption of Keplerian rotation, and the stellar parameters from Table \ref{table:stellar_parameters}. In case of missing error estimates, we use an error of $\pm$ 10$\%$ for the stellar mass, and $\pm$ 10$^{\circ}$ for the inclination.

We compare the R$_\mathrm{CO, 10\%}$ with the naively expected  cut-off of the inner disk, the dust sublimation radius at 1500 and 2000 K and, when available, the radius of the inner disk rim as traced by 2$\mu$m interferometry. We calculate the dust evaporation radius following
\begin{equation}
 R_\mathrm{subl} ~ = ~ 0.035 ~  \sqrt{Q_R} ~ \sqrt{\frac{L_\mathrm{\star}}{L_\mathrm{\odot}}} ~  \left(\frac{1500}{T_\mathrm{subl}}\right)^2 
\end{equation}
with Q$_R$ the ratio of the dust absorption efficiencies for radiation at color temperature T of the incident and re-emitted field. For a star of 10000 K, $\sqrt{Q_R}$ varies between 1 and 4.5 for grain sizes between respectively 1.00 and 0.01 $\mu$m \citep[cf. Figure 2 of ][]{2002ApJ...579..694M}. For our calculations we adopt $\sqrt{Q_R}$ = 2, but we note that  the choice for $\sqrt{Q_R}$ used in the literature for T Tauri stars varies between 1 and 2, and that this value increases with the color temperature of the central star. We show the HW10M radii, the dust sublimation radii for 1500 and 2000 K, the radius of the inner rim as traced by 2 $\mu$m interferometry, and the PAH-to-stellar-luminosity ratio in Table \ref{table:radii}.

\begin{table}
\small

\caption{Comparison of R$_\mathrm{CO, 10\%}$ to the location of the inner dust disk as calculated from the dust sublimation radius and, where available, as measured with IR interferometry.  The references for the 2 $\mu$m inner rim sizes in the fifth column are: $^a$  \citet{2006ApJ...647..444M}, IOTA, H band.  $^b$ \citet{2009ApJ...692..309E} (Keck, K band). $^c$ \citet{2008A&A...489.1157K}, VLTI/AMBER, K band.  $^d$ \citet{2010A&A...516A..48V}, VLTI/AMBER, K band  $^e$ \citet{2010A&A...511A..75B}, VLTI/AMBER, K band, $^f$ \citet{2014A&A...567A..51C}, VLTI/PIONIER H band. PAH to luminosity ratios in the last column are adapted from \citet{2010ApJ...718..558A} and \citet{2004A&A...426..151A}. The upper limits are 3 $\sigma$, a(b) represents a x 10$^b$, and n.s. means no spectrum is available. {\gc To calculate the error on R$_\mathrm{CO, 10\%}$ for sources with missing error estimates we adopt a 10\% error for the values of the stellar mass(\dag) and the inclination(\ddag)}.}\label{table:radii}

\centering
\renewcommand{\footnoterule}{}  
\def\arraystretch{1.2}
\tabcolsep0.15cm 
\noindent\begin{tabularx}{\columnwidth}{@{\extracolsep{\stretch{1}}}*{6}{l}@{}}
Name       & R$_\mathrm{CO, 10\%}$ & R$_\mathrm{subl}^{1500 K}$ & R$_\mathrm{subl}^{2000 K}$ & R$_\mathrm{dust}^{in, 2 \mu m}$  &  L$^\mathrm{PAH}_{6.2 \mu m}$ / L$_\ast$ \\
  \hline 
Group I    & [au]   &  [au] &  [au]  & [au]         &   \\
  \hline 

HD\,100546  & 6.8$^{\pm1.6}$   & 0.45  &  0.25  &  0.26$~^e$   & 1.30 $\pm$  0.13 (-3)   \\
HD\,97048   & 10.1$^{\pm1.8}$    & 0.36  &  0.20  &              & 1.29 $\pm$  0.01 (-3)       \\
HD\,179218  & 9.2$^{\pm 1.5}$    & 0.61  &  0.34  &  $>$3.1$~^a$ &  1.65  $\pm$  0.17 (-3)    \\
  \hline                                                  
Group II   &        &       &        &              &      \\
  \hline                                                  
HD\,101412  & 0.6$^{\pm0.1}$    & 0.35  &  0.20  &              & 2.57 $\pm$ 0.05 (-4) \\
HD\,190073  & 1.9$^{\pm2.8, \dag}$    & 0.64  &  0.36  &  0.62$~^b$   &  9.61 $\pm$ 2.80 (-5)\\
HD\,98922   & 4.2$^{\pm2.2, \ddag}$   & 2.09  &  1.18  &  1.25$~^c$   &  n.s.\\
HD\,95881   & 1.9$^{\pm0.6, \ddag}$    & 0.33  &  0.18  &  0.37$~ ^d$  &  9.53 $\pm$ 0.08 (-4)\\
HD\,150193  & 0.5$^{\pm0.2}$    & 0.28  &  0.15  &  0.71$~^a$   &  $<$ 5.3 (-5)\\
HD\,104237  & 0.3$^{\pm0.5, \dag}$    & 0.41  &  0.23  &  0.29$~^c$   &  $<$ 5.3 (-6)\\
  \hline                                                  
Not classified   &        &       &        &        &            \\
  \hline                                                  
HD\,141569  & 10.4$^{\pm1.9, \dag}$    & 0.25  &  0.14  &              & 8.70 $\pm$ 0.10 (-5)\\
HD\,135344B & 0.4$^{\pm0.2}$   & 0.22  &  0.13  &      0.2$~^f$        &  2.74 $\pm$ 0.12 (-4)\\

\hline
\end{tabularx}
\end{table}

 Inspection of Table \ref{table:radii} shows that the inner rim sizes determined from the hot dust loosely agree with a dust sublimation temperature between 1500 and 2000 K, and that the CO emission starts at the largest radii for flaring disks. In the group I (flaring) disks, the CO emission starts far beyond the calculated dust sublimation radius, at  $\approx$ 16-48 R$_\mathrm{subl}$. The disks with a group II (self-shadowed) dust disk but still flaring gas (HD\, 95881, HD\, 98922 and HD\, 101412, c.f. Section \ref{sec:PAH}), have intermediate CO inner radii, and the CO onset of the self-shadowed disks without flaring gas disks is approximately similar to R$_\mathrm{subl}$.

The outlier in this picture is HD\,190073, but we argue that it falls within the picture described, given the uncertainty in its stellar parameters.  For HD\,190073, the error on the inclination  of 23$^{+15}_{-23}$ degrees is large. \citet{2004ApJ...613.1049E} note that their data for HD\,190073 is also consistent with an almost face-on inclination. If we adopt an inclination of 12$^{\circ}$, The best fit CO inner radius becomes 0.35 au, the distance of the central star where $T_\mathrm{dust}$ = 2000K. 

\subsection{Keplerian fits to the line profiles}
\label{sec:COwhere?}

To investigate the spatial distribution of the CO emission as well as to find out to what extent the narrow (0.2$\arcsec$) spectrograph slit obscures the disk and influences the line profiles, we create a basic model of the CO emission. In this model, the gas is in Keplerian orbit in a flat disk with known inclination and PA around a star with known stellar mass and distance. The intensity of the emission decreases as $I(R) = I_\mathrm{in}\left(\frac{R}{R_\mathrm{in}}\right)^{-\alpha}$, with $\alpha$ = 2, $I_\mathrm{in}$ the intensity at the inner radius $R_\mathrm{in}$, and $R$ the radial distance from the star. $R_\mathrm{in}$ and $R_\mathrm{out}$ are the only free parameters in this model. The simulated disk is then convolved with the observed telluric PSF, and we project the slit over the simulated disk and calculate the resulting line profile. HD\,104237 and HD\,98922 have no known disk PAs and we assume those to be 0 degrees. Since the CO emission from those sources is not spatially resolved we assume no emission is cut off by the slit and thus that the observed line profile will not be a function of the slit PA. We show our best fits together with a sketch of the best fit disk+slit system in Figure \ref{fig:cosim}. Using our simple Keplerian model, we need outer disk radii $\geq$ 35 au for 8 of the 12 targets.

\begin{figure*}
   \centering
   \includegraphics[width=7.0in]{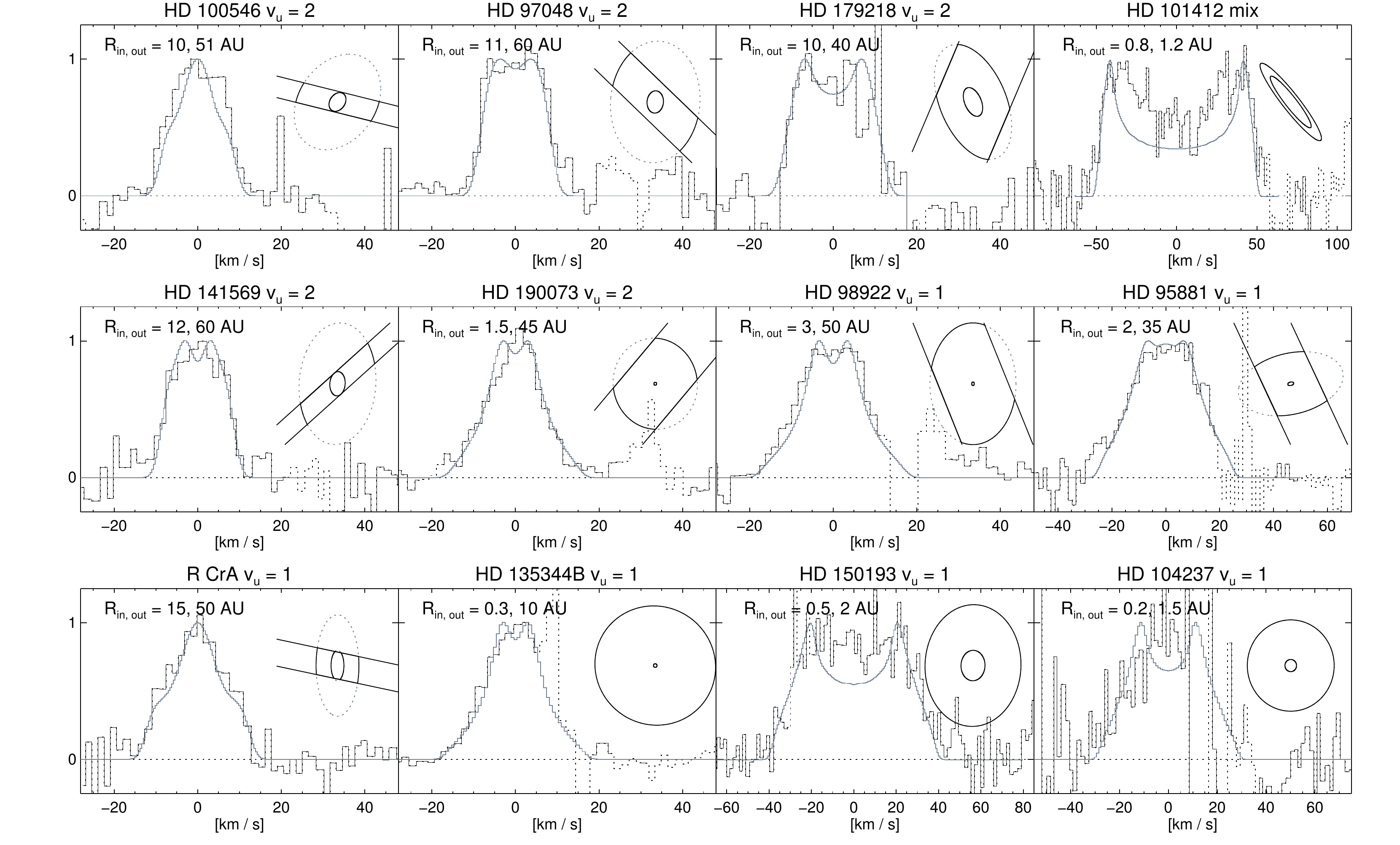}
   \caption{Normalized composite line profiles (black histogram) and the best model fit (gray line). Also plotted are the respective simulated disk+slit (north is up, east is left), and the best fit CO inner- and outer radius used in the model. The line profile for HD\,101412 is a mix of the $^{13}\mathrm{CO}$  v = 1-0 R23 and $^{12}\mathrm{CO}$  v = 4-3 R35 lines.\label{fig:cosim}}
        
   \end{figure*}

\subsection{A comparison to CO emission from T Tauri stars}

CO emission around T Tauri stars, the less massive siblings of HAeBe stars, is often detected at much smaller radii than in HAeBe stars. \citet{2003ApJ...589..931N} find an average FWHM of 70 km~s$^{-1}$ for 12 surveyed T Tauri stars, corresponding to inner radii of 0.02 - 0.05 au, and conclude that most emission comes from (close to) the co rotation radius. \citet{2009ApJ...699..330S} study 14 transitional (circumstellar disks with an optically thick outer zone but an inner region significantly depleted of small dust grains, of which 13 T Tauri stars and one HAeBe star) disks, and find that the disks with partially depleted inner disks most often have CO that extends to rather small ($\leq$1 au) radii, but compared to "classical" disks the CO emission radii are larger than that expected for dust sublimation.

CO emission in T Tauri stars thus probes distances as close as the co-rotation radius, which is also the case for most of the group II sources in our sample. Compared to T Tauri stars, CO emission around HAeBe group I stars originates from larger radii, analogous to the T Tauri stars with partially depleted inner disks.

\section{Discussion}
\label{sec:discussion}

\subsection{CO excitation mechanisms}\label{sec:co-excitation}
Our isothermal slab model successfully reproduces the emission characteristics in all but 3 sources. Only for the group I sources (HD\,100546, HD\,97048 and HD\,179218) do we detect a significant deviation between the rotational and vibrational temperatures. For those sources, we conclude that LTE excitation alone cannot be responsible for the CO excitation.

One other process that could play a role in CO excitation in these disks is the direct radiative pumping of the CO molecule by IR and UV photons.  
\textbf{UV pumping} \citep{1980ApJ...240..940K} is the process where UV radiation from the star or accreting material pumps CO in the ground electronic state to an excited electronic state, which then decays back into an excited vibrational band in the ground electronic state. The effect of UV pumping is to redistribute CO molecules over a large range of vibrational bands. The relative populations of these bands,  and thus the vibrational temperature,  is equal to or lower than the black body temperature of the stellar UV field (depending on the dilution of the stellar UV field which diminishes the influence of the UV pumping over the LTE de-excitation). 
\textbf{IR pumping}  \citep{1980ApJ...240..929S} is the process in which the IR radiation field coming from the star and the local 4.6 $\mu$m dust continuum directly pumps the fundamental CO lines.  This process is most efficient in pumping the v$_u$ = 1 lines (as compared to the higher bands), and plays its biggest role in HAeBe disks in the innermost disk.

Both processes have been used to explain CO emission from disks with non-LTE line populations. See \citep[e.g. ][]{2003ApJ...588..535B,2007ApJ...659..685B, 2009ApJ...702...85B, 2013ApJ...770...94B} for the use of UV pumping and \citet{2004ApJ...606L..73B} for the use or IR pumping.  The relative contribution of radiative pumping compared to collisions differs on a case by case basis and depends on the local temperature, density and radiation field. In general, if the CO emission originates from close to the star where temperatures and densities are high, IR pumping (by photons either coming from the star or the local dust continuum) plays a larger role, and mostly for the v=1-0 transitions. As the emission radius increases and temperatures and density drop, UV fluorescence and collisional excitation take over.  Modeling work by \citet{2013A&A...551A..49T} on CO excitation in Herbig Ae disks shows that UV fluorescence plays a larger role in disks of lower mass and in disks with an inner hole/gap. Even if UV and/or IR pumping dominate over the collisional excitation of CO vibrational bands, they are not expected to have a large impact on the rotational temperature of the CO molecules. The pure rotational collision rates are a factor of 10 - 100 larger than the ro-vibrational rates, and we expect the rotational levels within each vibrational band to reflect the local (LTE) temperature.

Detailed modeling of each individual source is beyond the scope of this paper. Rather, we generalize the above discussion as follows. We interpret the rotational temperature to reflect the local gas temperature. If the vibrational bands are excited up to a high level, UV fluorescence is most likely the dominant excitation mechanism and we then calculate the vibrational temperature from the vibrational bands above  v$_u$ = 1 of all detected bands (given the possible contribution of LTE excitation or IR pumping to that level). If we only detect CO coming from v$_u$ = 1 and 2, and the relative strength of these bands is not compatible with collisional excitation, we interpret this as the presence of IR pumping.

\subsubsection{$^{12}$CO temperature}
\label{sec:discussion-co-rot-temp}

\textbf{Rotational temperature:} Within the confidence intervals given in Table \ref{table:ex_properties}, there is a single excitation temperature for the $^{12}$CO vibrational bands within each source for all our targets except HD\,190073, where the v$_u$ = 2 excitation temperature is significantly  warmer than the v$_u$ = 1 excitation temperature.  Because of the comparable line widths (all disks except HD\,98922) and spatial extent (for the flaring disks), we assume the CO emission coming from all vibrational bands from each disk is emitted from the same surface and calculate one rotational temperature per source for those sources with detections in more than one vibrational band. We do this by summing the $\chi^2$ surfaces for all $^{12}$CO vibrational bands within each target, and list the resulting rotational temperatures in Table \ref{table:vibtemp}. 

The low best fit $^{12}$CO v$_u$ = 1 temperature of HD 97048 compared to the averaged temperature necessitates an emitting surface that is a factor of $\approx$ 1000 larger then that of the higher vibrational bands in this source. Because the CO emission in this source has been resolved \citep{2009A&A...500.1137V} we know that the emitting region in this source is similar for all bands. Most likely, our fitting method underestimates the temperature for the $^{12}$CO v$_u$ = 1 emission in HD 97048.

The $^{12}$CO v$_u$ = 1 temperature for HD 190073 is similarly low. For the respective best fit ($T_{\mathrm{rot}}$, $N$(CO)) parameters the emitting surface of the v$_u$ = 1 emission needs to be a factor of $\approx$ 1000  larger than that of the v$_u$ = 2  emission. This difference shrinks to a factor of $\approx$ 2 when considering a shared best fit $T_{\mathrm{rot}}$ and $N$(CO) obtained by summing the separate $\chi^2$ surfaces. The CO emission is not spatially resolved in this source, and since both best fit surfaces fit within our detection limit it is possible that the v$_u$ = 2  emission is constrained to a smaller, hotter region compared to the v$_u$ = 1  emission.

\textbf{Vibrational temperature:} In the vibrational populations of  HD\,100546 and HD\,179218 and, to a lesser extent, in that of HD\,97048, we observe a break at v = 2-1 (c.f. Figure \ref{fig:vib_12co}). The populations of the higher vibrational bands reflect a temperature higher than the rotational temperature of the gas. Both LTE excitation and IR pumping only affect the lower vibrational bands. UV pumping is required to explain the excitation of the higher vibrational bands. UV pumping alone can not explain the relative over population of the v$_u$ = 1 level. Both collisional excitation and IR pumping can serve to excite the CO  molecules up to the  v$_u$ = 1 level.

\subsubsection{$^{13}$CO temperature}
\label{sec:1213}
In all sources in which $^{13}$CO was detected, its rotational temperature is similar to or lower than the $^{12}$CO rotational temperature.
This temperature difference between the isotopes could e.g. be due to opacity effects and a vertical temperature gradient or to a different radial location of the emitting gas. Adding kinematic information can differentiate between the different  this degeneracy. In panel a of Figure \ref{fig:all-correlations} we compare the line widths of both isotopes, and show that the $^{13}$CO width in the flared disks is within error bars similar to the $^{12}$CO width, whereas the  $^{13}$CO lines in the self-shadowed disks are more narrow, and thus originate farther out. We also plot the ratio of the HW10M of both isotopes against the onset of the  $^{12}$CO emission (derived from the HW10M) in panel b of the same Figure. The magnitude of the radial difference is a function of the onset of the $^{12}$CO emission, and this difference decreases for increasing R$_\mathrm{CO, 10\%}$.  

This behavior regarding the relative location of the $^{12}$CO and $^{13}$CO lines can naturally be explained by the fact that $^{13}$CO needs a larger column of gas to be detected if these lines are at least partially optically thin. This column is then reached at a larger distance from the star, and so the  $^{13}$CO rotational temperatures should be lower and the lines should be narrower than $^{12}$CO, as observed. This difference becomes much smaller for disks with large CO inner radii, i.e. the flaring disks.

\subsection{CO kinematics}

Our disk + slit model (Figure \ref{fig:cosim}) yields good fits for the line wings of the disks, but predicts a double peaked emission line where a single peaked line profile is observed for HD\,98922, HD\,135344B, HD\,141569 and HD\,190073. The lack of low projected velocity CO gas in the models of HD\,98922, HD\,135344B, HD\,141569 and HD\,190073 can be remedied by extending the outer radius of the disks. However, because the spectrograph slit truncates large parts of the outer disk, an outer radius $>$ 100 au is needed to fill in the double peak. Given the quality of our data we expect to spatially resolve the CO emission in our disks outside $\approx$ 5 au for a disk at 100 parsec, but do not do so for all above mentioned targets except HD\,141569. 
This line core issue is also noted for a number of T Tauri stars by \citet{2011A&A...527A.119B} and \citet{2013ApJ...770...94B}. \citet{2011ApJ...733...84P} suggest that for T Tauri stars those centrally peaked line profiles are most likely a combination of emission from the inner part ($<$ a few au) of a circumstellar disk and a slowly moving disk wind, launched by either EUV emission or soft X-rays. It is unclear whether a similar mechanism could explain our HAeBe observations given the weaker winds expended for these types of objects. 
 
\begin{figure*}
        \raggedright
        \begin{subfigure}[b]{0.233\textwidth}
                \includegraphics[width=\textwidth]{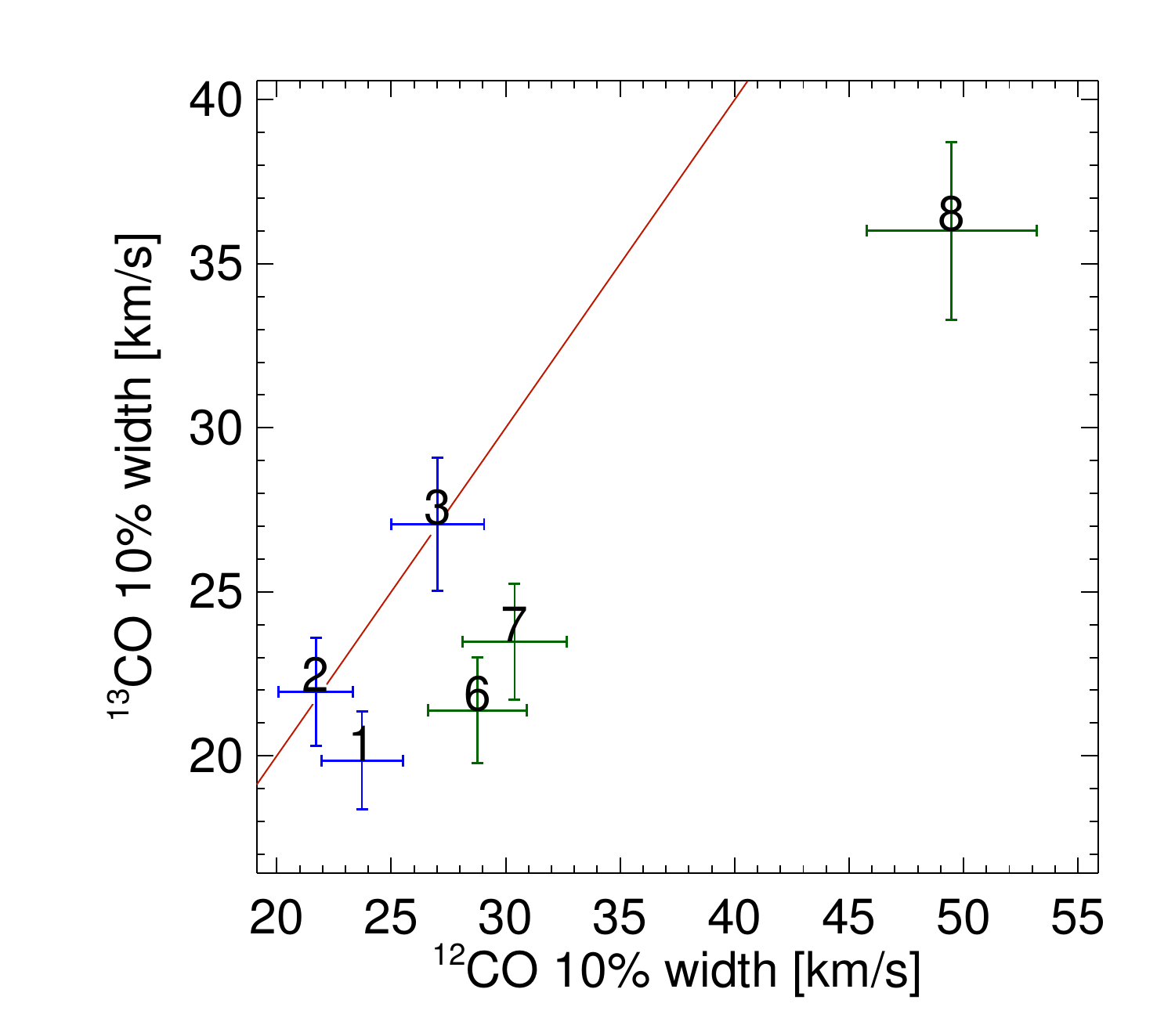}
         \label{fig:1213A}
        \end{subfigure}
        \quad
        \begin{subfigure}[b]{0.233\textwidth}
                \includegraphics[width=\textwidth]{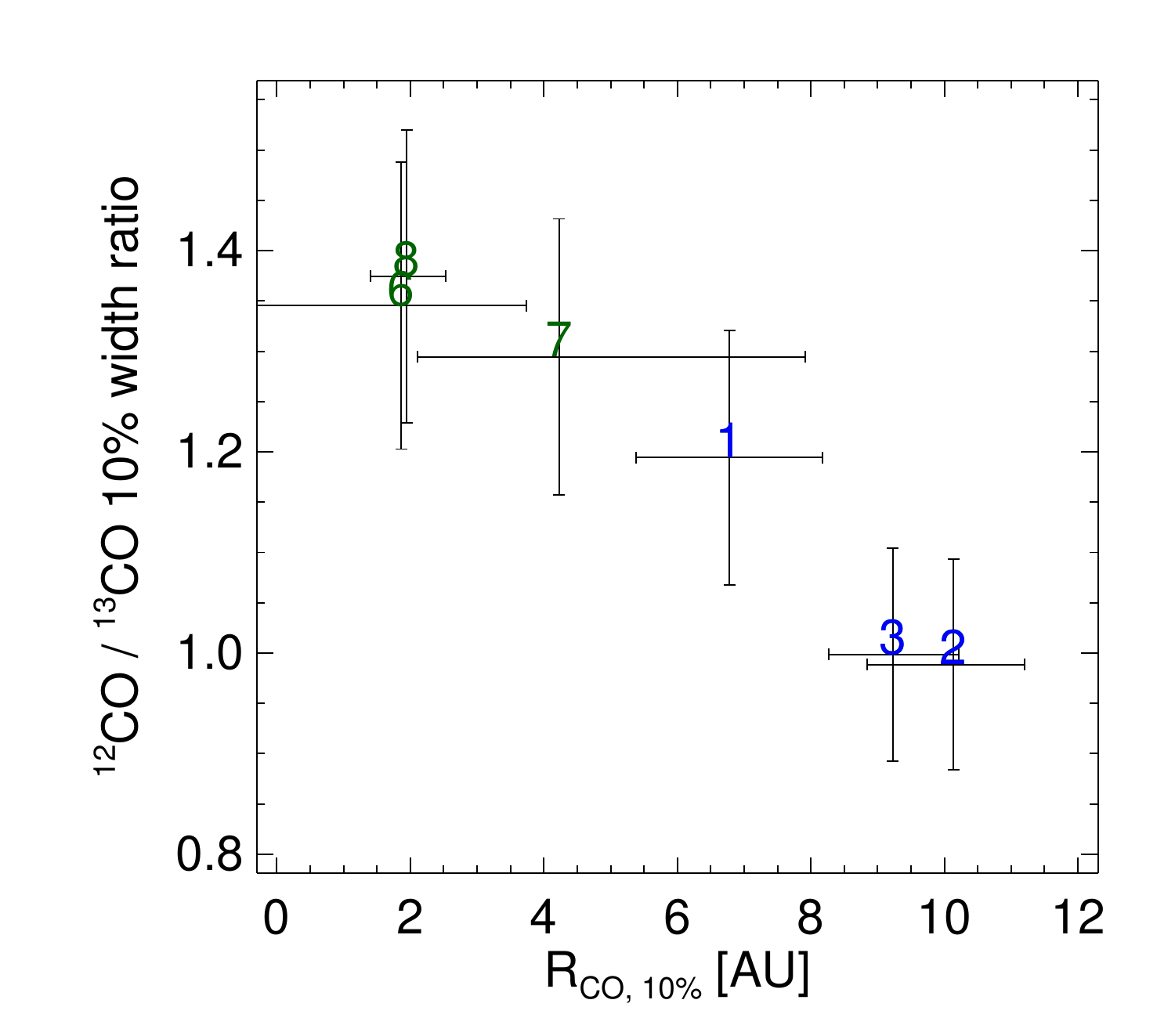}
         \label{fig:1213B}
        \end{subfigure}
\raggedright
       \begin{subfigure}[b]{0.233\textwidth}
                \includegraphics[width=\textwidth]{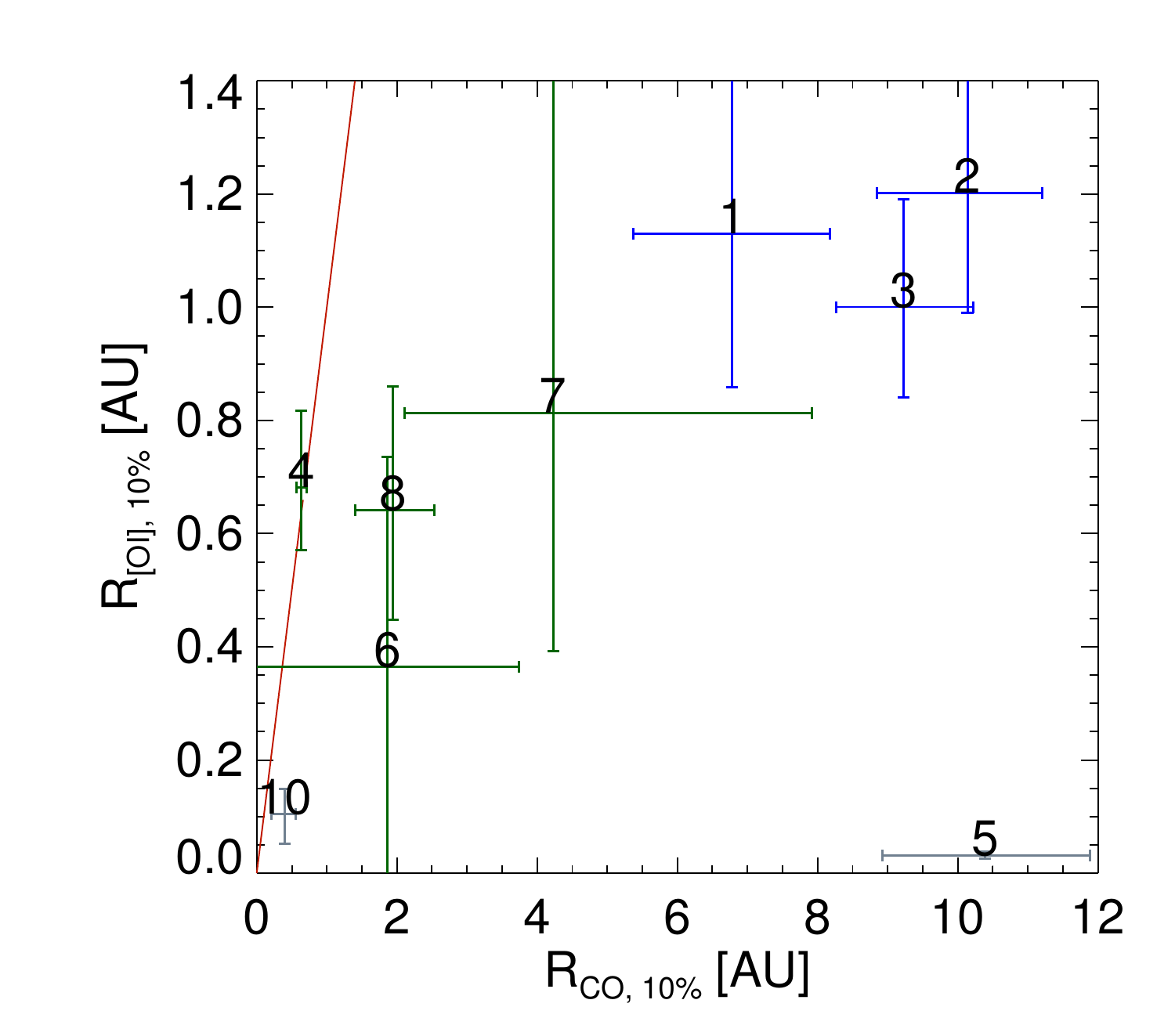}
         \label{fig:sixty_10}
        \end{subfigure}
     \quad
        \begin{subfigure}[b]{0.233\textwidth}
                \includegraphics[width=\textwidth]{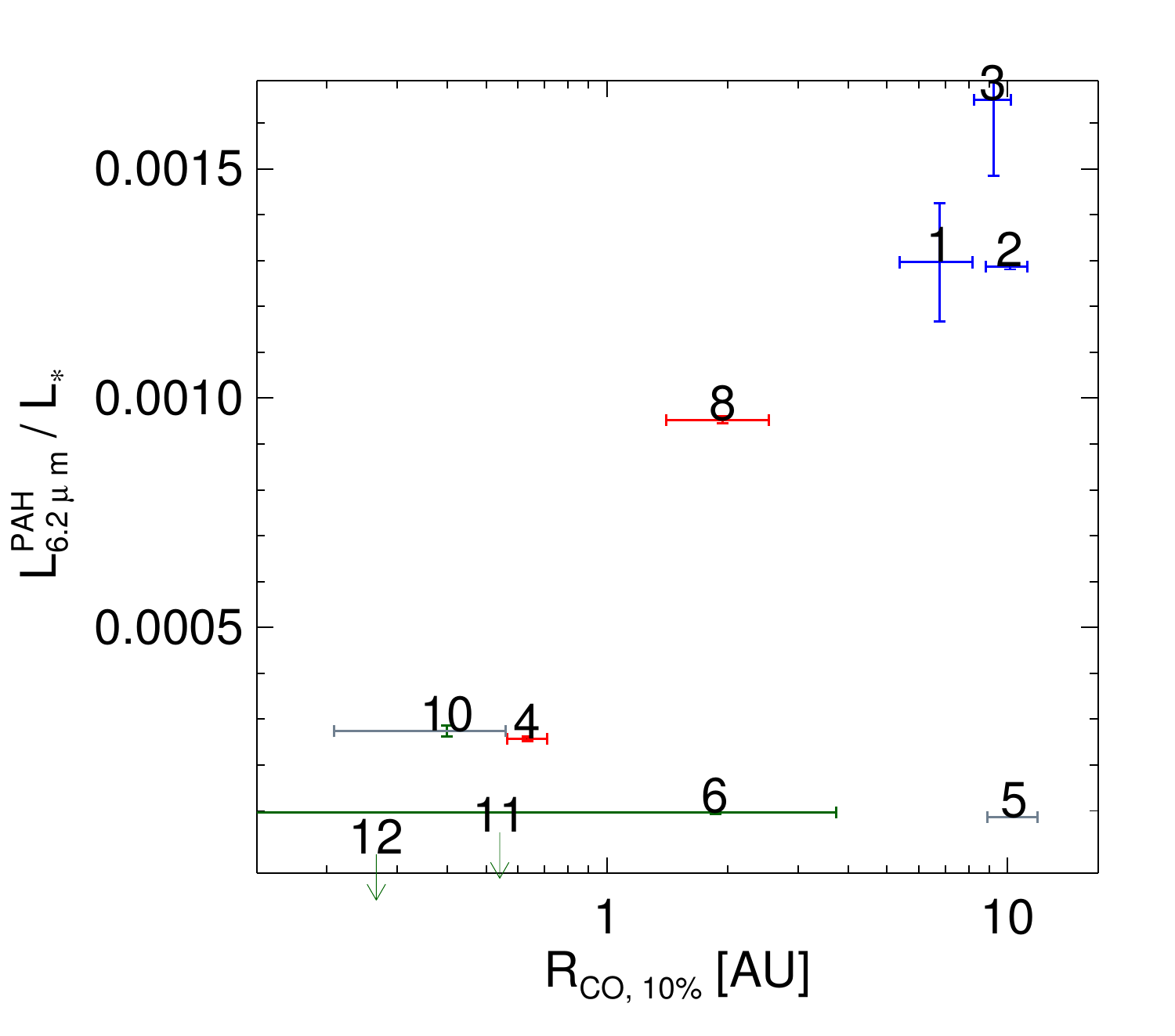}
         \label{fig:PAH2}
        \end{subfigure}
                \caption{(From left to right): \textbf{(a)} Correlation between the half line width at 10$\%$ of the maximum flux of  $^{12}$CO and $^{13}$CO emission. The 1:1 ratio is overplotted with a red line. \textbf{(b)} Correlation between the ratio of the 10$\%$ widths and the onset of the $^{12}$CO emission as derived from the half width at 10$\%$ of the maximum flux. \textbf{(c)} Correlation between the CO and [O\,{\sc i}] inner radii, the red line denotes the 1:1 ratio. \textbf{(d)} Correlation between the CO inner radius and the L$^\mathrm{PAH}_{6.2 \mu m}$ / L$_\ast$ ratio. Sources are named according to column 1 from Table \ref{table:stellar_parameters}. Flaring disks are marked in blue, self-shadowed disks in green and disks with an inner opacity hole are plotted in gray. }
                                \label{fig:all-correlations}
\end{figure*}

\subsection{CO compared to other gas tracers}

To get an as complete census of the gas around the studied HAeBes as possible, we complement the detected CO lines, H\,{\sc i} recombination lines Pf\,$\beta$ and Hu\,$\epsilon$ and in one case the sodium doublet, with two other gas tracers. Here we focus on [O\,{\sc i}] 6300 \AA ~  and Polycyclic Aromatic Hydrocarbon (PAH) emission, which both trace the PP disk surface at different radial regimes. [O\,{\sc i}] traces the gas up to the inner disk, between $\approx$ 0.1 - 50 au, and PAHs trace the outer disk, between $\approx$ 10 and 100 au. 

\subsubsection{CO and [O\,{\sc i}] emission}
\label{sec:gas diagnostics}

[O\,{\sc i}] 6300 \AA ~  emission in HAeBe stars is the by-product of photo-dissociation of OH molecules. It traces those regions where FUV radiation impinges on the (OH, H$_2$O) gas in the atmosphere of circumstellar disks, and is commonly observed in HAeBe stars \citep{2005A&A...436..209A}. Like fundamental CO emission, it traces the circumstellar disk from the innermost disk  out to tens of au. We interpret this emission as a tracer of the disk atmosphere, and show a comparison between the [O\,{\sc i}] and CO emission lines in Figure \ref{oi_co}. With the exception of HD\,101412, the [O\,{\sc i}] lines are much broader than the CO lines, with a HW10M ratio that is typically 2.5. We investigate this difference in line widths as a function of the HW10M radii in panel c of Figure \ref{fig:all-correlations}. The onset of the [O\,{\sc i}] emission is relatively constant at 0.8 $\pm$ 0.4 au, whereas the onset of the CO emission changes from less than 1 to 10 au. The presence of this [O\,{\sc i}] emission, together with the hydrogen recombination and sodium lines reported in Section \ref{sec:others}, shows that the lack of CO emission at small radii does not imply a scarcity of gas closer to the star. This point is also illustrated by the direct comparison of CO and [O\,{\sc i}] emission lines in Figure \ref{oi_co}, where, with the exception of HD\, 101412, the CO emission is about a factor of 2 narrower than the [O\,{\sc i}] line.

 \begin{figure*}
   \centering
   \includegraphics[width=7.0in]{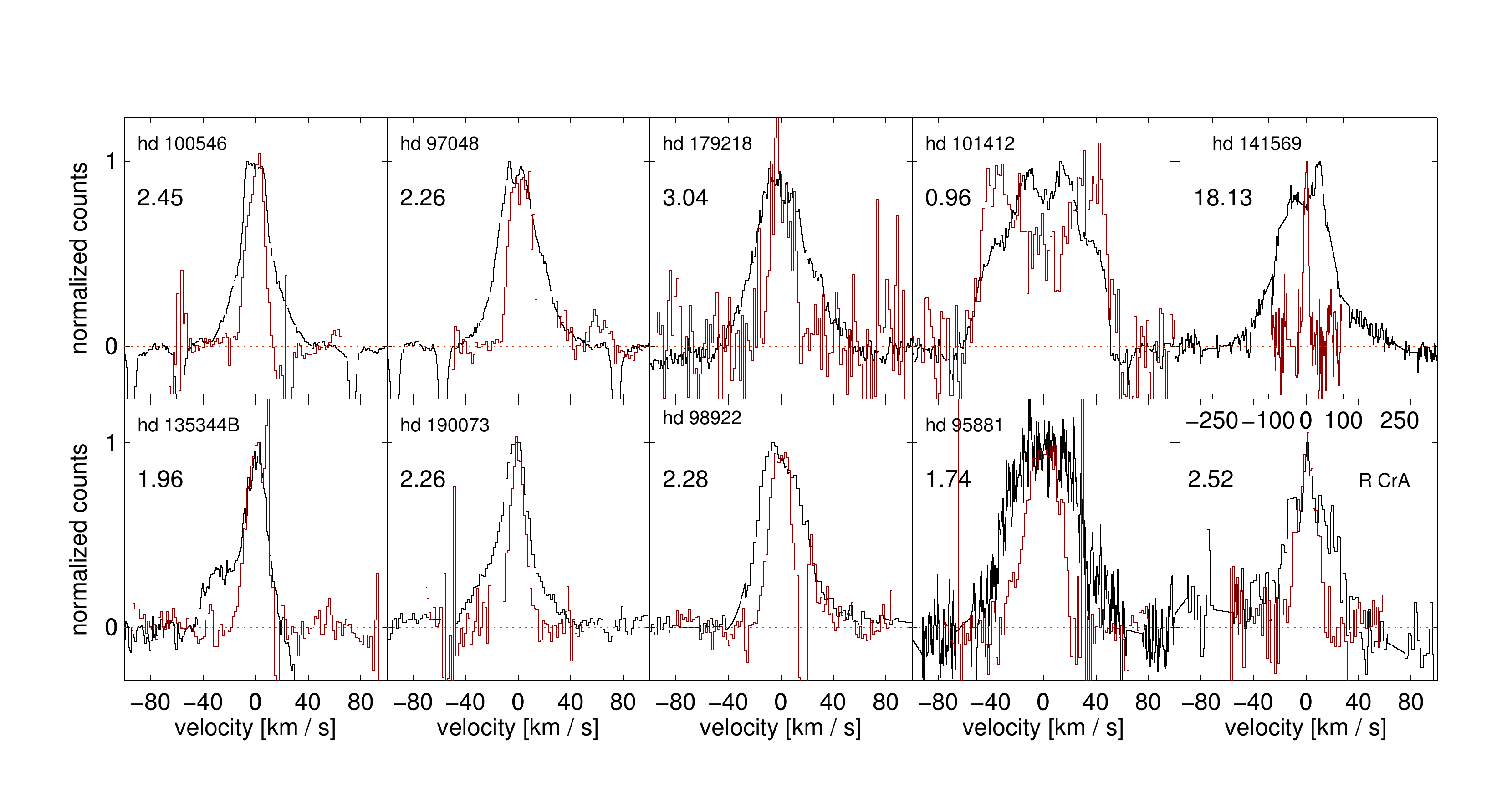}
   \caption{6300 \AA\, [O\,{\sc i}] emission (black histograms), and CO v = 1-0 J $\leq$ 15 average line profiles (red lines), together with their 10$\%$ width [O\,{\sc i}]/CO ratio. The CO emission of HD\,141569 is averaged over the  v = 2-1 lines, and because of problems with the telluric correction, we have used the high (J$_\mathrm{u} \geq$ 30) for R CrA and HD\,150193. The CO line profile plotted for HD\,101412 is a blend of the $^{13}$CO v = 1-0 R(23) and the $^{12}$CO v = 4-3 R(35) lines. We show [O\,{\sc i}] data of HD\,97048 and HD\,100546 from  \citet{2006A&A...449..267A}, of HD\,101412, HD\,135344B, and HD\,179218 from \citet{2008A&A...485..487V}, of HD\,190073 From R. van Lieshout and T. Bagnoli (private communication; HERMES spectrograph at the Mercator telescope), and from \citet{2005A&A...436..209A} for the other stars. We note that both the CO and [O\,{\sc i}] lines have been centered at 0 velocity, and the differing velocity scale for HD\,141569.}
         \label{oi_co}
   \end{figure*}

 \subsubsection{CO and PAH emission}
\label{sec:PAH}

PAHs are often detected and spatially resolved up to 100 au scales in the disks around HAeBe stars, and are a main contributor to the PP disk gas temperature in the upper layers of the disk due to the photo-electric heating of the gas \citep[see e.g.][]{2008ARA&A..46..289T, 2009A&A...501..383W}.  \citet{2001A&A...365..476M} first  suggested a possible correlation between the strength of the PAH features and the shape of the IR SED, which was later confirmed by \citet{2004A&A...426..151A, 2004A&A...427..179H, 2010ApJ...718..558A}. \citet{2010ApJ...718..558A} find that the PAH-to-stellar luminosity ratio is higher in disks with a flared geometry, but that a few of their sources with a flattened dust disk still show relatively strong PAH emission. This picture was modified in a recent paper by \citet{2014A&A...563A..78M}, who found that the PAH emission in transitional disks is dominated by the ionized PAH molecules residing in large disk gaps.

Because PAH molecules trace the gas disk, the presence of PAH emission in the IR spectra of HAeBe disks suggest that while the dust in these disks has already settled, the gas disk may still be flaring. This scenario of still flaring gas but settled dust has been previously suggested for two of our targets: HD\,101412 \citep[Section \ref{sec:HD101412} and ][]{2008A&A...491..809F} and HD\,95881 \citep{2010A&A...516A..48V}.  This scenario was also modeled and works if dust sedimentation is the dominant process in the disk. However, dust sedimentation does not explain the general trend of increasing PAH luminosity with increased flaring, but rather the opposite \citep{2007A&A...473..457D}. We also place HD\,98922 to the group of disks with settled dust but flaring gas based on the similarities in the dust (SED) and gas ([O\,{\sc i}], PAH, CO ro-vibrational) diagnostics between HD\,95881 and HD\,98922. 

We choose the 6.2 $\mu$m PAH band as a proxy for the PAH emission because the relative luminosity of this feature displays the strongest correlation with far-IR excess. We use the PAH-to-stellar luminosity ratio to eliminate uncertainties in distance. We show the PAH-to-stellar luminosity ratio based on Spitzer IRS \citep{2010ApJ...718..558A} and ISO-SWS \citep{2004A&A...426..151A} spectra, in Table \ref{table:radii}.

We plot the CO inner radius against the PAH-to-stellar luminosity ratio in panel d of Figure \ref{fig:all-correlations}. The self shadowed disks with no PAH detections have the smallest CO inner radii, and the inner CO radius increases with increasing PAH strength. The trend between disk-shape and PAH luminosity also holds for our sample, where the flaring disks  show the strongest PAH emission, the PAHs in the self shadowed disks are not detected or have very low strength, and the disks with still flaring gas and self-shadowed dust show intermediate PAH emission.

\subsubsection{The special case of HD\,101412}
\label{sec:HD101412}

HD\,101412 is the only HAeBe star in our sample for which we detect both  fundamental and first overtone CO emission, and where the CO line width is similar to the [O\,{\sc i}] line width.  The Spitzer-IRS spectrum of HD\,101412 has an unusual shape due to the presence of strong PAH emission bands, and its SED is typical for a self-shadowed disk.  However, it also displays some characteristics typical for a flaring disk, such as  extended PAH emission  \citep{2008A&A...491..809F}, and bright [O\,{\sc i}] emission \citep{2005A&A...436..209A} suggesting that this disk might be transitioning from flaring to self-shadowed. \cite{2008A&A...491..809F} and \cite{2008A&A...485..487V} have investigated the dust and gas components independently via high resolution spectroscopy of the [O\,{\sc i}] 6300 \AA\  line and mid-IR interferometry (VLTI/MIDI). The 8--12 $\mu$m emission has been resolved, and can be modeled by a ring between 0.4 and 1.9 au, and shows signs of asymmetry. The PAH feature at 11.3 $\mu$m is more extended than the continuum emission, and the [O\,{\sc i}] emission shows 2 components: One originating from the inner rim, and the other between 6 and several tens of au. Both [O\,{\sc i}] and PAH emission are thought to trace the part of the disk atmosphere that is directly exposed to the stellar UV field. Their conclusion is that the gas and dust in the disk around HD\,101412 are decoupled, with a gas disk that rises out of the shadow cast by the inner rim after $\approx$ 6 au.

  Based on the kinematics of the fundamental and first overtone emission (both have the same radial velocity as the central star, the line widths of the fundamental and first overtone emission are similar, but the line center is more filled in for the fundamental emission), the CO emission is constrained to the inner disk region.   We show the line profiles of [O\,{\sc i}], fundamental and first overtone CO emission in Figure \ref{fig:hd101412}. The comparable [O\,{\sc i}] and CO line widths demonstrate that the onset of their emission in disks around HAeBe stars \textit{can} be co-spatial. As discussed above, the dust and gas in the outer disk of HD\,101412 are decoupled, and the [O\,{\sc i}] emission originates from two components: A high projected velocity component from the hot inner rim, and a lower projected velocity component farther out. We test this interpretation to the first order by over plotting the (scaled) [O\,{\sc i}] emission from HD\,101412 in Figure \ref{fig:hd101412}. The high projected velocity components of both the [O\,{\sc i}] and the CO emission are indeed co-spatial, but the low projected velocity component seen in the [O\,{\sc i}] is \textit{not} detected in the CO emission. 

The detection of [O\,{\sc i}] and non-detection of CO emission farther out in the disk demonstrates that the same trend as observed at small radii - [O\,{\sc i}] becomes, for whatever reason (dust settling, CO depletion, photo-destruction of CO),  visible sooner in the inner disk - also holds farther out in the disk. HD\,101412 has a flaring gas ([O\,{\sc i}], PAH) disk but settled  dust disk. This suggests that the CO emission originates from deeper in the disk (closer to the dust, where it is either protected from dissociation and/or can be thermalized), and the [O\,{\sc i}] emission - as by-product of photo-dissociation - traces those regions of the disk where the stellar UV flux dissociates the OH gas.

\begin{figure}
  \includegraphics[width=\linewidth]{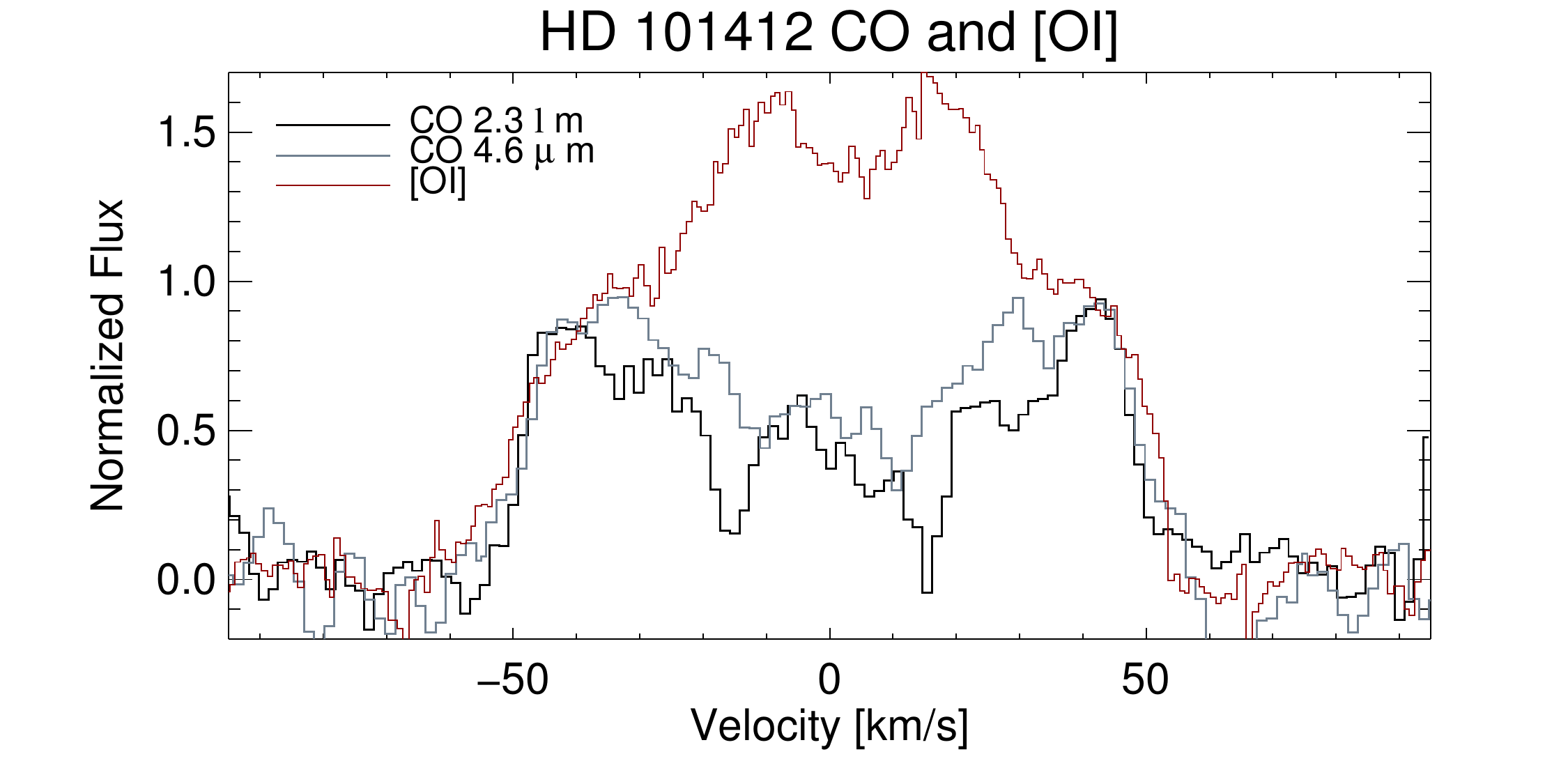}
  \caption{Composite line profiles of CO fundamental (gray) and first overtone (black) emission in HD\,101412, smoothed over 2 bins for clarity. Overplotted (red line) is the [O\,{\sc i}] emission, scaled vertically for the high velocity components to overlap.}
  \label{fig:hd101412}
\end{figure}

\subsection{Using CO ro-vibrational emission to find disk gaps}

It has recently been suggested by \cite{2013A&A...555A..64M} that all group I HAeBe stars have transitional (gapped) disks. The CO emission lines of all three group I sources in our sample show similar line shapes (narrow, flat topped) and dominant excitation condition (UV fluorescence). The disks around two of these objects show signs of harboring intermediate size gaps (HD\,100546 $\approx$ 13 au,  HD\,97048 $\approx$ 34${\pm 4}$ au, \cite{2013A&A...555A..64M}). CO emission from the inner wall of the outer disk, which is in full exposure of the stellar UV flux, would be a natural way to explain the narrow line profiles and the UV fluoresced CO molecules. Based on this similarity, {\gc we suggest that the disk around the 3$^{\mathrm{rd}}$ flaring disk,  HD\, 179218, harbors a similar gap. Given the larger distance to this source a gap  similar to the gap in the disk around HD\, 100546,  and possibly HD\, 97048, may well have been missed in modeling efforts based on the SED alone.}

Not all HAeBe disks with a disk gap show CO emission properties similar to the 3 group I sources discussed in this paper. The CO emission from AB Aur for example is dominated by the inner disk \citep{2003ApJ...588..535B,2004ApJ...606L..73B}, which extends out to $\approx$ 40 au. The UV field at the  inner rim of the outer disk at $\approx$ 300 au is likely too weak to play a significant role in exciting to CO molecules, and the gas at that location is likely so cold that the ro-vibrational CO emissivity is negligible. A similar case can be made for the disk around HD\,135344B.

 We suggest that the three group I sources in our sample trace a specific kind of transitional disks: those with a hole or an inner disk that is small enough so that its CO emission does not dominate the emission spectrum, and with an outer disk that is close enough to the star to be both warm enough to thermally excite the CO molecules and in contact with a strong enough (stellar) UV field to fluoresce the CO molecules.

\section{Conclusions}
\label{sec:conclusion}

 In this paper we present detections of ro-vibrational fundamental CO emission in 12 out of 13 surveyed HAeBe stars. We investigate the kinematics and temperature of this gas, and correlate these with disk dust properties (the amount of disk flaring), and other disk-gas tracers: PAH and [O\,{\sc i}] emission. Keeping in mind the modest sample size, we report the following trends between targets, and within the different isotopologues and vibrational bands for each target:

\vspace{2mm}

[1] CO fundamental emission is common in disks around HAeBe stars;

[2] CO first overtone emission is only detected in one out of 13 surveyed disks. {\gc This is consistent with the 7\% detection rate of CO overtone emission from HAeBe disks reported by \citet{2014arXiv1409.4897I}};

[3] The CO emission in the group I sources originates from larger distances than the CO in group II disks;

{\gc [4] The CO emission in R CrA is most likely originating from an outfow rather than from a disk.}

[5] The rotational temperature of the $^{13}$CO is lower than that of the  $^{12}$CO; 

[6] We detect a broadening of the emission lines as a function of excitation temperature in the disks around two group II sources: HD\,98922 and HD\,135344B. Most other group II sources are difficult to classify due to low S/N. The CO emission in the higher vibrational bands of the group I disks, however, does not show this trend; 

[7] The 4.6 $\mu$m dust continuum emission in all three group I disks, and the CO emission in two group I disks is spatially resolved; 

[8] The dominant excitation mechanism for the CO vibrational band populations is thermal or pumping by IR radiation for group II sources, and fluorescence for the group I sources. 

[9] We propose that the presence of fluoresced CO in Herbig Ae/Be disks is a proxy for the presence of moderately sized gaps in their disks.

[10] Following this recommendation, we suggest that the disk around HD\,179218 is a (pre) transitional disk (harbors a disk gap or hole) based on the CO line width and excitation mechanism.

\vspace{3mm}

These findings are  consistent with a picture in which the CO emission is dominated by the disk surface. In the group I disks the CO emitting surface extends up to large distances, but starts at radii larger than the dust sublimation radius, whereas in the group II disks the CO emanates in a narrow region closer to the dust sublimation radius. There is a strong correlation between the dominant mode of CO excitation and the topology of the circumstellar dust disk. The group I sources have line shapes that do not correlate with excitation energy; their higher vibrational bands are over populated compared to the expected values belonging to the rotational temperature of the gas, and their dust continuum and sometimes CO emission are spatially resolved up to tens of au. The self-shadowed dust disks (group II sources) in contrast do not show these signs. 

The difference between the observed CO emission properties of group I and group II sources are a direct reflection of differences in the excitation mechanism of the gas in the CO emitting region in both groups. This difference appears to be driven by a difference in radial location of the emitting region. Whereas in group II sources the CO emitting region is confined to the inner dust disk, in the group I sources the CO emitting region is more extended, and significant amounts of the observed CO emission originate from radii up to tens of au. The above picture is reminiscent of the interpretation by \citet{2013A&A...555A..64M, 2014A&A...563A..78M} for the presence of ionized PAH emission and spatially resolved Q-band emission in group I disks, where they argue that disk gaps are necessary to explain the observed phenomenology. In this picture, the fluorescent CO emission we have observed in the group I sources in our sample originates from the inner rim of the outer disk (in the cases where this rim is warm enough to thermally excite the rotational CO ladder). We thus suggest that the presence of fluoresced CO in Herbig Ae/Be disks is a proxy for the presence of moderately sized gaps in their disks.

\acknowledgements{GvdP offers his gratitude to B. Acke, R. van Lieshout and T. Bagnoli for providing the [O\,{\sc i}] spectra used in Figure \ref{oi_co}; to the Paranal staff for their assistance during the observations; to I. Kamp, A. Carmona, R. H. Bertelsen, W. F. Thi and the anonymous referee fotheir comments and discussions that have helped strengthen this work. Gvdp acknowledges support from the Millennium Science Initiative (Chilean Ministry of Economy) through grant Nucleus P10-022-F and acknowledges financial support provided by FONDECYT following grant 3140393.  This publication makes use of data products from the Two Micron All Sky Survey, which is a joint project of the University of Massachusetts and the Infrared Processing and Analysis Center/California Institute of Technology, funded by the National Aeronautics and Space Administration and the National Science Foundation. }

\newpage

\appendix

\section{Confidence intervals for $T_{ex}$ and $N$(CO) determination}

To compare our observations with the emission expected from an isothermal slab model with a given $T_{ex}$ and $N$(CO), we minimize the sum over all detections within one ro-vibrational transition following $\frac{1}{n_{obs}}$$\frac{(model - observation)^2} {error^2}$ in the parameter space  between 100 K $\leq$ $T_{ex}$ $\leq$ 2500 K and 10$^{16}$ cm$^{-2}$ $\leq$ $N$(CO) $\leq$ 10$^{20}$ cm$^{-2}$ for $^{12}$CO emission (Figure \ref{fig:cont}), and 100 K $\leq$ $T_{ex}$ $\leq$ 1200 K and 10$^{17}$ cm$^{-2}$ $\leq$ $N$(CO) $\leq$ 10$^{21}$ cm$^{-2}$ for $^{13}$CO emission (Figure \ref{fig:cont13}), in steps of $\Delta$$T_{ex}$ = 25 K and $\Delta$$^{10}$log($N$(CO)) = 0.1. We define our fitting error as to all solutions within a 1 $\sigma$ confidence interval of the best fit value. The best fit values are listed in Table \ref{table:ex_properties}.

   \begin{figure*}
   \centering
   \includegraphics[]{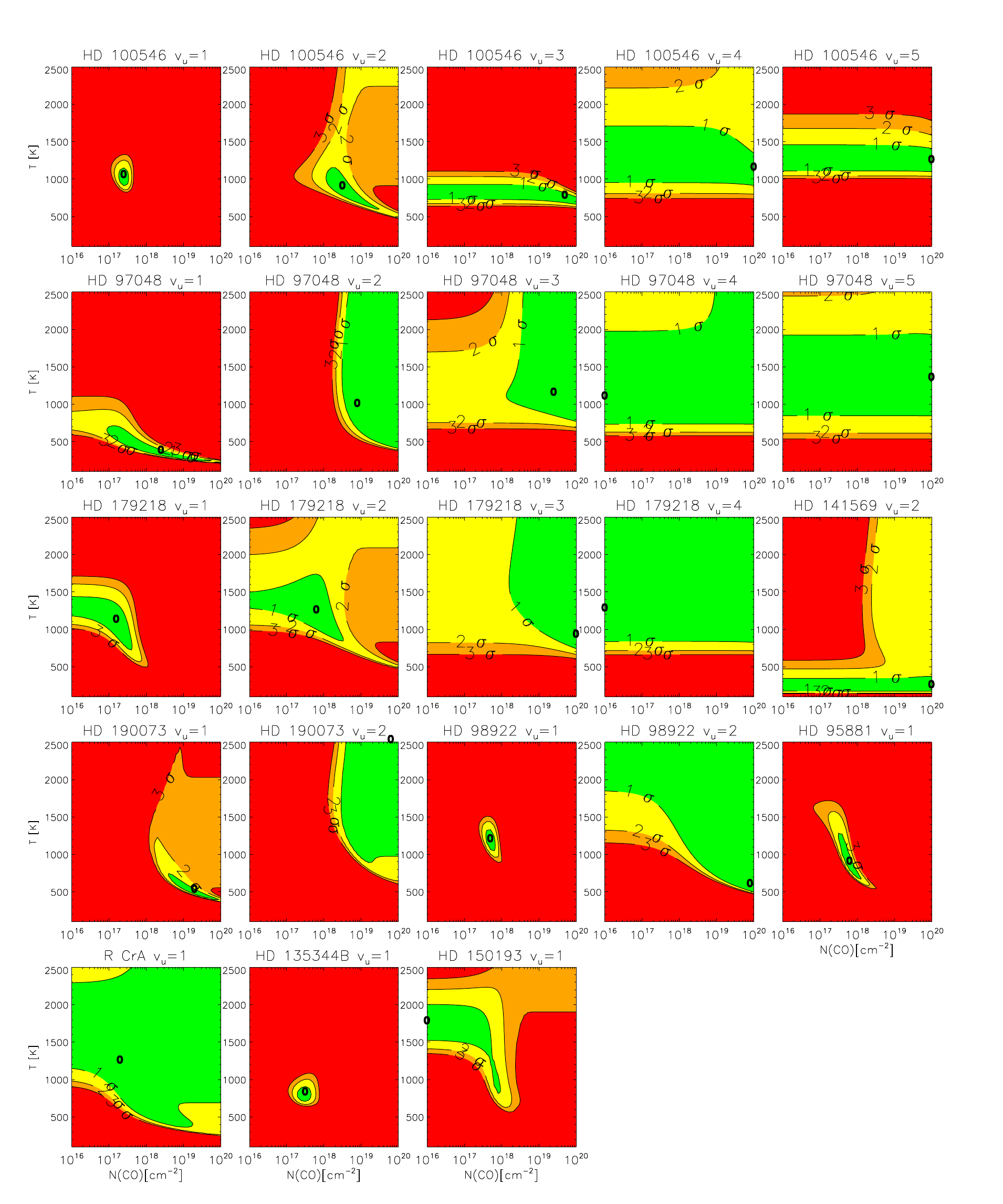}
   \caption{best fit contours for associated $^{12}$CO Boltzmann plots. Color contours denote deviations of 1 $\sigma$ (green), 2 $\sigma$ (yellow), 3 $\sigma$ (orange) and $>$ 3 $\sigma$ (red) confidence intervals}
          \label{fig:cont}
   \end{figure*}

    \begin{figure*}
   \centering
   \includegraphics[width=6.5in]{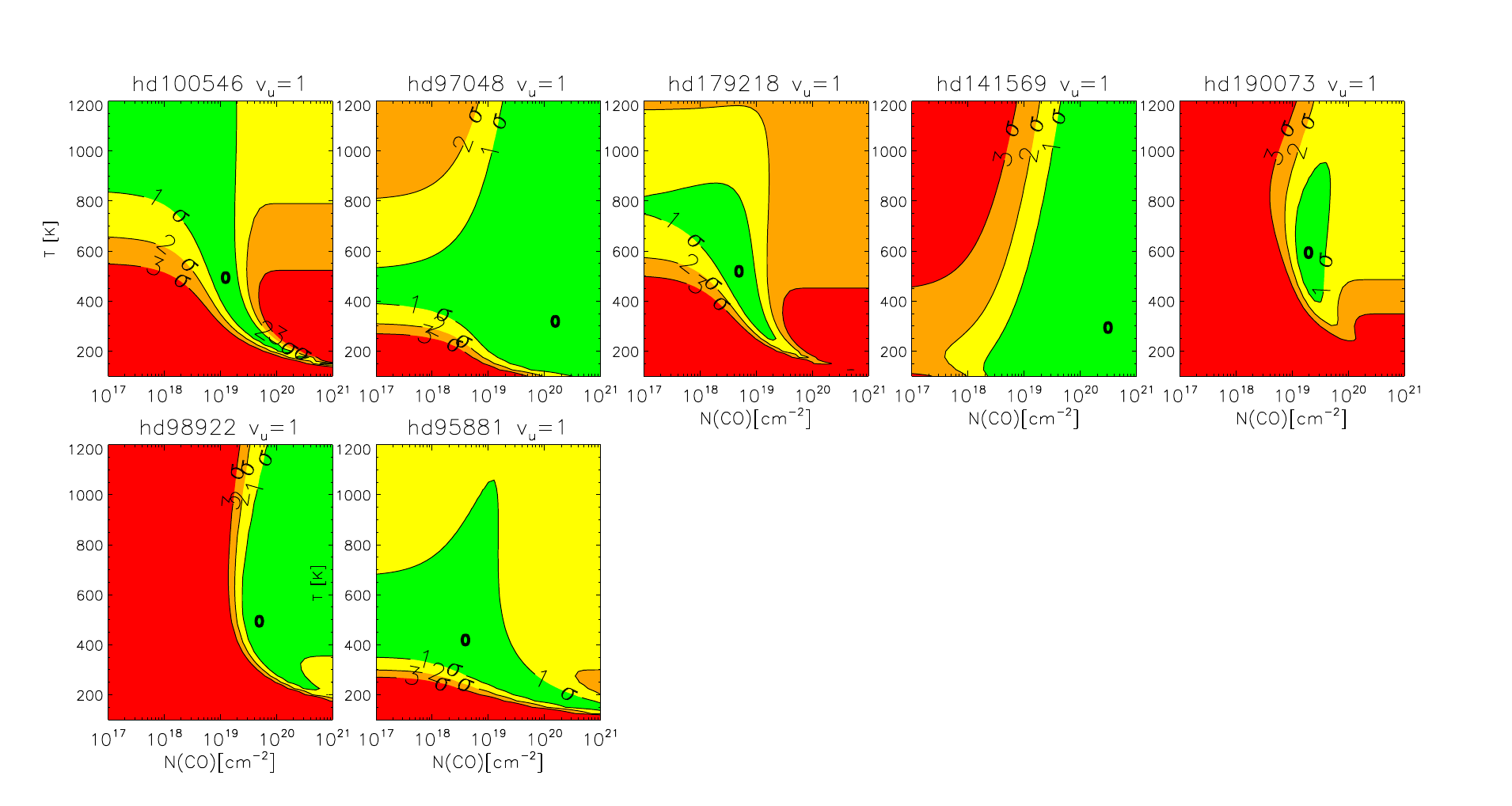}
   \caption{best fit contours for associated $^{13}$CO Boltzmann plots. Color contours denote deviations of 1 $\sigma$ (green), 2 $\sigma$ (yellow), 3 $\sigma$ (orange) and $>$ 3 $\sigma$ (red) confidence intervals}
          \label{fig:cont13}
   \end{figure*}

\clearpage   
\section{Line flux tables}

We list the fluxes and upper limits of the CO v=2-0 R03 lines in Table \ref{table:2mic} and those of the Pfund $\beta$ hydrogen recombination line in Table \ref{tab:pfb}. We were unable to identify this line in the spectrum of HD\,101412 due to the many strong and overlapping ro-vibrational CO lines in the line region (c.g. Figure \ref{fig:4662_5034_all}).

We flux-calibrate the CO overtone spectra using the Ks magnitudes at  2.17 $\mu$m as published in the 2MASS catalog \citep{2006AJ....131.1163S}, and the Pfund $\beta$ and CO fundamental spectra with the 4.77 $\mu$m flux, as determined with a spline interpolation to the M band data points, from \citep[e.g.][]{1985SAAOC...9...55K, 1992ApJ...397..613H, 2001A&A...380..609D}.

 \begin{table}
\caption{Continuum flux values used for calculating the line fluxes listed in Tables \ref{table:2mic} - \ref{default-last}}. 
\centering
\noindent\begin{tabularx}{\columnwidth}{@{\extracolsep{\stretch{1}}}*{3}{l}@{}}
Target & F$_{continuum, 2.17 \mu m}$ & F$_{continuum, 4.77 \mu m}$ \\
 \hline
 & [erg cm$^{-2}$ $\mu$m$^{-1}$  s$^{-1}$] & [erg cm$^{-2}$ $\mu$m$^{-1}$ s$^{-1}$]\\
\hline
HD\,100546 & $2.8 \times 10^{-9}$  &  6.9 $\times$ $10^{-10}$\\
HD\,97048  & $2.0 \times 10^{-9}$  &    3.1  $\times$  $10^{-10}$\\
HD\,179218 & $3.1 \times 10^{-8}$  &   4.7  $\times$  $10^{-10}$\\
HD\,101412 & $2.2 \times 10^{-9}$  &   2.0 $\times$  $10^{-10}$\\
HD\,141569 & $2.0 \times 10^{-9}$  &  5.5  $\times$  $10^{-11}$\\
HD\,190073 & $8.0 \times 10^{-10}$ &   5.8 $\times$  $10^{-10}$\\
HD\,98922  &  &   2.7  $\times$  $10^{-9}$\\
R CrA      & $1.7 \times 10^{-9}$  &    8.1 $\times$  $10^{-9}$\\
HD\,135344B & $1.8 \times 10^{-9}$  &    3.6  $\times$  $10^{-10}$\\
HD\,150193 & $2.9 \times 10^{-9}$  &    6.9  $\times$  $10^{-10}$\\
HD\,104237 &   &    1.1  $\times$  $10^{-9}$\\
HD\,95881  & $4.4 \times 10^{-10}$ &   6.6  $\times$  $10^{-10}$\\
\end{tabularx}
\label{tab:contflux}
\end{table}

  \begin{table}
\caption{Line fluxes and upper limits of the CO v=2-0 R03 line at 2337.4128 nm. Detection limits are 3 $\sigma$ assuming a line width of 20 km s$^{-1}$ and the continuum fluxes listed in Table \ref{tab:contflux}. We note that there are no CO overtone observations for HD\,98922, HD\,104237 and HD\,142666} 
\centering
\noindent\begin{tabularx}{\columnwidth}{@{\extracolsep{\stretch{1}}}*{3}{l}@{}}
 Target & Line ID & F$_{line}$ \\
 \hline
 &  & [erg cm$^{-2}$ s$^{-1}$]\\
\hline
HD\,100546 & $^{12}$CO v=2-0 R03  & $< 8.1 \times 10^{-15}$  \\
HD\,97048  & $^{12}$CO v=2-0 R03  & $< 1.2 \times 10^{-14}$  \\
HD\,179218 & $^{12}$CO v=2-0 R03  & $< 1.2 \times 10^{-13}$  \\
HD\,101412 & $^{12}$CO v=2-0 R03   & $4.6 \times 10^{-14}$  \\
HD\,141569 & $^{12}$CO v=2-0 R03   & $< 1.04 \times 10^{-14}$  \\
HD\,190073 & $^{12}$CO v=2-0 R03  & $< 5.5 \times 10^{-15}$  \\
HD\,95881  & $^{12}$CO v=2-0 R03  & $< 2.2 \times 10^{-15}$  \\
R CrA      & $^{12}$CO v=2-0 R03   & $-1.0 \times 10^{-14}$  \\
HD\,135344 & $^{12}$CO v=2-0 R03   & $< 8.9 \times 10^{-15}$  \\
HD\,150193 & $^{12}$CO v=2-0 R03   & $< 1.3 \times 10^{-14}$  \\
\hline
\end{tabularx}
\label{table:2mic}
\end{table}

\begin{table}[htdp]
\caption{Pfund $\beta$ line fluxes and FWHM values for the programme stars. Line fluxes are calculated using the continuum fluxes listed in Table \ref{tab:contflux}. We note that for HD\,95881  and HD\,150193 (marked with a "$^\star$"),  the Pfund $\beta$ line there are lines visible in the raw spectra and have a respective FWHM of $\approx$ 100 and 180 km s$^{-1}$ and a maximum intensity of $\approx$ 20 \% and 15 \% above the continuum. Because of the presence of Pfund $\beta$ emission in the spectrum of the telluric standard star we are unable to calculate the line fluxes, and therefore do not list these values in the Table.}
\begin{center}
\noindent\begin{tabularx}{\columnwidth}{@{\extracolsep{\stretch{1}}}*{5}{l}@{}}

Target & FWHM & error & Line flux & error\\
 & [km s$^{-1}$] & [km s$^{-1}$] & [erg cm$^{-2}$ s$^{-1}$ ] & [erg cm$^{-2}$ s$^{-1}$ ]\\
\hline

HD\,100546  & 218 & 17 & 8.75 $\times$ 10$^{-13}$ & 1.31 $\times$ 10$^{-13}$\\
HD\,97048   & 149 & 10 & 5.71 $\times$ 10$^{-13}$ & 9.99 $\times$ 10$^{-14}$\\
HD\,179218  & 150 & 10 & 8.19 $\times$ 10$^{-13}$ & 9.35 $\times$ 10$^{-14}$\\
HD\,101412  &  - & - &  $<$ 8.7 $\times$ 10$^{-14}$ &  \\
HD\,141569  & 420 & 60 & 1.19 $\times$ 10$^{-13}$ & 3.32 $\times$ 10$^{-14}$\\
HD\,190073  & 92 & 9 & 3.41 $\times$ 10$^{-13}$ & 8.68 $\times$ 10$^{-14}$\\
HD\,98922   & 140 & 12 & 1.07 $\times$ 10$^{-12}$ & 2.67 $\times$ 10$^{-13}$\\
HD\,95881$^\star$ & -  & - &- &- \\
R CrA      & 230 & 40 & 2.82 $\times$ 10$^{-12}$ & 6.45 $\times$ 10$^{-13}$\\
HD\,135344B & 142 & 10 & 8.06 $\times$ 10$^{-14}$ & 4.03 $\times$ 10$^{-15}$\\
HD\,150193$^\star$ &  - &- & - &-\\
HD\,104237  & 117 & 14 & 1.08 $\times$ 10$^{-12}$ & 1.14 $\times$ 10$^{-13}$\\
HD\,142666  & 325 & 30 & 5.82 $\times$ 10$^{-14}$ & 1.77 $\times$ 10$^{-13}$\\
\hline
\end{tabularx}
\end{center}
\label{tab:pfb}
\end{table}%

\begin{table}[htdp]
\caption{Line flux and error values used in this manuscript. Line equivalent widths are converted to line fluxes using the conversion values listed in Table \ref{tab:contflux}}
\begin{center}
\begin{tabular}{llll}

Target & line id & Line flux & Error\\
 &  & [erg cm$^{-2}$ s$^{-1}$] & [erg cm$^{-2}$ s$^{-1}$]\\
\hline
HD 100546 & $^{12}$CO v(1-0) R08 & 1.18$\times$ 10$^{-13}$ & 4.94$\times$ 10$^{-15}$ \\
 & $^{12}$CO v(1-0) R07 & 1.18$\times$ 10$^{-13}$ & 4.70$\times$ 10$^{-15}$ \\
 & $^{12}$CO v(1-0) R06 & 1.30$\times$ 10$^{-13}$ & 5.11$\times$ 10$^{-15}$ \\
 & $^{12}$CO v(1-0) R06 & 1.22$\times$ 10$^{-13}$ & 5.01$\times$ 10$^{-15}$ \\
 & $^{12}$CO v(1-0) R03 & 1.29$\times$ 10$^{-13}$ & 4.27$\times$ 10$^{-15}$ \\
 & $^{12}$CO v(1-0) R02 & 9.54$\times$ 10$^{-14}$ & 3.48$\times$ 10$^{-15}$ \\
 & $^{12}$CO v(1-0) R01 & 7.50$\times$ 10$^{-14}$ & 3.19$\times$ 10$^{-15}$ \\
 & $^{12}$CO v(1-0) P02 & 5.55$\times$ 10$^{-14}$ & 3.00$\times$ 10$^{-15}$ \\
 & $^{12}$CO v(1-0) P05 & 1.12$\times$ 10$^{-13}$ & 4.25$\times$ 10$^{-15}$ \\
 & $^{12}$CO v(1-0) P07 & 1.24$\times$ 10$^{-13}$ & 6.33$\times$ 10$^{-15}$ \\
 & $^{12}$CO v(1-0) P08 & 1.34$\times$ 10$^{-13}$ & 5.18$\times$ 10$^{-15}$ \\
 & $^{12}$CO v(1-0) P08 & 1.41$\times$ 10$^{-13}$ & 4.85$\times$ 10$^{-15}$ \\
 & $^{12}$CO v(1-0) P11 & 1.45$\times$ 10$^{-13}$ & 4.98$\times$ 10$^{-15}$ \\
 & $^{12}$CO v(1-0) P11 & 1.34$\times$ 10$^{-13}$ & 4.89$\times$ 10$^{-15}$ \\
 & $^{12}$CO v(1-0) P12 & 1.52$\times$ 10$^{-13}$ & 5.89$\times$ 10$^{-15}$ \\
 & $^{12}$CO v(1-0) P14 & 1.50$\times$ 10$^{-13}$ & 6.38$\times$ 10$^{-15}$ \\
 & $^{12}$CO v(1-0) P30 & 7.09$\times$ 10$^{-14}$ & 3.06$\times$ 10$^{-15}$ \\
 & $^{12}$CO v(1-0) P32 & 5.07$\times$ 10$^{-14}$ & 8.10$\times$ 10$^{-15}$ \\
 & $^{12}$CO v(1-0) P32 & 5.07$\times$ 10$^{-14}$ & 7.12$\times$ 10$^{-15}$ \\
 & $^{12}$CO v(1-0) P36 & 4.84$\times$ 10$^{-14}$ & 1.32$\times$ 10$^{-14}$ \\
 & $^{12}$CO v(1-0) P36 & 3.82$\times$ 10$^{-14}$ & 9.41$\times$ 10$^{-15}$ \\
 & $^{12}$CO v(1-0) P37 & 3.83$\times$ 10$^{-14}$ & 5.42$\times$ 10$^{-15}$ \\
\hline
HD 100546 & $^{13}$CO v(1-0) R18 & 1.07$\times$ 10$^{-14}$ & 1.22$\times$ 10$^{-15}$ \\
 & $^{13}$CO v(1-0) R15 & 1.08$\times$ 10$^{-14}$ & 1.92$\times$ 10$^{-15}$ \\
 & $^{13}$CO v(1-0) R15 & 1.61$\times$ 10$^{-14}$ & 8.01$\times$ 10$^{-16}$ \\
 & $^{13}$CO v(1-0) R13 & 1.58$\times$ 10$^{-14}$ & 7.92$\times$ 10$^{-16}$ \\
 & $^{13}$CO v(1-0) R12 & 1.33$\times$ 10$^{-14}$ & 2.35$\times$ 10$^{-15}$ \\
 & $^{13}$CO v(1-0) R12 & 1.75$\times$ 10$^{-14}$ & 2.64$\times$ 10$^{-15}$ \\
 & $^{13}$CO v(1-0) R09 & 1.58$\times$ 10$^{-14}$ & 2.04$\times$ 10$^{-15}$ \\
 & $^{13}$CO v(1-0) P23 & 1.07$\times$ 10$^{-14}$ & 7.33$\times$ 10$^{-16}$ \\
\hline
HD 100546 & $^{12}$CO v(2-1) R17 & 3.64$\times$ 10$^{-14}$ & 1.93$\times$ 10$^{-15}$ \\
 & $^{12}$CO v(2-1) R16 & 5.13$\times$ 10$^{-14}$ & 4.46$\times$ 10$^{-15}$ \\
 & $^{12}$CO v(2-1) R14 & 3.97$\times$ 10$^{-14}$ & 2.39$\times$ 10$^{-15}$ \\
 & $^{12}$CO v(2-1) R14 & 4.35$\times$ 10$^{-14}$ & 3.75$\times$ 10$^{-15}$ \\
 & $^{12}$CO v(2-1) R12 & 5.42$\times$ 10$^{-14}$ & 7.79$\times$ 10$^{-15}$ \\
 & $^{12}$CO v(2-1) R12 & 6.18$\times$ 10$^{-14}$ & 2.64$\times$ 10$^{-15}$ \\
 & $^{12}$CO v(2-1) R11 & 5.47$\times$ 10$^{-14}$ & 2.20$\times$ 10$^{-15}$ \\
 & $^{12}$CO v(2-1) R10 & 5.90$\times$ 10$^{-14}$ & 2.35$\times$ 10$^{-15}$ \\
 & $^{12}$CO v(2-1) R10 & 6.03$\times$ 10$^{-14}$ & 2.30$\times$ 10$^{-15}$ \\
 & $^{12}$CO v(2-1) R09 & 5.14$\times$ 10$^{-14}$ & 2.75$\times$ 10$^{-15}$ \\
 & $^{12}$CO v(2-1) R06 & 5.61$\times$ 10$^{-14}$ & 2.63$\times$ 10$^{-15}$ \\
 & $^{12}$CO v(2-1) R06 & 5.56$\times$ 10$^{-14}$ & 2.36$\times$ 10$^{-15}$ \\
 & $^{12}$CO v(2-1) R05 & 4.62$\times$ 10$^{-14}$ & 2.09$\times$ 10$^{-15}$ \\
 & $^{12}$CO v(2-1) R04 & 3.45$\times$ 10$^{-14}$ & 2.02$\times$ 10$^{-15}$ \\
 & $^{12}$CO v(2-1) P25 & 3.54$\times$ 10$^{-14}$ & 5.15$\times$ 10$^{-15}$ \\
 & $^{12}$CO v(2-1) P25 & 3.65$\times$ 10$^{-14}$ & 2.46$\times$ 10$^{-15}$ \\
 & $^{12}$CO v(2-1) P26 & 3.64$\times$ 10$^{-14}$ & 3.46$\times$ 10$^{-15}$ \\
 & $^{12}$CO v(2-1) P28 & 2.66$\times$ 10$^{-14}$ & 2.36$\times$ 10$^{-15}$ \\
 & $^{12}$CO v(2-1) P28 & 3.06$\times$ 10$^{-14}$ & 3.99$\times$ 10$^{-15}$ \\
\hline
HD 100546 & $^{12}$CO v(3-2) R25 & 2.11$\times$ 10$^{-14}$ & 1.63$\times$ 10$^{-15}$ \\
 & $^{12}$CO v(3-2) R23 & 2.21$\times$ 10$^{-14}$ & 1.11$\times$ 10$^{-15}$ \\
 & $^{12}$CO v(3-2) R23 & 2.03$\times$ 10$^{-14}$ & 1.03$\times$ 10$^{-15}$ \\
 & $^{12}$CO v(3-2) R20 & 2.12$\times$ 10$^{-14}$ & 1.19$\times$ 10$^{-15}$ \\
 & $^{12}$CO v(3-2) R18 & 2.64$\times$ 10$^{-14}$ & 1.56$\times$ 10$^{-15}$ \\
 & $^{12}$CO v(3-2) R17 & 3.31$\times$ 10$^{-14}$ & 1.90$\times$ 10$^{-15}$ \\
 & $^{12}$CO v(3-2) R14 & 3.70$\times$ 10$^{-14}$ & 1.50$\times$ 10$^{-15}$ \\
 & $^{12}$CO v(3-2) R14 & 3.90$\times$ 10$^{-14}$ & 1.44$\times$ 10$^{-15}$ \\
 & $^{12}$CO v(3-2) R11 & 3.93$\times$ 10$^{-14}$ & 1.59$\times$ 10$^{-15}$ \\
 & $^{12}$CO v(3-2) R08 & 3.86$\times$ 10$^{-14}$ & 2.96$\times$ 10$^{-15}$ \\
 & $^{12}$CO v(3-2) R08 & 3.49$\times$ 10$^{-14}$ & 2.09$\times$ 10$^{-15}$ \\
 & $^{12}$CO v(3-2) R05 & 3.19$\times$ 10$^{-14}$ & 3.00$\times$ 10$^{-15}$ \\
\hline
\end{tabular}
\end{center}
\label{tab:default11}
\end{table}%

\begin{table}[htdp]
\caption{Table \ref{tab:default11} continued}
\begin{center}
\begin{tabular}{llll}
Target & line id & Line flux & Error\\
 &  & [erg cm$^{-2}$ s$^{-1}$] & [erg cm$^{-2}$ s$^{-1}$]\\
\hline
 & $^{12}$CO v(3-2) R01 & 2.08$\times$ 10$^{-14}$ & 4.52$\times$ 10$^{-15}$ \\
 & $^{12}$CO v(3-2) P21 & 2.07$\times$ 10$^{-14}$ & 1.70$\times$ 10$^{-15}$ \\
 & $^{12}$CO v(3-2) P21 & 2.19$\times$ 10$^{-14}$ & 3.91$\times$ 10$^{-15}$ \\
\hline
HD 100546 & $^{12}$CO v(4-3) R32 & 1.20$\times$ 10$^{-14}$ & 1.77$\times$ 10$^{-15}$ \\
 & $^{12}$CO v(4-3) R27 & 1.84$\times$ 10$^{-14}$ & 1.62$\times$ 10$^{-15}$ \\
 & $^{12}$CO v(4-3) R27 & 1.48$\times$ 10$^{-14}$ & 5.04$\times$ 10$^{-15}$ \\
 & $^{12}$CO v(4-3) R23 & 1.82$\times$ 10$^{-14}$ & 2.14$\times$ 10$^{-15}$ \\
 & $^{12}$CO v(4-3) R23 & 2.28$\times$ 10$^{-14}$ & 1.86$\times$ 10$^{-15}$ \\
 & $^{12}$CO v(4-3) R19 & 2.38$\times$ 10$^{-14}$ & 3.20$\times$ 10$^{-15}$ \\
 & $^{12}$CO v(4-3) R19 & 2.21$\times$ 10$^{-14}$ & 2.19$\times$ 10$^{-15}$ \\
 & $^{12}$CO v(4-3) R16 & 2.45$\times$ 10$^{-14}$ & 3.45$\times$ 10$^{-15}$ \\
 & $^{12}$CO v(4-3) R15 & 2.44$\times$ 10$^{-14}$ & 3.77$\times$ 10$^{-15}$ \\
 & $^{12}$CO v(4-3) R13 & 2.65$\times$ 10$^{-14}$ & 2.47$\times$ 10$^{-15}$ \\
 & $^{12}$CO v(4-3) R12 & 3.10$\times$ 10$^{-14}$ & 3.53$\times$ 10$^{-15}$ \\
 & $^{12}$CO v(4-3) R11 & 2.99$\times$ 10$^{-14}$ & 3.57$\times$ 10$^{-15}$ \\
 & $^{12}$CO v(4-3) R10 & 2.79$\times$ 10$^{-14}$ & 8.46$\times$ 10$^{-15}$ \\
 & $^{12}$CO v(4-3) R08 & 2.64$\times$ 10$^{-14}$ & 2.71$\times$ 10$^{-15}$ \\
 & $^{12}$CO v(4-3) P17 & 2.41$\times$ 10$^{-14}$ & 3.41$\times$ 10$^{-15}$ \\
 & $^{12}$CO v(4-3) P17 & 2.31$\times$ 10$^{-14}$ & 2.53$\times$ 10$^{-15}$ \\
\hline
HD 100546 & $^{12}$CO v(5-4) R39 & 1.00$\times$ 10$^{-14}$ & 4.68$\times$ 10$^{-15}$ \\
 & $^{12}$CO v(5-4) R35 & 9.75$\times$ 10$^{-15}$ & 3.92$\times$ 10$^{-15}$ \\
 & $^{12}$CO v(5-4) R34 & 6.59$\times$ 10$^{-15}$ & 3.78$\times$ 10$^{-15}$ \\
 & $^{12}$CO v(5-4) R33 & 8.96$\times$ 10$^{-15}$ & 4.76$\times$ 10$^{-15}$ \\
 & $^{12}$CO v(5-4) R33 & 6.06$\times$ 10$^{-15}$ & 3.23$\times$ 10$^{-15}$ \\
 & $^{12}$CO v(5-4) R27 & 1.69$\times$ 10$^{-14}$ & 7.18$\times$ 10$^{-15}$ \\
 & $^{12}$CO v(5-4) P06 & 2.58$\times$ 10$^{-14}$ & 7.45$\times$ 10$^{-15}$ \\
 & $^{12}$CO v(5-4) P07 & 3.15$\times$ 10$^{-14}$ & 5.50$\times$ 10$^{-15}$ \\
 & $^{12}$CO v(5-4) P07 & 1.94$\times$ 10$^{-14}$ & 1.39$\times$ 10$^{-15}$ \\
 & $^{12}$CO v(5-4) P09 & 1.34$\times$ 10$^{-14}$ & 1.50$\times$ 10$^{-15}$ \\
 & $^{12}$CO v(5-4) P09 & 2.07$\times$ 10$^{-14}$ & 1.36$\times$ 10$^{-15}$ \\
\hline
HD 97048 & $^{12}$CO v(1-0) R08 & 1.06$\times$ 10$^{-13}$ & 1.42$\times$ 10$^{-14}$ \\
 & $^{12}$CO v(1-0) R07 & 9.97$\times$ 10$^{-14}$ & 8.45$\times$ 10$^{-15}$ \\
 & $^{12}$CO v(1-0) R07 & 9.53$\times$ 10$^{-14}$ & 7.90$\times$ 10$^{-15}$ \\
 & $^{12}$CO v(1-0) R06 & 1.15$\times$ 10$^{-13}$ & 9.49$\times$ 10$^{-15}$ \\
 & $^{12}$CO v(1-0) R06 & 1.05$\times$ 10$^{-13}$ & 8.57$\times$ 10$^{-15}$ \\
 & $^{12}$CO v(1-0) P05 & 1.21$\times$ 10$^{-13}$ & 4.99$\times$ 10$^{-15}$ \\
 & $^{12}$CO v(1-0) P05 & 1.31$\times$ 10$^{-13}$ & 1.84$\times$ 10$^{-14}$ \\
 & $^{12}$CO v(1-0) P07 & 1.18$\times$ 10$^{-13}$ & 9.91$\times$ 10$^{-15}$ \\
 & $^{12}$CO v(1-0) P07 & 1.01$\times$ 10$^{-13}$ & 8.47$\times$ 10$^{-15}$ \\
 & $^{12}$CO v(1-0) P08 & 1.23$\times$ 10$^{-13}$ & 1.01$\times$ 10$^{-14}$ \\
 & $^{12}$CO v(1-0) P08 & 1.22$\times$ 10$^{-13}$ & 1.02$\times$ 10$^{-14}$ \\
 & $^{12}$CO v(1-0) P09 & 1.07$\times$ 10$^{-13}$ & 8.99$\times$ 10$^{-15}$ \\
 & $^{12}$CO v(1-0) P11 & 1.35$\times$ 10$^{-13}$ & 1.13$\times$ 10$^{-14}$ \\
 & $^{12}$CO v(1-0) P11 & 1.11$\times$ 10$^{-13}$ & 9.33$\times$ 10$^{-15}$ \\
 & $^{12}$CO v(1-0) P12 & 1.40$\times$ 10$^{-13}$ & 7.85$\times$ 10$^{-15}$ \\
 & $^{12}$CO v(1-0) P14 & 1.21$\times$ 10$^{-13}$ & 2.87$\times$ 10$^{-14}$ \\
 & $^{12}$CO v(1-0) P30 & 3.45$\times$ 10$^{-14}$ & 9.24$\times$ 10$^{-15}$ \\
 & $^{12}$CO v(1-0) P32 & 2.60$\times$ 10$^{-14}$ & 6.68$\times$ 10$^{-15}$ \\
 & $^{12}$CO v(1-0) P36 & 1.34$\times$ 10$^{-14}$ & 7.66$\times$ 10$^{-15}$ \\
\hline
HD 97048 & $^{13}$CO v(1-0) R15 & 2.02$\times$ 10$^{-14}$ & 2.07$\times$ 10$^{-15}$ \\
 & $^{13}$CO v(1-0) R13 & 1.91$\times$ 10$^{-14}$ & 1.52$\times$ 10$^{-15}$ \\
 & $^{13}$CO v(1-0) R12 & 2.04$\times$ 10$^{-14}$ & 2.74$\times$ 10$^{-15}$ \\
 & $^{13}$CO v(1-0) R12 & 1.71$\times$ 10$^{-14}$ & 1.68$\times$ 10$^{-15}$ \\
 & $^{13}$CO v(1-0) R09 & 2.35$\times$ 10$^{-14}$ & 3.77$\times$ 10$^{-15}$ \\
 & $^{13}$CO v(1-0) R09 & 1.46$\times$ 10$^{-14}$ & 1.98$\times$ 10$^{-15}$ \\
 & $^{13}$CO v(1-0) R06 & 3.06$\times$ 10$^{-14}$ & 3.51$\times$ 10$^{-15}$ \\
 & $^{13}$CO v(1-0) R05 & 2.93$\times$ 10$^{-14}$ & 2.31$\times$ 10$^{-15}$ \\
 & $^{13}$CO v(1-0) R04 & 1.57$\times$ 10$^{-14}$ & 1.75$\times$ 10$^{-15}$ \\
\hline
HD 97048 & $^{12}$CO v(2-1) R17 & 4.12$\times$ 10$^{-14}$ & 3.39$\times$ 10$^{-15}$ \\
 & $^{12}$CO v(2-1) R14 & 3.97$\times$ 10$^{-14}$ & 3.05$\times$ 10$^{-15}$ \\
\hline
\end{tabular}
\end{center}
\label{default1}
\end{table}%

\begin{table}[htdp]
\caption{Table \ref{tab:default11} continued}
\begin{center}
\begin{tabular}{llll}
Target & line id & Line flux & Error\\
 &  & [erg cm$^{-2}$ s$^{-1}$] & [erg cm$^{-2}$ s$^{-1}$]\\
\hline
 & $^{12}$CO v(2-1) R13 & 4.31$\times$ 10$^{-14}$ & 2.50$\times$ 10$^{-15}$ \\
 & $^{12}$CO v(2-1) R12 & 4.36$\times$ 10$^{-14}$ & 2.27$\times$ 10$^{-15}$ \\
 & $^{12}$CO v(2-1) R11 & 4.05$\times$ 10$^{-14}$ & 2.19$\times$ 10$^{-15}$ \\
 & $^{12}$CO v(2-1) R10 & 3.83$\times$ 10$^{-14}$ & 1.78$\times$ 10$^{-15}$ \\
 & $^{12}$CO v(2-1) R09 & 4.05$\times$ 10$^{-14}$ & 1.62$\times$ 10$^{-15}$ \\
 & $^{12}$CO v(2-1) R06 & 3.85$\times$ 10$^{-14}$ & 1.58$\times$ 10$^{-15}$ \\
 & $^{12}$CO v(2-1) R06 & 3.89$\times$ 10$^{-14}$ & 1.37$\times$ 10$^{-15}$ \\
 & $^{12}$CO v(2-1) R05 & 4.98$\times$ 10$^{-14}$ & 2.39$\times$ 10$^{-15}$ \\
 & $^{12}$CO v(2-1) R04 & 3.91$\times$ 10$^{-14}$ & 2.24$\times$ 10$^{-15}$ \\
 & $^{12}$CO v(2-1) R04 & 4.03$\times$ 10$^{-14}$ & 1.92$\times$ 10$^{-15}$ \\
 & $^{12}$CO v(2-1) P02 & 3.20$\times$ 10$^{-14}$ & 3.59$\times$ 10$^{-15}$ \\
 & $^{12}$CO v(2-1) P02 & 3.03$\times$ 10$^{-14}$ & 4.25$\times$ 10$^{-15}$ \\
 & $^{12}$CO v(2-1) P07 & 4.57$\times$ 10$^{-14}$ & 1.39$\times$ 10$^{-14}$ \\
\hline
HD 97048 & $^{12}$CO v(3-2) R26 & 2.65$\times$ 10$^{-14}$ & 2.12$\times$ 10$^{-15}$ \\
 & $^{12}$CO v(3-2) R22 & 3.06$\times$ 10$^{-14}$ & 2.99$\times$ 10$^{-15}$ \\
 & $^{12}$CO v(3-2) R20 & 2.20$\times$ 10$^{-14}$ & 1.68$\times$ 10$^{-15}$ \\
 & $^{12}$CO v(3-2) R18 & 3.15$\times$ 10$^{-14}$ & 2.81$\times$ 10$^{-15}$ \\
 & $^{12}$CO v(3-2) R18 & 2.77$\times$ 10$^{-14}$ & 2.08$\times$ 10$^{-15}$ \\
 & $^{12}$CO v(3-2) R17 & 2.25$\times$ 10$^{-14}$ & 1.64$\times$ 10$^{-15}$ \\
 & $^{12}$CO v(3-2) R14 & 3.23$\times$ 10$^{-14}$ & 2.46$\times$ 10$^{-15}$ \\
 & $^{12}$CO v(3-2) R14 & 2.25$\times$ 10$^{-14}$ & 5.89$\times$ 10$^{-15}$ \\
 & $^{12}$CO v(3-2) R11 & 3.16$\times$ 10$^{-14}$ & 2.89$\times$ 10$^{-15}$ \\
 & $^{12}$CO v(3-2) R08 & 3.99$\times$ 10$^{-14}$ & 4.93$\times$ 10$^{-15}$ \\
 & $^{12}$CO v(3-2) R08 & 4.68$\times$ 10$^{-14}$ & 3.79$\times$ 10$^{-15}$ \\
 & $^{12}$CO v(3-2) R05 & 1.90$\times$ 10$^{-14}$ & 2.30$\times$ 10$^{-15}$ \\
\hline
HD 97048 & $^{12}$CO v(4-3) R27 & 1.11$\times$ 10$^{-14}$ & 5.01$\times$ 10$^{-15}$ \\
 & $^{12}$CO v(4-3) R23 & 2.04$\times$ 10$^{-14}$ & 10.00$\times$ 10$^{-15}$ \\
 & $^{12}$CO v(4-3) R16 & 3.39$\times$ 10$^{-14}$ & 3.55$\times$ 10$^{-15}$ \\
 & $^{12}$CO v(4-3) R15 & 4.14$\times$ 10$^{-14}$ & 7.01$\times$ 10$^{-15}$ \\
 & $^{12}$CO v(4-3) R13 & 3.16$\times$ 10$^{-14}$ & 3.64$\times$ 10$^{-15}$ \\
 & $^{12}$CO v(4-3) R11 & 2.13$\times$ 10$^{-14}$ & 1.08$\times$ 10$^{-14}$ \\
 & $^{12}$CO v(4-3) R09 & 1.81$\times$ 10$^{-14}$ & 2.56$\times$ 10$^{-15}$ \\
 & $^{12}$CO v(4-3) R08 & 3.70$\times$ 10$^{-14}$ & 3.10$\times$ 10$^{-15}$ \\
 & $^{12}$CO v(4-3) P15 & 1.99$\times$ 10$^{-14}$ & 6.46$\times$ 10$^{-15}$ \\
 & $^{12}$CO v(4-3) P17 & 2.53$\times$ 10$^{-14}$ & 8.85$\times$ 10$^{-15}$ \\
\hline
HD 97048 & $^{12}$CO v(5-4) R33 & 1.16$\times$ 10$^{-14}$ & 4.73$\times$ 10$^{-15}$ \\
 & $^{12}$CO v(5-4) R15 & 2.13$\times$ 10$^{-14}$ & 1.11$\times$ 10$^{-14}$ \\
 & $^{12}$CO v(5-4) P06 & 3.15$\times$ 10$^{-14}$ & 1.63$\times$ 10$^{-14}$ \\
 & $^{12}$CO v(5-4) P09 & 2.60$\times$ 10$^{-14}$ & 1.58$\times$ 10$^{-14}$ \\
\hline
HD 179218 & $^{12}$CO v(1-0) R06 & 2.96$\times$ 10$^{-14}$ & 4.07$\times$ 10$^{-15}$ \\
 & $^{12}$CO v(1-0) R03 & 2.06$\times$ 10$^{-14}$ & 2.91$\times$ 10$^{-15}$ \\
 & $^{12}$CO v(1-0) P10 & 2.45$\times$ 10$^{-14}$ & 3.39$\times$ 10$^{-15}$ \\
 & $^{12}$CO v(1-0) P12 & 3.31$\times$ 10$^{-14}$ & 1.44$\times$ 10$^{-14}$ \\
 & $^{12}$CO v(1-0) P30 & 1.48$\times$ 10$^{-14}$ & 1.13$\times$ 10$^{-15}$ \\
 & $^{12}$CO v(1-0) P32 & 1.17$\times$ 10$^{-14}$ & 9.78$\times$ 10$^{-16}$ \\
 & $^{12}$CO v(1-0) P36 & 8.56$\times$ 10$^{-15}$ & 1.17$\times$ 10$^{-15}$ \\
\hline
HD 179218 & $^{13}$CO v(1-0) R21 & 5.53$\times$ 10$^{-15}$ & 7.27$\times$ 10$^{-16}$ \\
 & $^{13}$CO v(1-0) R15 & 8.43$\times$ 10$^{-15}$ & 1.01$\times$ 10$^{-15}$ \\
 & $^{13}$CO v(1-0) R12 & 1.07$\times$ 10$^{-14}$ & 9.01$\times$ 10$^{-16}$ \\
 & $^{13}$CO v(1-0) R12 & 9.88$\times$ 10$^{-15}$ & 9.16$\times$ 10$^{-16}$ \\
 & $^{13}$CO v(1-0) R03 & 8.04$\times$ 10$^{-15}$ & 2.40$\times$ 10$^{-15}$ \\
 & $^{13}$CO v(1-0) P21 & 6.19$\times$ 10$^{-15}$ & 5.37$\times$ 10$^{-16}$ \\
 & $^{13}$CO v(1-0) P21 & 8.17$\times$ 10$^{-15}$ & 9.10$\times$ 10$^{-16}$ \\
\hline
HD 179218 & $^{12}$CO v(2-1) R14 & 1.16$\times$ 10$^{-14}$ & 7.95$\times$ 10$^{-16}$ \\
 & $^{12}$CO v(2-1) R13 & 1.09$\times$ 10$^{-14}$ & 6.59$\times$ 10$^{-16}$ \\
 & $^{12}$CO v(2-1) R09 & 1.17$\times$ 10$^{-14}$ & 9.08$\times$ 10$^{-16}$ \\
 & $^{12}$CO v(2-1) R08 & 1.08$\times$ 10$^{-14}$ & 8.24$\times$ 10$^{-16}$ \\
 & $^{12}$CO v(2-1) R06 & 1.08$\times$ 10$^{-14}$ & 7.12$\times$ 10$^{-16}$ \\
 & $^{12}$CO v(2-1) R04 & 5.93$\times$ 10$^{-15}$ & 7.27$\times$ 10$^{-16}$ \\
 & $^{12}$CO v(2-1) P25 & 8.30$\times$ 10$^{-15}$ & 1.44$\times$ 10$^{-15}$ \\
\hline
\end{tabular}
\end{center}
\label{default6}
\end{table}%

\begin{table}[htdp]
\caption{Table \ref{tab:default11} continued}
\begin{center}
\begin{tabular}{llll}
Target & line id & Line flux & Error\\
 &  & [erg cm$^{-2}$ s$^{-1}$] & [erg cm$^{-2}$ s$^{-1}$]\\
\hline
 & $^{12}$CO v(2-1) P26 & 1.08$\times$ 10$^{-14}$ & 2.54$\times$ 10$^{-15}$ \\
 & $^{12}$CO v(2-1) P28 & 8.30$\times$ 10$^{-15}$ & 1.01$\times$ 10$^{-15}$ \\
 & $^{12}$CO v(2-1) P28 & 5.80$\times$ 10$^{-15}$ & 6.15$\times$ 10$^{-16}$ \\
\hline
HD 179218 & $^{12}$CO v(3-2) R23 & 9.22$\times$ 10$^{-15}$ & 7.57$\times$ 10$^{-16}$ \\
 & $^{12}$CO v(3-2) R20 & 1.11$\times$ 10$^{-14}$ & 1.90$\times$ 10$^{-15}$ \\
 & $^{12}$CO v(3-2) R20 & 1.03$\times$ 10$^{-14}$ & 8.03$\times$ 10$^{-16}$ \\
 & $^{12}$CO v(3-2) R18 & 8.96$\times$ 10$^{-15}$ & 7.59$\times$ 10$^{-16}$ \\
 & $^{12}$CO v(3-2) R14 & 1.24$\times$ 10$^{-14}$ & 3.19$\times$ 10$^{-15}$ \\
 & $^{12}$CO v(3-2) P21 & 9.09$\times$ 10$^{-15}$ & 8.27$\times$ 10$^{-16}$ \\
 & $^{12}$CO v(3-2) P21 & 9.88$\times$ 10$^{-15}$ & 5.48$\times$ 10$^{-16}$ \\
\hline
HD 179218 & $^{12}$CO v(4-3) R27 & 6.32$\times$ 10$^{-15}$ & 2.80$\times$ 10$^{-15}$ \\
 & $^{12}$CO v(4-3) R26 & 8.04$\times$ 10$^{-15}$ & 3.29$\times$ 10$^{-15}$ \\
 & $^{12}$CO v(4-3) R23 & 8.56$\times$ 10$^{-15}$ & 4.85$\times$ 10$^{-15}$ \\
 & $^{12}$CO v(4-3) R23 & 1.00$\times$ 10$^{-14}$ & 3.72$\times$ 10$^{-15}$ \\
 & $^{12}$CO v(4-3) P17 & 9.62$\times$ 10$^{-15}$ & 1.57$\times$ 10$^{-15}$ \\
 & $^{12}$CO v(4-3) P20 & 7.64$\times$ 10$^{-15}$ & 1.26$\times$ 10$^{-15}$ \\
\hline
HD 141569 & $^{13}$CO v(1-0) R09 & 2.12$\times$ 10$^{-14}$ & 8.18$\times$ 10$^{-15}$ \\
 & $^{13}$CO v(1-0) R04 & 1.91$\times$ 10$^{-14}$ & 7.44$\times$ 10$^{-15}$ \\
 & $^{13}$CO v(1-0) R00 & 2.42$\times$ 10$^{-14}$ & 1.06$\times$ 10$^{-14}$ \\
\hline
HD 141569 & $^{12}$CO v(2-1) R14 & 2.42$\times$ 10$^{-14}$ & 8.32$\times$ 10$^{-15}$ \\
 & $^{12}$CO v(2-1) R11 & 5.85$\times$ 10$^{-14}$ & 1.41$\times$ 10$^{-14}$ \\
 & $^{12}$CO v(2-1) R10 & 5.30$\times$ 10$^{-14}$ & 8.53$\times$ 10$^{-15}$ \\
 & $^{12}$CO v(2-1) R07 & 7.13$\times$ 10$^{-14}$ & 1.01$\times$ 10$^{-14}$ \\
 & $^{12}$CO v(2-1) R06 & 6.46$\times$ 10$^{-14}$ & 9.17$\times$ 10$^{-15}$ \\
 & $^{12}$CO v(2-1) R04 & 8.34$\times$ 10$^{-14}$ & 1.19$\times$ 10$^{-14}$ \\
 & $^{12}$CO v(2-1) R02 & 6.73$\times$ 10$^{-14}$ & 9.88$\times$ 10$^{-15}$ \\
 & $^{12}$CO v(2-1) P01 & 3.52$\times$ 10$^{-14}$ & 8.77$\times$ 10$^{-15}$ \\
\hline
HD 190073 & $^{13}$CO v(1-0) R23 & 1.03$\times$ 10$^{-14}$ & 9.97$\times$ 10$^{-16}$ \\
 & $^{13}$CO v(1-0) R22 & 1.19$\times$ 10$^{-14}$ & 6.65$\times$ 10$^{-16}$ \\
 & $^{13}$CO v(1-0) R21 & 1.40$\times$ 10$^{-14}$ & 9.13$\times$ 10$^{-16}$ \\
 & $^{13}$CO v(1-0) R19 & 1.59$\times$ 10$^{-14}$ & 7.05$\times$ 10$^{-16}$ \\
 & $^{13}$CO v(1-0) R19 & 1.49$\times$ 10$^{-14}$ & 6.02$\times$ 10$^{-16}$ \\
 & $^{13}$CO v(1-0) R18 & 1.67$\times$ 10$^{-14}$ & 8.95$\times$ 10$^{-16}$ \\
 & $^{13}$CO v(1-0) R16 & 1.45$\times$ 10$^{-14}$ & 6.10$\times$ 10$^{-16}$ \\
 & $^{13}$CO v(1-0) R16 & 1.07$\times$ 10$^{-14}$ & 1.72$\times$ 10$^{-15}$ \\
 & $^{13}$CO v(1-0) R15 & 1.63$\times$ 10$^{-14}$ & 7.12$\times$ 10$^{-16}$ \\
 & $^{13}$CO v(1-0) R13 & 2.08$\times$ 10$^{-14}$ & 8.07$\times$ 10$^{-16}$ \\
 & $^{13}$CO v(1-0) R12 & 1.74$\times$ 10$^{-14}$ & 7.12$\times$ 10$^{-16}$ \\
 & $^{13}$CO v(1-0) R12 & 1.99$\times$ 10$^{-14}$ & 8.52$\times$ 10$^{-16}$ \\
 & $^{13}$CO v(1-0) R11 & 1.34$\times$ 10$^{-14}$ & 7.13$\times$ 10$^{-16}$ \\
 & $^{13}$CO v(1-0) R10 & 1.77$\times$ 10$^{-14}$ & 6.58$\times$ 10$^{-16}$ \\
 & $^{13}$CO v(1-0) R10 & 1.58$\times$ 10$^{-14}$ & 5.98$\times$ 10$^{-16}$ \\
 & $^{13}$CO v(1-0) R09 & 1.40$\times$ 10$^{-14}$ & 2.76$\times$ 10$^{-15}$ \\
 & $^{13}$CO v(1-0) R09 & 1.62$\times$ 10$^{-14}$ & 4.26$\times$ 10$^{-15}$ \\
 & $^{13}$CO v(1-0) R06 & 1.11$\times$ 10$^{-14}$ & 1.21$\times$ 10$^{-15}$ \\
 & $^{13}$CO v(1-0) R05 & 1.21$\times$ 10$^{-14}$ & 8.57$\times$ 10$^{-16}$ \\
 & $^{13}$CO v(1-0) R05 & 2.36$\times$ 10$^{-14}$ & 8.52$\times$ 10$^{-15}$ \\
 & $^{13}$CO v(1-0) R03 & 2.36$\times$ 10$^{-14}$ & 1.22$\times$ 10$^{-15}$ \\
 & $^{13}$CO v(1-0) R03 & 1.49$\times$ 10$^{-14}$ & 9.33$\times$ 10$^{-16}$ \\
 & $^{13}$CO v(1-0) P01 & 9.22$\times$ 10$^{-15}$ & 9.67$\times$ 10$^{-16}$ \\
 & $^{13}$CO v(1-0) P03 & 1.09$\times$ 10$^{-14}$ & 1.08$\times$ 10$^{-15}$ \\
 & $^{13}$CO v(1-0) P21 & 2.32$\times$ 10$^{-14}$ & 1.14$\times$ 10$^{-15}$ \\
 & $^{13}$CO v(1-0) P21 & 1.29$\times$ 10$^{-14}$ & 7.43$\times$ 10$^{-16}$ \\
 & $^{13}$CO v(1-0) P23 & 1.57$\times$ 10$^{-14}$ & 7.04$\times$ 10$^{-16}$ \\
 & $^{13}$CO v(1-0) P23 & 1.19$\times$ 10$^{-14}$ & 7.24$\times$ 10$^{-16}$ \\
\hline
HD 190073 & $^{12}$CO v(1-0) R08 & 3.54$\times$ 10$^{-14}$ & 1.79$\times$ 10$^{-15}$ \\
 & $^{12}$CO v(1-0) R06 & 3.62$\times$ 10$^{-14}$ & 1.58$\times$ 10$^{-15}$ \\
 & $^{12}$CO v(1-0) R06 & 4.76$\times$ 10$^{-14}$ & 1.80$\times$ 10$^{-15}$ \\
 & $^{12}$CO v(1-0) R04 & 3.48$\times$ 10$^{-14}$ & 1.62$\times$ 10$^{-15}$ \\
 & $^{12}$CO v(1-0) R04 & 3.86$\times$ 10$^{-14}$ & 1.59$\times$ 10$^{-15}$ \\
\hline
\end{tabular}
\end{center}
\label{default4}
\end{table}%

\begin{table}[htdp]
\caption{Table \ref{tab:default11} continued}
\begin{center}
\begin{tabular}{llll}
Target & line id & Line flux & Error\\
 &  & [erg cm$^{-2}$ s$^{-1}$] & [erg cm$^{-2}$ s$^{-1}$]\\
\hline
 & $^{12}$CO v(1-0) R02 & 3.97$\times$ 10$^{-14}$ & 1.87$\times$ 10$^{-15}$ \\
 & $^{12}$CO v(1-0) R01 & 3.37$\times$ 10$^{-14}$ & 1.31$\times$ 10$^{-15}$ \\
 & $^{12}$CO v(1-0) P02 & 3.98$\times$ 10$^{-14}$ & 1.43$\times$ 10$^{-15}$ \\
 & $^{12}$CO v(1-0) P02 & 5.65$\times$ 10$^{-14}$ & 2.54$\times$ 10$^{-15}$ \\
 & $^{12}$CO v(1-0) P05 & 3.33$\times$ 10$^{-14}$ & 2.88$\times$ 10$^{-15}$ \\
 & $^{12}$CO v(1-0) P06 & 6.21$\times$ 10$^{-14}$ & 3.40$\times$ 10$^{-15}$ \\
 & $^{12}$CO v(1-0) P08 & 4.59$\times$ 10$^{-14}$ & 2.88$\times$ 10$^{-15}$ \\
 & $^{12}$CO v(1-0) P08 & 7.63$\times$ 10$^{-14}$ & 1.21$\times$ 10$^{-14}$ \\
 & $^{12}$CO v(1-0) P09 & 8.74$\times$ 10$^{-14}$ & 1.24$\times$ 10$^{-14}$ \\
 & $^{12}$CO v(1-0) P30 & 5.55$\times$ 10$^{-14}$ & 1.83$\times$ 10$^{-15}$ \\
 & $^{12}$CO v(1-0) P32 & 4.69$\times$ 10$^{-14}$ & 1.62$\times$ 10$^{-15}$ \\
 & $^{12}$CO v(1-0) P32 & 4.33$\times$ 10$^{-14}$ & 1.49$\times$ 10$^{-15}$ \\
 & $^{12}$CO v(1-0) P36 & 3.25$\times$ 10$^{-14}$ & 1.49$\times$ 10$^{-15}$ \\
 & $^{12}$CO v(1-0) P36 & 3.07$\times$ 10$^{-14}$ & 1.33$\times$ 10$^{-15}$ \\
 & $^{12}$CO v(1-0) P37 & 3.40$\times$ 10$^{-14}$ & 1.38$\times$ 10$^{-15}$ \\
 & $^{12}$CO v(1-0) P38 & 3.44$\times$ 10$^{-14}$ & 3.04$\times$ 10$^{-15}$ \\
\hline
HD 190073 & $^{12}$CO v(2-1) R13 & 2.03$\times$ 10$^{-14}$ & 8.11$\times$ 10$^{-16}$ \\
 & $^{12}$CO v(2-1) R12 & 2.11$\times$ 10$^{-14}$ & 8.38$\times$ 10$^{-16}$ \\
 & $^{12}$CO v(2-1) R11 & 2.07$\times$ 10$^{-14}$ & 9.18$\times$ 10$^{-16}$ \\
 & $^{12}$CO v(2-1) R06 & 2.52$\times$ 10$^{-14}$ & 8.43$\times$ 10$^{-16}$ \\
 & $^{12}$CO v(2-1) R06 & 1.87$\times$ 10$^{-14}$ & 7.00$\times$ 10$^{-16}$ \\
 & $^{12}$CO v(2-1) P25 & 1.49$\times$ 10$^{-14}$ & 7.91$\times$ 10$^{-16}$ \\
 & $^{12}$CO v(2-1) P25 & 1.88$\times$ 10$^{-14}$ & 3.43$\times$ 10$^{-15}$ \\
 & $^{12}$CO v(2-1) P27 & 1.82$\times$ 10$^{-14}$ & 9.49$\times$ 10$^{-16}$ \\
 & $^{12}$CO v(2-1) P28 & 1.83$\times$ 10$^{-14}$ & 8.80$\times$ 10$^{-16}$ \\
 & $^{12}$CO v(2-1) P31 & 1.69$\times$ 10$^{-14}$ & 1.44$\times$ 10$^{-15}$ \\
 & $^{12}$CO v(2-1) P31 & 1.99$\times$ 10$^{-14}$ & 5.13$\times$ 10$^{-15}$ \\
 & $^{12}$CO v(2-1) P32 & 1.63$\times$ 10$^{-14}$ & 2.56$\times$ 10$^{-15}$ \\
\hline
HD 98922 & $^{13}$CO v(1-0) R16 & 8.83$\times$ 10$^{-15}$ & 6.84$\times$ 10$^{-16}$ \\
 & $^{13}$CO v(1-0) R15 & 1.00$\times$ 10$^{-14}$ & 5.15$\times$ 10$^{-16}$ \\
 & $^{13}$CO v(1-0) R12 & 1.09$\times$ 10$^{-14}$ & 5.60$\times$ 10$^{-16}$ \\
 & $^{13}$CO v(1-0) R10 & 1.26$\times$ 10$^{-14}$ & 7.33$\times$ 10$^{-16}$ \\
 & $^{13}$CO v(1-0) R05 & 1.49$\times$ 10$^{-14}$ & 2.56$\times$ 10$^{-15}$ \\
 & $^{13}$CO v(1-0) R03 & 1.23$\times$ 10$^{-14}$ & 1.42$\times$ 10$^{-15}$ \\
 & $^{13}$CO v(1-0) P01 & 1.09$\times$ 10$^{-14}$ & 3.01$\times$ 10$^{-15}$ \\
 & $^{13}$CO v(1-0) P21 & 1.30$\times$ 10$^{-14}$ & 4.52$\times$ 10$^{-15}$ \\
 & $^{13}$CO v(1-0) P21 & 8.70$\times$ 10$^{-15}$ & 2.46$\times$ 10$^{-15}$ \\
 & $^{13}$CO v(1-0) P25 & 1.16$\times$ 10$^{-14}$ & 1.98$\times$ 10$^{-15}$ \\
 & $^{13}$CO v(1-0) P25 & 8.56$\times$ 10$^{-15}$ & 2.51$\times$ 10$^{-15}$ \\
 & $^{13}$CO v(1-0) P26 & 8.70$\times$ 10$^{-15}$ & 2.69$\times$ 10$^{-15}$ \\
 & $^{13}$CO v(1-0) P27 & 1.25$\times$ 10$^{-14}$ & 3.69$\times$ 10$^{-15}$ \\
\hline
HD 98922 & $^{12}$CO v(1-0) R07 & 5.96$\times$ 10$^{-14}$ & 1.90$\times$ 10$^{-15}$ \\
 & $^{12}$CO v(1-0) R06 & 5.84$\times$ 10$^{-14}$ & 1.86$\times$ 10$^{-15}$ \\
 & $^{12}$CO v(1-0) R03 & 4.52$\times$ 10$^{-14}$ & 1.92$\times$ 10$^{-15}$ \\
 & $^{12}$CO v(1-0) R02 & 4.80$\times$ 10$^{-14}$ & 1.83$\times$ 10$^{-15}$ \\
 & $^{12}$CO v(1-0) R01 & 4.73$\times$ 10$^{-14}$ & 2.92$\times$ 10$^{-15}$ \\
 & $^{12}$CO v(1-0) P02 & 3.04$\times$ 10$^{-14}$ & 1.03$\times$ 10$^{-15}$ \\
 & $^{12}$CO v(1-0) P05 & 3.87$\times$ 10$^{-14}$ & 1.33$\times$ 10$^{-15}$ \\
 & $^{12}$CO v(1-0) P05 & 3.73$\times$ 10$^{-14}$ & 1.59$\times$ 10$^{-15}$ \\
 & $^{12}$CO v(1-0) P06 & 4.43$\times$ 10$^{-14}$ & 3.04$\times$ 10$^{-15}$ \\
 & $^{12}$CO v(1-0) P08 & 4.88$\times$ 10$^{-14}$ & 1.49$\times$ 10$^{-15}$ \\
 & $^{12}$CO v(1-0) P08 & 4.20$\times$ 10$^{-14}$ & 1.40$\times$ 10$^{-15}$ \\
 & $^{12}$CO v(1-0) P12 & 6.15$\times$ 10$^{-14}$ & 3.50$\times$ 10$^{-15}$ \\
 & $^{12}$CO v(1-0) P30 & 4.16$\times$ 10$^{-14}$ & 1.95$\times$ 10$^{-15}$ \\
 & $^{12}$CO v(1-0) P31 & 3.18$\times$ 10$^{-14}$ & 5.42$\times$ 10$^{-15}$ \\
 & $^{12}$CO v(1-0) P34 & 3.54$\times$ 10$^{-14}$ & 8.92$\times$ 10$^{-15}$ \\
 & $^{12}$CO v(1-0) P36 & 2.16$\times$ 10$^{-14}$ & 9.44$\times$ 10$^{-16}$ \\
 & $^{12}$CO v(1-0) P36 & 3.00$\times$ 10$^{-14}$ & 1.28$\times$ 10$^{-15}$ \\
 & $^{12}$CO v(1-0) P37 & 2.06$\times$ 10$^{-14}$ & 9.63$\times$ 10$^{-16}$ \\
 & $^{12}$CO v(1-0) P38 & 2.33$\times$ 10$^{-14}$ & 1.12$\times$ 10$^{-15}$ \\
\hline
\end{tabular}
\end{center}
\label{default18}
\end{table}%

\begin{table}[htdp]
\caption{Table \ref{tab:default11} continued}
\begin{center}
\begin{tabular}{llll}
Target & line id & Line flux & Error\\
 &  & [erg cm$^{-2}$ s$^{-1}$] & [erg cm$^{-2}$ s$^{-1}$]\\
\hline
HD 98922 & $^{12}$CO v(2-1) R13 & 1.17$\times$ 10$^{-14}$ & 1.96$\times$ 10$^{-15}$ \\
 & $^{12}$CO v(2-1) R06 & 2.25$\times$ 10$^{-14}$ & 10.00$\times$ 10$^{-15}$ \\
 & $^{12}$CO v(2-1) P25 & 1.37$\times$ 10$^{-14}$ & 2.10$\times$ 10$^{-15}$ \\
 & $^{12}$CO v(2-1) P27 & 1.04$\times$ 10$^{-14}$ & 1.60$\times$ 10$^{-15}$ \\
 & $^{12}$CO v(2-1) P28 & 1.33$\times$ 10$^{-14}$ & 4.74$\times$ 10$^{-15}$ \\
 & $^{12}$CO v(2-1) P28 & 1.38$\times$ 10$^{-14}$ & 4.92$\times$ 10$^{-15}$ \\
\hline
HD 95881 & $^{12}$CO v(1-0) R06 & 5.07$\times$ 10$^{-14}$ & 4.01$\times$ 10$^{-15}$ \\
 & $^{12}$CO v(1-0) R05 & 5.48$\times$ 10$^{-14}$ & 2.99$\times$ 10$^{-15}$ \\
 & $^{12}$CO v(1-0) P08 & 5.65$\times$ 10$^{-14}$ & 2.17$\times$ 10$^{-15}$ \\
 & $^{12}$CO v(1-0) P11 & 4.16$\times$ 10$^{-14}$ & 3.30$\times$ 10$^{-15}$ \\
 & $^{12}$CO v(1-0) P32 & 4.07$\times$ 10$^{-14}$ & 1.84$\times$ 10$^{-15}$ \\
 & $^{12}$CO v(1-0) P36 & 1.74$\times$ 10$^{-14}$ & 1.70$\times$ 10$^{-15}$ \\
 & $^{12}$CO v(1-0) P37 & 3.31$\times$ 10$^{-14}$ & 2.12$\times$ 10$^{-15}$ \\
\hline
HD 95881 & $^{13}$CO v(1-0) R16 & 8.56$\times$ 10$^{-15}$ & 1.61$\times$ 10$^{-15}$ \\
 & $^{13}$CO v(1-0) R13 & 1.00$\times$ 10$^{-14}$ & 6.15$\times$ 10$^{-16}$ \\
 & $^{13}$CO v(1-0) R10 & 1.38$\times$ 10$^{-14}$ & 2.12$\times$ 10$^{-15}$ \\
 & $^{13}$CO v(1-0) R10 & 9.88$\times$ 10$^{-15}$ & 1.81$\times$ 10$^{-15}$ \\
 & $^{13}$CO v(1-0) R09 & 1.40$\times$ 10$^{-14}$ & 2.11$\times$ 10$^{-15}$ \\
 & $^{13}$CO v(1-0) R04 & 1.05$\times$ 10$^{-14}$ & 1.89$\times$ 10$^{-15}$ \\
 & $^{13}$CO v(1-0) R03 & 1.12$\times$ 10$^{-14}$ & 2.13$\times$ 10$^{-15}$ \\
 & $^{13}$CO v(1-0) P25 & 5.27$\times$ 10$^{-15}$ & 2.74$\times$ 10$^{-15}$ \\
\hline
R CrA & $^{12}$CO v(1-0) P04 & 4.05$\times$ 10$^{-14}$ & 5.65$\times$ 10$^{-15}$ \\
 & $^{12}$CO v(1-0) P05 & 4.11$\times$ 10$^{-14}$ & 5.95$\times$ 10$^{-15}$ \\
 & $^{12}$CO v(1-0) P05 & 7.02$\times$ 10$^{-14}$ & 3.33$\times$ 10$^{-14}$ \\
 & $^{12}$CO v(1-0) P06 & 4.37$\times$ 10$^{-14}$ & 1.73$\times$ 10$^{-14}$ \\
 & $^{12}$CO v(1-0) P07 & 2.71$\times$ 10$^{-14}$ & 3.21$\times$ 10$^{-15}$ \\
 & $^{12}$CO v(1-0) P07 & 4.52$\times$ 10$^{-14}$ & 1.34$\times$ 10$^{-14}$ \\
 & $^{12}$CO v(1-0) P08 & 3.93$\times$ 10$^{-14}$ & 6.33$\times$ 10$^{-15}$ \\
 & $^{12}$CO v(1-0) P08 & 3.95$\times$ 10$^{-14}$ & 1.62$\times$ 10$^{-14}$ \\
 & $^{12}$CO v(1-0) P11 & 4.77$\times$ 10$^{-14}$ & 4.96$\times$ 10$^{-15}$ \\
 & $^{12}$CO v(1-0) P11 & 4.10$\times$ 10$^{-14}$ & 2.34$\times$ 10$^{-14}$ \\
 & $^{12}$CO v(1-0) P12 & 4.10$\times$ 10$^{-14}$ & 1.49$\times$ 10$^{-14}$ \\
 & $^{12}$CO v(1-0) P13 & 4.51$\times$ 10$^{-14}$ & 7.01$\times$ 10$^{-15}$ \\
 & $^{12}$CO v(1-0) P14 & 4.61$\times$ 10$^{-14}$ & 1.67$\times$ 10$^{-14}$ \\
 & $^{12}$CO v(1-0) P30 & 3.16$\times$ 10$^{-14}$ & 1.22$\times$ 10$^{-14}$ \\
 & $^{12}$CO v(1-0) P31 & 2.60$\times$ 10$^{-14}$ & 7.24$\times$ 10$^{-15}$ \\
 & $^{12}$CO v(1-0) P32 & 1.98$\times$ 10$^{-14}$ & 9.90$\times$ 10$^{-15}$ \\
 & $^{12}$CO v(1-0) P32 & 1.95$\times$ 10$^{-14}$ & 4.27$\times$ 10$^{-15}$ \\
 & $^{12}$CO v(1-0) P34 & 1.66$\times$ 10$^{-14}$ & 4.13$\times$ 10$^{-15}$ \\
 & $^{12}$CO v(1-0) P36 & 1.67$\times$ 10$^{-14}$ & 5.25$\times$ 10$^{-15}$ \\
 & $^{12}$CO v(1-0) P37 & 2.07$\times$ 10$^{-14}$ & 4.88$\times$ 10$^{-15}$ \\
 & $^{12}$CO v(1-0) P37 & 2.11$\times$ 10$^{-14}$ & 5.39$\times$ 10$^{-15}$ \\
\hline
HD 135344B & $^{12}$CO v(1-0) R08 & 6.77$\times$ 10$^{-14}$ & 1.08$\times$ 10$^{-14}$ \\
 & $^{12}$CO v(1-0) R07 & 5.63$\times$ 10$^{-14}$ & 3.47$\times$ 10$^{-15}$ \\
 & $^{12}$CO v(1-0) R07 & 5.24$\times$ 10$^{-14}$ & 1.79$\times$ 10$^{-15}$ \\
 & $^{12}$CO v(1-0) R06 & 4.56$\times$ 10$^{-14}$ & 3.82$\times$ 10$^{-15}$ \\
 & $^{12}$CO v(1-0) R06 & 5.01$\times$ 10$^{-14}$ & 1.86$\times$ 10$^{-15}$ \\
 & $^{12}$CO v(1-0) R05 & 4.85$\times$ 10$^{-14}$ & 2.36$\times$ 10$^{-15}$ \\
 & $^{12}$CO v(1-0) R05 & 4.64$\times$ 10$^{-14}$ & 1.68$\times$ 10$^{-15}$ \\
 & $^{12}$CO v(1-0) R03 & 3.35$\times$ 10$^{-14}$ & 2.46$\times$ 10$^{-15}$ \\
 & $^{12}$CO v(1-0) R02 & 4.14$\times$ 10$^{-14}$ & 2.06$\times$ 10$^{-15}$ \\
 & $^{12}$CO v(1-0) R02 & 4.16$\times$ 10$^{-14}$ & 1.40$\times$ 10$^{-15}$ \\
 & $^{12}$CO v(1-0) R01 & 3.41$\times$ 10$^{-14}$ & 2.22$\times$ 10$^{-15}$ \\
 & $^{12}$CO v(1-0) R01 & 3.28$\times$ 10$^{-14}$ & 1.27$\times$ 10$^{-15}$ \\
 & $^{12}$CO v(1-0) P02 & 2.98$\times$ 10$^{-14}$ & 2.29$\times$ 10$^{-15}$ \\
 & $^{12}$CO v(1-0) P02 & 2.98$\times$ 10$^{-14}$ & 1.12$\times$ 10$^{-15}$ \\
 & $^{12}$CO v(1-0) P04 & 6.98$\times$ 10$^{-14}$ & 4.23$\times$ 10$^{-15}$ \\
 & $^{12}$CO v(1-0) P06 & 5.77$\times$ 10$^{-14}$ & 2.71$\times$ 10$^{-14}$ \\
 & $^{12}$CO v(1-0) P10 & 4.39$\times$ 10$^{-14}$ & 1.25$\times$ 10$^{-14}$ \\
 & $^{12}$CO v(1-0) P10 & 7.66$\times$ 10$^{-14}$ & 4.43$\times$ 10$^{-15}$ \\
\hline
\end{tabular}
\end{center}
\label{default17}
\end{table}%

\begin{table}[htdp]
\caption{Table \ref{tab:default11} continued}
\begin{center}
\begin{tabular}{llll}
Target & line id & Line flux & Error\\
 &  & [erg cm$^{-2}$ s$^{-1}$] & [erg cm$^{-2}$ s$^{-1}$]\\
\hline
 & $^{12}$CO v(1-0) P11 & 5.05$\times$ 10$^{-14}$ & 8.29$\times$ 10$^{-15}$ \\
 & $^{12}$CO v(1-0) P30 & 1.84$\times$ 10$^{-14}$ & 3.00$\times$ 10$^{-15}$ \\
 & $^{12}$CO v(1-0) P31 & 1.81$\times$ 10$^{-14}$ & 3.13$\times$ 10$^{-15}$ \\
 & $^{12}$CO v(1-0) P32 & 2.53$\times$ 10$^{-14}$ & 8.65$\times$ 10$^{-15}$ \\
 & $^{12}$CO v(1-0) P32 & 1.23$\times$ 10$^{-14}$ & 5.11$\times$ 10$^{-15}$ \\
 & $^{12}$CO v(1-0) P36 & 1.69$\times$ 10$^{-14}$ & 6.42$\times$ 10$^{-15}$ \\
 & $^{12}$CO v(1-0) P38 & 1.38$\times$ 10$^{-14}$ & 5.31$\times$ 10$^{-15}$ \\
\hline
HD 150193 & $^{12}$CO v(1-0) R06 & 6.83$\times$ 10$^{-14}$ & 2.58$\times$ 10$^{-14}$ \\
 & $^{12}$CO v(1-0) P08 & 8.22$\times$ 10$^{-14}$ & 1.94$\times$ 10$^{-14}$ \\
 & $^{12}$CO v(1-0) P30 & 9.61$\times$ 10$^{-14}$ & 4.48$\times$ 10$^{-15}$ \\
 & $^{12}$CO v(1-0) P32 & 6.50$\times$ 10$^{-14}$ & 2.52$\times$ 10$^{-15}$ \\
 & $^{12}$CO v(1-0) P32 & 7.06$\times$ 10$^{-14}$ & 3.95$\times$ 10$^{-15}$ \\
 & $^{12}$CO v(1-0) P36 & 5.57$\times$ 10$^{-14}$ & 4.94$\times$ 10$^{-15}$ \\
\hline
HD 104237 & $^{12}$CO v(1-0) R08 & 8.21$\times$ 10$^{-14}$ & 1.52$\times$ 10$^{-14}$ \\
 & $^{12}$CO v(1-0) R07 & 7.76$\times$ 10$^{-14}$ & 9.49$\times$ 10$^{-15}$ \\
 & $^{12}$CO v(1-0) R06 & 5.38$\times$ 10$^{-14}$ & 9.10$\times$ 10$^{-15}$ \\
 & $^{12}$CO v(1-0) R05 & 4.89$\times$ 10$^{-14}$ & 3.08$\times$ 10$^{-14}$ \\
 & $^{12}$CO v(1-0) P06 & 5.71$\times$ 10$^{-14}$ & 8.84$\times$ 10$^{-15}$ \\
 & $^{12}$CO v(1-0) P08 & 8.47$\times$ 10$^{-14}$ & 1.48$\times$ 10$^{-14}$ \\
 & $^{12}$CO v(1-0) P10 & 7.60$\times$ 10$^{-14}$ & 1.88$\times$ 10$^{-14}$ \\
\hline        
\end{tabular}
\end{center}
\label{default-last}
\end{table}%

\end{document}